\documentclass[10pt,superscriptaddress,amsmath,amssymb,aps,prx,longbibliography,twocolumn,footinbib]{revtex4-2}

\usepackage{upgreek}
\usepackage{graphicx}
\usepackage{nccmath}
\usepackage{braket}
\usepackage{bm}
\usepackage{color}
\usepackage{xcolor}
\usepackage[mathlines]{lineno}
\usepackage{dcolumn}
\usepackage{amsthm}
\usepackage{ragged2e}
\usepackage{enumitem}
\usepackage{mathtools}
\usepackage{dsfont}

\usepackage{hyperref}
\hypersetup{colorlinks,breaklinks,
  urlcolor=[rgb]{0,0,0.64},
  linkcolor=[rgb]{0,0,0.64},
  citecolor=[rgb]{0,0,0.64},
  filecolor=[rgb]{0,0,0.64}
}

\newcommand{\Harvard}{Department of Physics, Harvard University, Cambridge, Massachusetts 02138, USA}

\newcommand{\InnsbruckTh}{Institute for Theoretical Physics, University of Innsbruck, 6020 Innsbruck, Austria}
\newcommand{\InnsbruckQO}{Institute for Quantum Optics and Quantum Information of the Austrian Academy of Sciences, 6020 Innsbruck, Austria}

\newcommand{\QuEra}{QuEra Computing Inc., 1284 Soldiers Field Road, Boston, MA, 02135, USA}

\newcommand{\ITAMP}{ITAMP, Harvard-Smithsonian Center for Astrophysics, Cambridge, MA 02138, USA}

\newcommand{\PlanQC}{PlanQC GmbH, 85748 Garching, Germany}

\newcommand{\CSIC}{Instituto de F\'isica Te\'orica UAM-CSIC, C. Nicol\'as Cabrera 13-15, Cantoblanco, 28049 Madrid, Spain}

\usepackage{mystyle}

\newcommand{\sidefigcaption}[2][]{%
  \refstepcounter{figure}%
  \ifx\relax#1\relax\else\label{#1}\fi
  {\small\textbf{FIG.~\thefigure.}~#2}%
}

\newcommand{\tr}{\,\mathrm{tr}}

\newcommand{\beginsupplement}{%
  \setcounter{table}{0}
  \renewcommand{\thetable}{S\arabic{table}}%
  \setcounter{figure}{0}
  \renewcommand{\thefigure}{S\arabic{figure}}%
  \setcounter{equation}{0}
  \renewcommand{\theequation}{S\arabic{equation}}%
}

\begin{document}

\title{Inverse Quantum Simulation for Quantum Material Design}

\begin{abstract}
    Quantum simulation provides a powerful route for exploring  many-body phenomena beyond the capabilities of classical computation. Existing approaches typically proceed in the forward direction: a model Hamiltonian is specified, implemented on a programmable quantum platform, and its phase diagram and properties are explored. 
    Here we present a quantum algorithmic 
    framework for inverse quantum simulation, enabling quantum material design with desired properties.
    Target material characteristics are encoded as a cost function, which is minimized on quantum hardware to prepare a many-body state with the desired properties in quantum memory. 
    Hamiltonian learning is then used to reconstruct a low-energy Hamiltonian for which this state is an approximate ground state, yielding a physically interpretable model that can guide experimental synthesis. 
    As illustrative applications, we outline how the method can be used to search for high-temperature superconductors within the fermionic Hubbard model, enhancing $d$-wave correlations over a broad range of dopings and temperatures, design 
    quantum phases  by stabilizing a topological order through continuous Hamiltonian modifications, and optimize dynamical properties relevant for photochemistry and frequency- and momentum-resolved condensed-matter data.
    These results extend the scope of quantum simulators from exploring quantum many-body systems to designing and discovering new quantum materials.
\end{abstract}

\author{Christian~Kokail}
\thanks{These authors contributed equally to this work\\
\href{mailto:christian.kokail@cfa.harvard.edu}{christian.kokail@cfa.harvard.edu}\\
\href{mailto:p_dolgirev@g.harvard.edu}{p\_dolgirev@g.harvard.edu}}
\affiliation{\ITAMP}
\affiliation{\Harvard}
\affiliation{\QuEra}

\author{Pavel~E.~Dolgirev}
\thanks{These authors contributed equally to this work\\
\href{mailto:christian.kokail@cfa.harvard.edu}{christian.kokail@cfa.harvard.edu}\\
\href{mailto:p_dolgirev@g.harvard.edu}{p\_dolgirev@g.harvard.edu}}
\affiliation{\Harvard}
\affiliation{\QuEra}

\author{Rick~van~Bijnen}
\affiliation{\InnsbruckTh}
\affiliation{\InnsbruckQO}
\affiliation{\PlanQC}

\author{Daniel Gonz\'alez-Cuadra}
\affiliation{\Harvard}
\affiliation{\CSIC}

\author{Mikhail~D.~Lukin}
\affiliation{\Harvard}

\author{Peter~Zoller}
\affiliation{\InnsbruckTh}
\affiliation{\InnsbruckQO}

\date{\today}

\maketitle

\begin{figure*}
  \centering
  \begin{minipage}[t]{0.6\textwidth}
    \vspace*{0pt}
    \includegraphics[width=\linewidth]{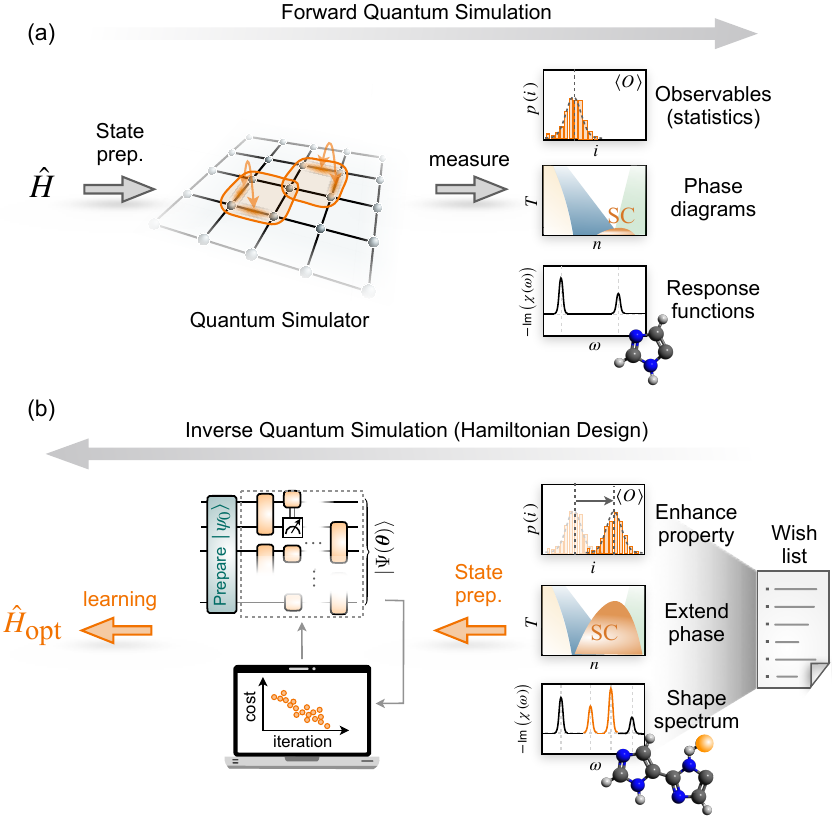}
  \end{minipage}\hspace{1em}
  \begin{minipage}[t]{0.25\textwidth}
    \vspace*{0pt}
    \justifying
    \sidefigcaption[fig::Fig1]{\textbf{Designing many-body Hamiltonians with tailored target properties.}
    (a) Left-to-right: Conventional forward quantum simulation, which proceeds from a model Hamiltonian to ground-state preparation on quantum hardware, followed by measurements of static or dynamical properties.
    (b) Right-to-left: Inverse approach, where a list of desired target properties is encoded into a cost function minimized via a variational quantum circuit $\ket{\psi(\boldsymbol{\theta})}$.
    Application of Hamiltonian learning yields a many-body Hamiltonian $\hat{H}_{\rm opt}$ that has $\ket{\psi(\boldsymbol{\theta})}$ as its approximate ground state and thus realizes the prescribed material preferences.
    }\label{fig::Fig1}
  \end{minipage}
\end{figure*}

\section*{Introduction}
Quantum simulation has emerged as a central application of quantum information processing, enabling exploration of many-body phenomena beyond the reach of classical computation. In forward simulation [Fig.~\ref{fig::Fig1}(a)], one specifies a Hamiltonian \(\hat{H}\), prepares its ground state or induces nonequilibrium dynamics via a quench, and measures the resulting observables. This paradigm has delivered profound insights into strongly correlated phases, thermalization processes, and topological order across a wide range of experimental platforms, including neutral atoms~\cite{Bernien2017, Semeghini2021, Ebadi256Nature, Léséleuc2019, HartkeScience23, ImpertroNatPhys2025, MuqingNature2025, juliàfarré2025hybridquantumclassicalanalogsimulation}, trapped ions~\cite{RevModPhys.93.025001, JoshiKokailNature2023, IqbalNature2024}, and superconducting qubits~\cite{google2025observation, Satzinger2021}.

In the context of quantum material design, quantum chemistry or drug discovery, however, forward simulation faces intrinsic limitations.  
Physically realizing complex interacting Hamiltonians and preparing their ground states can be resource-intensive, with no guarantee that the resulting phases will exhibit the intended characteristics.
Efforts toward \textit{inverse design} -- starting from desired material properties and finding parent Hamiltonians -- on classical computers have leveraged modern optimization and learning 
algorithms~\cite{aspuru2018, ma15051811, Inui2023}, but remain constrained by the exponential scaling of correlated quantum systems. 
For instance, faithfully representing the underlying quantum state in quantum materials whose behavior is dominated by strong fermionic correlations is computationally prohibitive \cite{PhysRevLett.94.170201, PhysRevLett.98.140506}, calling for an intrinsically quantum approach.

Here we introduce an \textit{inverse quantum simulation} (IQS) framework [Fig.~\ref{fig::Fig1}(b)] that directly addresses these challenges, extending the scope of quantum simulation from exploring known models to \textit{designing} quantum materials with prescribed properties. The procedure begins by encoding the targeted properties -- a wish list of desired characteristics -- into a cost functional \(\mathcal{C}[\psi]\) [see Fig.~\ref{fig::Fig1}(b), right], defined over the many-body state \(\ket{\psi}\). This functional is minimized on a programmable quantum device, yielding an optimal state retained within the device’s quantum memory. Several quantum-native optimization strategies~\cite{Volpe2025QuantumOptimizationReview}, including variational approaches, are compatible with this step.

From the optimized state, the protocol proceeds by employing methods from quantum learning theory \cite{gebhartreview} to infer a \emph{geometrically local parent Hamiltonian} \(\hat{H}_{\mathrm{opt}}\) [Methods]. The optimization of \(\mathcal{C}\) is constrained such that the resulting state approximates the ground state of \(\hat{H}_{\mathrm{opt}}\), thereby defining a corresponding low-temperature quantum phase and its properties. The framework leverages the capacity of quantum hardware to generate and sample entangled many-body states beyond classical reach \cite{BergamaschiProc2024}, providing direct access to correlation functions essential for faithful Hamiltonian reconstruction. Importantly, the learned Hamiltonian need not be unique, allowing for multiple distinct physical realizations that reproduce the desired material characteristics.
We illustrate this method using three distinct many-body examples [see Methods for further detailed discussion of the IQS protocol]: (i) designing extensions to the Hubbard Hamiltonian that enhance $d$-wave pairing in fermionic lattice systems, (ii) expanding topological phases through learned continuous Hamiltonian deformations, and (iii) constructing Hamiltonians that generate tailored frequency-resolved features relevant for molecular design and condensed-matter inference. In these examples, we show that IQS applies across both programmable analog and digital quantum platforms \cite{alam2025_1, alam2025_2, GonzalezPNAS}, including early fault-tolerant architectures \cite{bluvstein2025architectural} where deep circuits exploring the landscape of the cost function $\cal C$ can traverse phase transitions and access genuine quantum phases. 

By translating a list of target properties into a physical Hamiltonian, IQS provides a blueprint for quantum material design, by offering a direct route from abstract material requirements to experimentally relevant synthesis guidelines. In this way, programmable quantum simulators may evolve from tools for simulating existing materials, to devices geared at discovering entirely new ones -- potentially addressing important challenges such as identifying novel high-temperature superconductors.

\begin{figure*}[t!]
    \centering
    \includegraphics[width=0.9\linewidth]{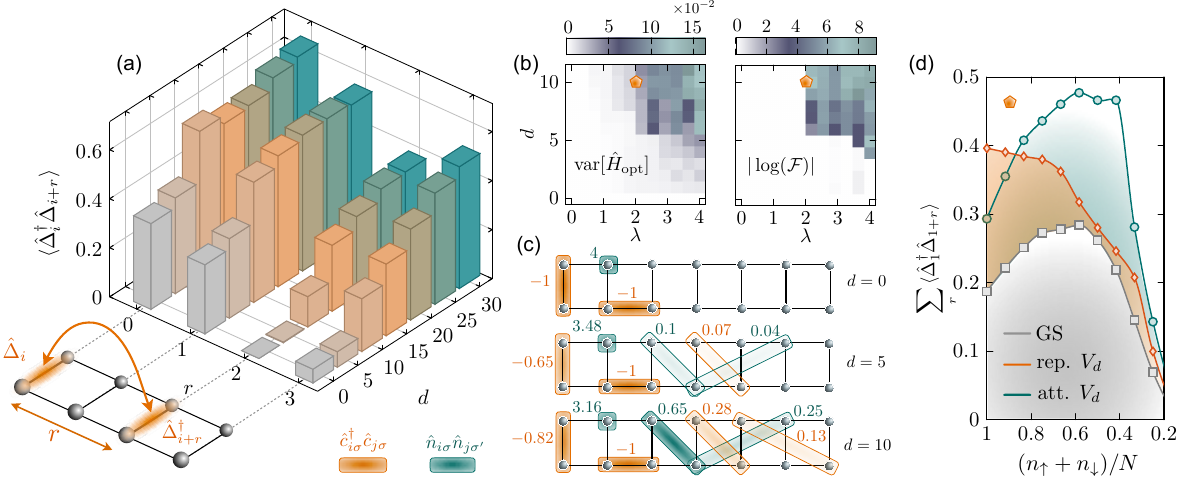}
    \caption{ \textbf{Designing a fermionic Hamiltonian with enhanced $d$-wave correlations.}  
    (a) Spatial profile of the $d$-wave correlation function $\langle \hat{\Delta}^\dagger_{1}\hat{\Delta}_{1 + r}\rangle$ in a four-site Hubbard ladder for circuits optimized at various depths $d$, where $\hat\Delta_i$ denotes the singlet annihilation operator on the rung at site $i$.
    Deeper circuits, targeting $-\sum_{r} \langle \hat{\Delta}^\dagger_{1}\hat{\Delta}_{1 + r}\rangle$, naturally yield stronger $d$-wave correlations.
    (b) Learning maps. 
    Shown are the variance (left) of the learned Hamiltonian $\hat{H}_{\rm opt}$ and the fidelity (right) with respect to the ground state of $\hat{H}_{\rm opt}$, across $(\lambda,d)$. 
     The cost function is $C[\lambda] = \langle \hat{H}_0\rangle - \lambda \sum_{r} \langle \hat{\Delta}^\dagger_{1}\hat{\Delta}_{1 + r}\rangle$, where $\hat{H}_0$ is the reference Hubbard model with $t_x=t_y=-1$ (defining the energy unit) and $U=4$; see panel (c). 
     Learning degrades at large $\lambda$ or $d$ as the Hamiltonian ansatz lacks sufficient non-locality; extending it with additional non-local terms can restore performance. 
    (c) Learned translationally invariant Hamiltonians for $d=5$ and $d=10$. 
    Off-diagonal hoppings and off-site density–density interactions are identified as the dominant terms enhancing $d$-wave correlations; here, $\lambda=2$ and the interactions are restricted to be repulsive.
    (d) $d$-wave correlations remain enhanced across the full doping range of a $2\times 24$-site ladder (relative to the ground state of $\hat{H}_0$), despite using the learned Hamiltonian $\hat{H}_{\rm opt}$ (pentagon symbols in (b)) obtained on a four-site ladder at quarter filling.  
    Enforcing all off-site density–density interactions to be repulsive (orange) still yields appreciable enhancement; allowing them to be attractive (green) can lead to a stronger effect.
    }
    \label{fig::Fig2}
\end{figure*}

\section*{Enhancing $d$-wave pair correlations in fermionic lattice models}
As a first demonstration, we apply IQS to systems operating on fermions, such as neutral-atom optical lattices, which naturally realize 2D Hubbard-type models.
Throughout this section, we focus on implementations with native fermions, while noting that the same techniques apply directly to digital qubit platforms via efficient fermion-to-qubit mappings~\cite{maskara2025fastsimulationfermionsreconfigurable}, which are particularly relevant for deep circuits.
The corresponding Hubbard Hamiltonian is given by 
\begin{align} \label{eqn::main_HU}
    \hat{H}_0 =  -\sum_{\braket{ij}} t \left( \hat{c}_{i \sigma}^{\dagger} \hat{c}_{j \sigma} + \text{H.c.} \right) + U \sum_i \hat{n}_{i \uparrow} \hat{n}_{i \downarrow},
\end{align}
where $\hat{c}_{i\sigma}$ annihilates a fermion at site $i$ and spin $\sigma\in\{\uparrow, \downarrow\}$, $t$ denotes the (nearest-neighbor) hopping amplitudes, and $\hat{n}_{i\sigma}\equiv\hat{c}^\dagger_{i\sigma}\hat{c}_{i\sigma}$.
This paradigmatic model is believed to capture the essential physics of high-$T_c$ cuprate superconductors \cite{PhysRevLett.110.216405}, yet the corresponding superconducting phase has remained experimentally elusive -- most likely because the accessible temperatures remain far above those required for robust fermion pairing.
Within the IQS framework, a natural objective is to identify a Hamiltonian -- closely related to Eq.~\eqref{eqn::main_HU} and composed of experimentally native terms such as extended hoppings and density–density interactions -- in which superconducting correlations are significantly enhanced and thus more readily observable.
Beyond facilitating experimental efforts, pursuing this objective advances the broader goal of searching for and understanding high-$T_c$ superconductors.

To explore this idea, we employ a shallow-depth variational optimization scheme, in which a short sequence of Hubbard-like quench unitaries (see also Ref.~\cite{3nx4-bnyy}) is applied to the ground state $\ket{\psi_0}$ of $\hat H_0$ to minimize a suitably defined cost function [Fig.~\ref{fig::Fig1}(b) and Methods]. 
The experimental preparation of $\ket{\psi_0}$ would mark a defining application of quantum hardware, providing access to classically intractable fermionic many-body states; such states have not yet been realized, and, in practice, accessible Gibbs states may provide a viable alternative within the IQS protocol, as demonstrated below.

Given the computational intractability of simulating 2D Hubbard models~\cite{jiang2019superconductivity, xu2024coexistence}, we focus on two-leg ladder systems that provide a controlled setting for our numerical analyses while retaining essential features of $d$-wave superconductivity~\cite{NOACK1996281, PhysRevB.92.195139, PhysRevB.83.054508}; moreover, such analyses are directly relevant to ladder cuprates~\cite{chen2021anomalously,PhysRevX.15.021049,scheie2025cooper}.
In this context, the analog of $d$-wave pairing is probed through the pseudo pair correlation function $\langle \hat{\Delta}^\dagger_{i}\hat{\Delta}_{j}\rangle$, where $\hat{\Delta}_i = \hat{c}_{(i,1),\uparrow}\hat{c}_{(i,2),\downarrow} - \hat{c}_{(i,1),\downarrow}\hat{c}_{(i,2),\uparrow}$ annihilates a singlet on rung $i$ (the second index $\lambda \in \{1,2\}$ in $(i,\lambda)$ labels the ladder leg).

Based on these definitions, a natural first step is to define a cost function that directly amplifies \(d\)-wave correlations by promoting the delocalization of fermionic singlets,  
\(\mathcal{C}(\bm{\theta}) = - \sum_{r} \langle \hat{\Delta}^\dagger_{1}\hat{\Delta}_{1+r} \rangle_{\bm{\theta}}\).  
While circuit optimization indeed amplifies superconducting behavior and improves with increasing circuit depth $d$ [Fig.~\ref{fig::Fig2}(a)], the corresponding Hamiltonians inferred from the optimized states become progressively nonlocal,
motivating the introduction of a regularization term.
To steer the reconstruction towards the original Hubbard Hamiltonian $\hat{H}_0$, while utilizing the fact that $\ket{\psi_0}$ is its ground state, we modify the cost function to minimize
\begin{align}
    \mathcal{C}[\bm\theta;\lambda]
    = \langle \hat{H}_0 \rangle_{\bm\theta}
    - \lambda \sum_{r} \langle \hat{\Delta}^\dagger_{1}\hat{\Delta}_{1+r} \rangle_{\bm\theta}.
\end{align}
To assess learning performance, we quantify the fidelity of the optimized states as well as the variance of the reconstructed Hamiltonians, 
\(\mathrm{var}[\hat{H}_{\rm opt}] = \langle \hat{H}_{\rm opt}^2\rangle_{\bm\theta} - \langle \hat{H}_{\rm opt}\rangle_{\bm\theta}^2\). 
Scanning these quantities over \((\lambda,d)\) yields \emph{learning maps} [Fig.~\ref{fig::Fig2}(b)] that reveal the trade-off between \(d\)-wave enhancement and locality.
Large \(\lambda\) or \(d\) produce strong pairing but require increasingly nonlocal ansatz terms \(\hat{h}_i\), 
whereas small \(\lambda\) or \(d\) keep \(\hat{H}_{\rm opt}\) Hubbard-like with only modest enhancement.
Extending the ansatz to include longer-range terms systematically improves performance and shifts the poor-learning region in Fig.~\ref{fig::Fig2}(b) to larger \(\lambda\) and \(d\) [SI Sec.~\ref{sec::HL_additional}].

Our central result [Fig.~\ref{fig::Fig2}(c)] identifies the Hamiltonian terms that enhance superconducting-like behavior when density-density interactions are constrained to be strictly repulsive [Methods].  
Remarkably, even in this repulsive regime we observe a progressive enhancement of $d$-wave-like correlations, providing direct evidence that repulsion alone can promote superconductivity.  
In this setting, diagonal density-density interactions proved more relevant than off-diagonal hoppings, whose relative sign with respect to the bare hoppings is negative, consistent with previous studies of the Hubbard model \cite{PhysRevLett.87.047003}.  
Adding further nonlocal interactions amplifies $d$-wave correlations [Fig.~\ref{fig::Fig2}(c)], a particularly exciting result for cavity heterojunctions, which naturally generate such non-localities but whose role for superconductivity remains unsettled.  
By contrast, when density-density interactions are allowed to be attractive (see also recent experimental work on ladder cuprates Ref.~\cite{chen2021anomalously,PhysRevX.15.021049,scheie2025cooper}), the enhancement of superconducting behavior is readily explained by the delocalization of bound singlets. In this case, hopping terms become significant, together with off-site density interactions such as $\sim\hat{n}_{i,1}\hat{n}_{i,2}$.  

To further demonstrate superconducting phase extension, we apply large-scale DMRG simulations to $2\times24$ ladders using the Hamiltonian learned at quarter filling on the eight-site case [diamond in Fig.~\ref{fig::Fig2}(b)].  
This analysis shows that $d$-wave correlations are enhanced across the full doping range [Fig.~\ref{fig::Fig2}(d)], and can be further strengthened when attractive off-site density–density interactions are included.

While the ground state \(\ket{\psi_0}\) has yet to be realized experimentally, accessible high-temperature Gibbs states already display strong correlations, as evidenced by recent observations of the onset of antiferromagnetic order and stripe phases.
When applied to such states [SI Sec.~\ref{sec::finite_T}], the IQS framework shows that circuit optimization even at a relatively high temperature [$T \simeq 0.32t$] yields \(d\)-wave correlations comparable to those of the zero-temperature Hubbard model $\hat{H}_0$, while the corresponding learned Hamiltonian outperforms \(\hat{H}_0\) across the entire temperature range [SI Sec.~\ref{sec::HL_additional}]. 
These results indicate that a co-design strategy linking the IQS protocol and available quantum hardware is readily suitable for modern fermionic quantum simulators [see SI Sec.~\ref{sec::HL_additional} for further discussion of noise estimates].

\begin{figure*}[t!]
    \centering
    \includegraphics[width=0.85\linewidth]{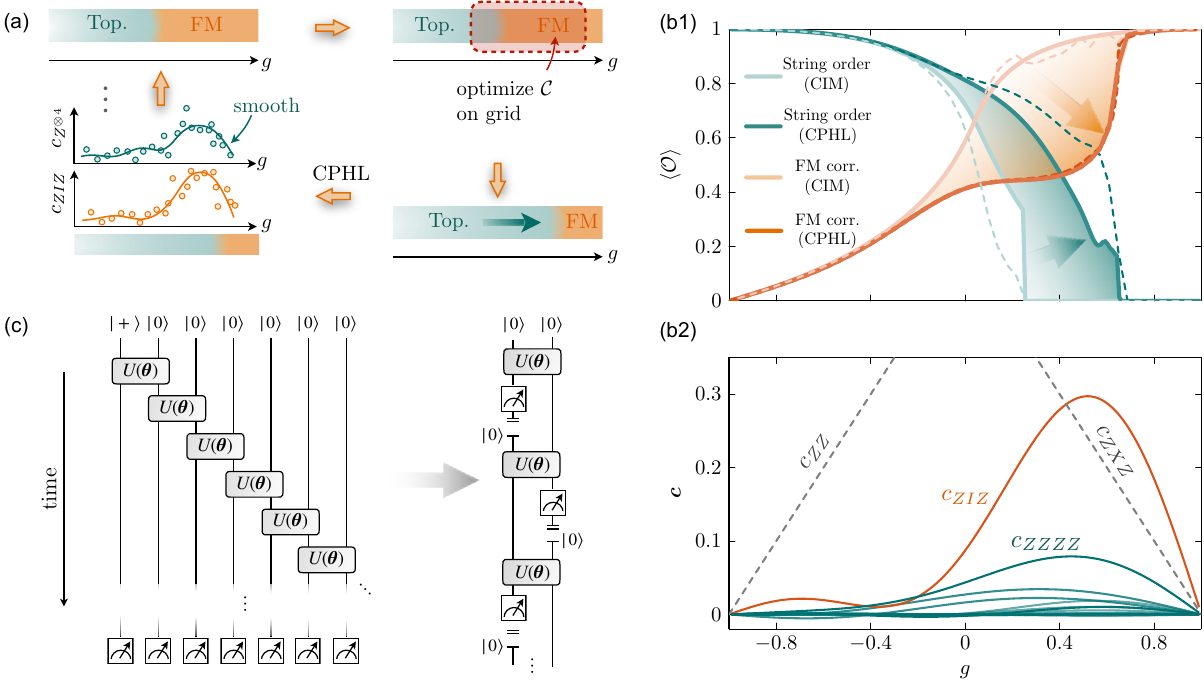}
    \caption{ \textbf{ Quantum phase design.} (a) Schematic of the CPHL algorithm.
    Starting from the phase diagram of a many-body system, the goal is to extend a target (possibly fragile) quantum phase.
    A region in parameter space is selected and discretized into a grid of points.
    At each point, a variational circuit is optimized with a tailored cost function designed to enhance the target phase.
    From the optimized circuits, Hamiltonian coefficients are learned for each grid point and subsequently smoothed into a continuous manifold.
    The phase diagram is updated with the resulting continuous Hamiltonian family, and the procedure is iterated until convergence. 
    (b1) Application to the CIM in Eq.~\eqref{eqn::main_CIM}, a minimal 1D setting exhibiting a topological phase transition as the tuning parameter $g$ is varied from the topological phase ($g = -1$) to the ferromagnetic phase ($g = 1$). The CPHL algorithm successfully extends the topological phase at the expense of competing ferromagnetic correlations. 
    (b2) The key stabilizing terms are identified as the AFM next-nearest-neighbor coupling $ZIZ$ and quartic interaction $ZZZZ$, both of which suppress ferromagnetism.
    (c) Variational staircase circuit with tunable two-qubit unitary $U(\boldsymbol{\theta})$ captures both ends of the CIM phase diagram.
    The circuit can be naturally implemented using periodic teleportation and qubit reset.
    Dashed lines in panel (b1) show results from circuit simulations optimized to maximize fidelities with the exact ground states, indicating that the topological phase extension could, in principle, be achieved through the staircase circuit alone.
     }
    \label{fig::Fig_cphl}
\end{figure*}

\begin{figure*}
    \centering
    \includegraphics[width=0.8\linewidth]{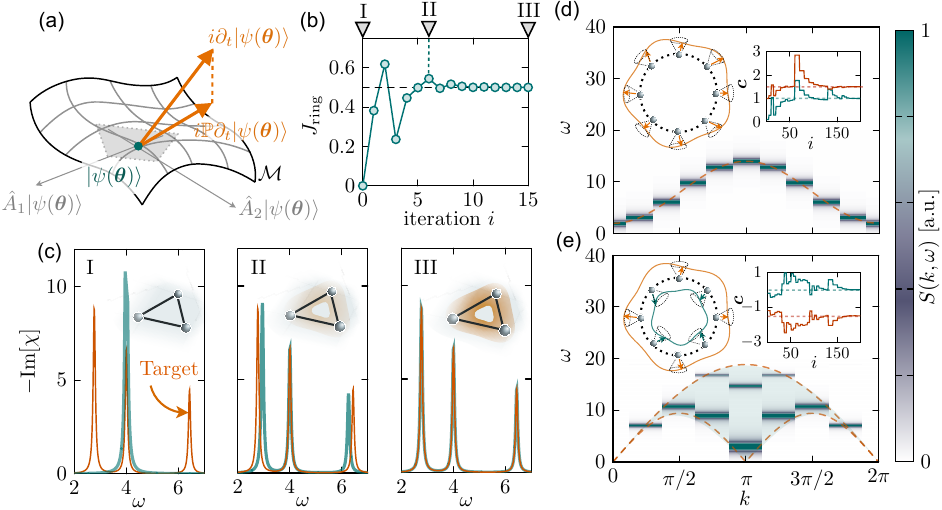}
    \caption{ \textbf{Design of linear dynamical properties.} 
    (a) Tangent space of the variational manifold ${\cal M}$ capturing weak time-dependent perturbations around $\ket{\psi(\boldsymbol{\theta})}$, used to evaluate frequency-resolved dynamical cost functions. On quantum hardware, this reduces to preparing $\ket{\psi(\boldsymbol{\theta})}$ and measuring a fixed set of Hermitian operators (see Methods).  
    (b)-(c) Benchmark with three spins-$1/2$, where the task is to learn a single-parameter Hamiltonian containing the ring-exchange interaction $J_{\rm ring}$ from a target response function (red spectrum in (c)).
    Starting from an initial guess with a single (incorrect) peak (green spectrum in (c)), the algorithm rapidly converges (b) to the correct three-peak structure and recovers the target $J_{\rm ring}$ (dashed line).
    (d)-(e) Applications to frequency- and momentum-resolved condensed-matter data. (d) The dynamical spin structure factor of a ferromagnetic chain, ${\cal S}_{\rm FM}(\omega, k)$, exhibits sharp peaks at the magnon (left inset) dispersion. 
    Given only this spectrum, the algorithm recovers the original two-parameter ferromagnetic Hamiltonian (right inset). 
    (e) For the AFM Heisenberg chain, the dynamical structure factor, ${\cal S}_{\rm AFM}(\omega, k)$, exhibits a continuum of excitations (shaded region) bounded by lower and upper branches (dashed orange lines). 
    For a finite chain, ${\cal S}_{\rm AFM}$ is discrete but still fills the region between the branches. 
    Learning the correct Hamiltonian (right inset) from such spectra provides a stringent benchmark of the framework. 
    }
    \label{fig::Figure_dyn}
\end{figure*}

\section*{Designing quantum phases of matter}
In the preceding example, IQS was applied using shallow circuits generated by Hubbard-type quench dynamics to enhance $d$-wave-like correlations at a fixed point in the phase diagram, i.e., at fixed doping and temperature. IQS based on shallow circuits is natural when the initial state already resides in the vicinity of the desired phase, allowing short-depth evolution to amplify its characteristic correlations and to explore the immediate neighborhood of the initial Hamiltonian $\hat{H}_0$. Here we address a particularly relevant situation arising in quantum material design in which the initial state belongs to a different quantum phase than the intended target state. In this case, state preparation requires circuits that can generate long-range entanglement and drive the system across phase boundaries, for instance via deep circuits, the use of long-range interactions, or measurement and feed-forward. 

This perspective motivates a broader goal for IQS to \textit{continuously} extend a target -- possibly fragile -- quantum phase of matter.
We note that Hamiltonians learned at different points of the phase diagram are not necessarily smoothly connected; however, IQS can be generalized to take into account not just a single point but rather a continuous region in the phase diagram to enhance the desired property across it and to infer a continuous Hamiltonian family of the form: 
\begin{align}
    \hat{H}[\bm g] = \hat{H}_0[\bm g] + \sum_i c_i(\bm g)\, \hat{h}_i ,
    \label{eqn::main_cphl_gen}
\end{align}
where $\bm g$ denotes the set of tunable parameters (e.g., applied magnetic or electric field, density, interaction strength, etc.), $\hat{H}_0[\bm g]$ is the bare Hamiltonian generating the original phase diagram, and the learned coefficients $c_i(\bm{g})$ are constrained to vary continuously across the region. 
To this end, we introduce a grid covering the chosen region and optimize \emph{deep} variational circuits capable of crossing topological phase transitions [see Fig.~\ref{fig::Fig_cphl}(c)] at each point [Fig.~\ref{fig::Fig_cphl}(a)].
A natural choice of cost function is
\begin{align}
    \mathcal{C}[\bm \theta; \bm g; \lambda] = \langle \hat{H}[{\bm g}] \rangle_{\bm \theta }
    - \lambda \langle \hat{\mathcal{O}} \rangle_{\bm \theta },
\end{align}
where $\hat{\mathcal{O}}$ represents, for instance, the order parameter of the target phase and $\lambda$ is an optimization hyperparameter.
Learning on this grid yields a discrete set of coefficients, which we approximate by a continuous manifold. 
This smoothing step improves learning robustness by eliminating irregularities from individual grid points. 
The resulting continuous Hamiltonian then serves as the updated starting point, and the procedure is iterated until convergence [Fig.~\ref{fig::Fig_cphl}(a); see also SI Sec.~\ref{sup::cphl}]. 
We refer to this extension as \emph{continuous-phase Hamiltonian learning} (CPHL).

As an illustration, we apply CPHL to the cluster Ising model (CIM) \cite{PhysRevA.84.022304}, 
a 1D system of $N$ spin-$\tfrac{1}{2}$ particles with Hamiltonian
\begin{align}
    \hat{H}_0[g] = -  \sum_{i }\Big[ \frac{1 - g}{2} \hat{Z}_{i-1} \hat{X}_i \hat{Z}_{i + 1} 
    + \frac{1 + g}{2} \hat{Z}_i \hat{Z}_{i + 1}\Big],
    \label{eqn::main_CIM}
\end{align}
where $\hat{X}_i,\, \hat{Y}_i, \, \hat{Z}_i$ are Pauli operators at site $i$. 
CIM provides a minimal setting that hosts a topological phase at $g=-1$, where the ground state is the cluster state maximizing the non-local string order parameter 
$\hat{\cal O} = (-1)^N \hat{Z}_1 \hat{Y}_2 \Big[ \prod_{i = 3}^{N-2} \hat{X}_i \Big] \hat{Y}_{N-1} \hat{Z}_N$, 
and a ferromagnetic phase at $g=1$, for which $\langle \hat{\cal O} \rangle = 0$ [Fig.~\ref{fig::Fig_cphl}(b1)].
CPHL learns continuous Hamiltonian modifications as in Eq.\eqref{eqn::main_cphl_gen}, with the additional constraint $c_i(\pm 1)=0$ to fix the reference points $g = \pm 1$, which appreciably extends the topological phase [Fig.\ref{fig::Fig_cphl}(b1)].
The most relevant learned terms are couplings that suppress ferromagnetism -- most prominently the antiferromagnetic next-nearest-neighbor interaction $\hat{Z}\hat{I}\hat{Z}$ and quartic interaction $\hat{Z}\hat{Z}\hat{Z}\hat{Z}$ --
while leaving topological correlations essentially intact, thereby stabilizing the topological phase [Fig.~\ref{fig::Fig_cphl}(b2) and SI Sec.~\ref{sup::cphl}]. 
The learning procedure is robust, interpretable, and weakly dependent on the system size $N$ [SI Sec.~\ref{sup::cphl}]. 

Both ends of the CIM phase diagram -- the cluster and ferromagnetic states -- can be represented by depth-$N$ staircase circuits \cite{PhysRevResearch.4.L022020}, as illustrated in Fig.~\ref{fig::Fig_cphl}(c). These circuits can generate long-range entanglement across the entire chain and, when suitably parameterized, can traverse topological phase transitions, thereby capturing the full CIM/CPHL extended phase diagram.
Utilizing the fact that such 1D circuits are classically simulable and can be trained efficiently without barren plateaus \cite{PhysRevA.109.L050402, slattery2021quantumcircuitstwodimensionalisometric}, we benchmark their performance in Fig.~\ref{fig::Fig_cphl}(b1) by optimizing their fidelity with DMRG ground states [SI Sec.~\ref{sup::cphl}]. The same circuit architecture can be realized experimentally through a holographic protocol [Fig.~\ref{fig::Fig_cphl}(b1)] involving sequential measurements and qubit reset. This quantum hardware approach naturally extends to 2D isoTNSs, for which the computation of local expectation values has been shown to be BQP-complete~\cite{PRXQuantum.6.020310}.


\section*{Designing Tailored Response Functions}
Material design goes beyond static properties discussed so far to include the optimization of dynamical, frequency-resolved linear-response functions. 
Relevant applications include: (i) enhancing a chosen spectral component, for example in photochemistry, (ii) reconstructing a Hamiltonian from a measured molecular spectrum, reminiscent of nuclear magnetic resonance inference, and (iii) learning many-body translationally invariant Hamiltonians from frequency- and momentum-resolved condensed-matter data obtained from probes such as neutron scattering, angle-resolved photoemission spectroscopy, Brillouin light scattering, or resonant inelastic X-ray scattering, etc.

IQS is naturally suited to such tasks, as quantum hardware enables efficient evaluation of linear-response functions $\chi(\omega)$ \cite{HuangVariationalResponse, PhysRevResearch.2.033043}. 
In this work, we are proposing an approach in which the system’s response to weak time-dependent perturbations around $\ket{\psi(\boldsymbol{\theta})}$ stored in quantum memory is captured variationally by constraining the dynamics to a manifold ${\cal M}$ within the Hilbert space [Fig.~\ref{fig::Figure_dyn}(a)].
The McLachlan time-dependent variational principle \cite{PRXQuantum.2.030307, yuan2019theory} then projects the Schr\"odinger evolution onto ${\cal M}$ [orange projection in Fig.~\ref{fig::Figure_dyn}(a)], and linearizing this projected dynamics yields $\chi(\omega)$.
Because the tangent vectors of ${\cal M}$ can be computed analytically [Fig.~\ref{fig::Figure_dyn}(a) and Methods], $\chi(\omega)$ can be efficiently obtained by preparing $\ket{\psi(\boldsymbol{\theta})}$ and measuring a fixed set of Hermitian operators, followed by classical post-processing (see Methods).

In direct analogy to the static case, the objective can be framed as minimizing a cost function [SI Sec.~\ref{sup::dyn}] -- for instance, matching a given molecular spectrum ${\rm Im}[\chi_{\rm tar}(\omega)]$ by maximizing its overlap with the variational spectral profile ${\rm Im}[\chi_{\rm var}(\omega)]$:
\begin{align}
    {\cal C} = - \int \frac{{\rm d}\omega}{2\pi}\, {\rm Im}[\chi_{\rm var}(\omega)]\, {\rm Im}[\chi_{\rm tar}(\omega)].
\end{align}
Optimizing linear response functions in this way requires knowledge of the underlying Hamiltonian, which is obtained via the Hamiltonian learning step as before. 
We refer to this frequency-resolved extension of our framework as \emph{spectral Hamiltonian learning} (SHL). 

Figure~\ref{fig::Figure_dyn}(b,c) benchmarks SHL on a simple system of three spin-$\tfrac{1}{2}$ particles, where the goal is to learn a single-parameter Hamiltonian containing a ring-exchange interaction $J_{\rm ring}\, \hat{\bm \sigma}_1 \cdot (\hat{\bm \sigma}_2 \times \hat{\bm \sigma}_3)$ from a given spectrum. 
Starting from an initial guess that yields a single-peak response, the algorithm rapidly converges to the correct three-peak structure and recovers the target value of $J_{\rm ring}$ [Fig.~\ref{fig::Figure_dyn}(b)].

SHL can also infer the intrinsic translationally invariant many-body Hamiltonian directly from the spectrum of collective modes. 
The simplest example is magnons in a 1D ferromagnetic spin chain [Fig.~\ref{fig::Figure_dyn}(d)], corresponding to spin-flip excitations at finite momentum $k$. 
Figure~\ref{fig::Figure_dyn}(d) mimics a neutron-scattering experiment, which measures the momentum- and frequency-resolved dynamical structure factor ${\cal S}_{\rm FM}(k,\omega)$, peaked along the magnon dispersion $\omega_m(k)$ [orange dashed line]. 
Given only ${\cal S}_{\rm FM}(k,\omega)$, SHL successfully reconstructs the two-parameter Hamiltonian [right inset]. 
This ferromagnetic case serves as a benchmark, enabling comparison between our numerical routines and analytics [SI Sec.~\ref{sup::dyn}].  
A more challenging case arises in the Heisenberg AFM chain in Fig.~\ref{fig::Figure_dyn}(e), where -- as known from the Bethe ansatz -- a spin-flip  excitation (carrying spin-1) fractionalizes into a continuum of spin-$\tfrac{1}{2}$ excitations [left inset], producing a much richer dynamical structure factor ${\cal S}_{\rm AFM}(k,\omega)$. 
For $B_z=0$ in the thermodynamic limit, the two-spinon continuum is bounded by [orange dashed lines] $\omega_{\mathrm{LB}}(k) = - 2\pi J \, |\sin k|$ and $\omega_{\mathrm{UB}}(k) = - 4\pi J \, \big|\sin(k/2)\big|$.
In practice, for finite system sizes we compute ${\cal S}_{\rm AFM}(k,\omega)$ using the Kubo formula, which yields a set of discrete peaks distributed between the two branches [Fig.~\ref{fig::Figure_dyn}(e)]. 
From this complex dynamical structure factor, SHL nevertheless successfully reconstructs the intrinsic Hamiltonian [right inset], providing a stringent benchmark of the method.

\section*{Discussion and outlook}
Our work establishes a route for programmable quantum simulators to evolve from tools for exploring established models into platforms for designing and discovering new quantum materials.
We illustrated our IQS framework across a range of systems and applications, from enhancing $d$-wave correlations in the doping-temperature phase diagram, to extending a topological phase, and to shaping dynamical response functions.
All proposed protocols are experimentally accessible on current quantum hardware, including both analog and early fault-tolerant architectures.

Near-term quantum simulators are poised to access large 2D systems, creating opportunities to investigate and design 2D quantum materials -- a central frontier in condensed-matter physics of atomically thin heterostructures.
Deploying IQS on large-scale quantum devices and in the context of real materials could reveal models and parameter regimes key to the design and understanding of high-$T_c$ superconductors.
On early fault-tolerant platforms, constant-entropy simulation of isoTNSs may enable quantum-hardware exploration of complex topological magnetism, including clarifying the role of $\Gamma$-terms in stabilizing Kitaev spin liquids.
Applied to experimental data, 
SHL could enable quantum hardware to reconstruct effective Hamiltonians and reveal the microscopic interactions governing correlated materials.
 Extending the approach to open, driven quantum systems and nonlinear response functions would substantially broaden the scope of SHL -- from the design of photonic materials to the engineering of steady states with tailored properties.
Finally, a promising avenue for our IQS framework is to design classically difficult distributions of target observables efficiently~\cite{huang2025generativequantumadvantageclassical}.

\section*{Acknowledgments} The authors thank R.~Ott, N.~Maskara, S.~Ostermann, A.~W.~Young, M.~Kalinowski, H. Yu, D. Mark, T.~Cochran, and M.~Mitrano for fruitful discussions.
C.K. acknowledges support
from the NSF through a grant for the ITAMP at Harvard University.
Parts of this work were completed while P.E.D. was a Visiting Researcher at QuEra Computing Inc. D.G.-C. acknowledges financial support through the Ramón y Cajal Program (RYC2023-044201-I), financed by MICIU/AEI/10.13039/501100011033 and by the FSE+. Work at Harvard was supported by the U.S. Department of Energy (DOE Quantum Systems Accelerator Center, grant number DE-AC02-05CH11231, and the QUACQ program, grant number DE-SC0025572),  the Center for Ultracold Atoms (a NSF Physics Frontiers Center, PHY-1734011), the National Science Foundation (grant numbers PHY-2012023 and  CCF-2313084). 

\section*{Author contributions}
C.K., P.Z., P.E.D., and R.v.B. conceived the project. P.E.D. and C.K. developed all theoretical analysis and led manuscript preparation, with critical input from all co-authors. All authors contributed to discussions of results and refinement of the framework. M.D.L. and P.Z. supervised the project. 

\section*{Declarations}
M.D.L. is a Chief Scientist, co-founder and shareholder of QuEra Computing. Authors affiliated with QuEra Computing are employees or interns at QuEra Computing at the time of their contributions. Raw data and analysis code are available upon reasonable request.

\bibliography{QC}

@article{Volpe2025QuantumOptimizationReview,
  author    = {Volpe, Deborah and Orlandi, Giacomo and Turvani, Giovanna},
  title     = {Improving the Solving of Optimization Problems: A Comprehensive Review of Quantum Approaches},
  journal   = {Quantum Reports},
  volume    = {7},
  number    = {1},
  pages     = {1--19},
  year      = {2025},
  issn      = {2624-960X},
  doi       = {10.3390/quantum7010003}
}

@article{shi2018variational,
  title={Variational study of fermionic and bosonic systems with non-Gaussian states: Theory and applications},
  author={Shi, Tao and Demler, Eugene and Cirac, J Ignacio},
  journal={Annals of Physics},
  volume={390},
  pages={245--302},
  year={2018},
  publisher={Elsevier}
}

@article{grimsley2019adaptive,
  title={An adaptive variational algorithm for exact molecular simulations on a quantum computer},
  author={Grimsley, Harper R and Economou, Sophia E and Barnes, Edwin and Mayhall, Nicholas J},
  journal={Nat. Commun.},
  volume={10},
  number={1},
  pages={3007},
  year={2019},
  publisher={Nature Publishing Group UK London},
doi = {10.1038/s41467-019-10988-2},
url = {https://doi.org/10.1038/s41467-019-10988-2}
}

@article{PRXQuantum.2.030307,
  title = {Adaptive Variational Quantum Dynamics Simulations},
  author = {Yao, Yong-Xin and Gomes, Niladri and Zhang, Feng and Wang, Cai-Zhuang and Ho, Kai-Ming and Iadecola, Thomas and Orth, Peter P.},
  journal = {PRX Quantum},
  volume = {2},
  issue = {3},
  pages = {030307},
  numpages = {14},
  year = {2021},
  month = {Jul},
  publisher = {American Physical Society},
  doi = {10.1103/PRXQuantum.2.030307},
  url = {https://link.aps.org/doi/10.1103/PRXQuantum.2.030307}
}

@article{yuan2019theory,
  title={Theory of variational quantum simulation},
  author={Yuan, Xiao and Endo, Suguru and Zhao, Qi and Li, Ying and Benjamin, Simon C},
  journal={Quantum},
  volume={3},
  pages={191},
  year={2019},
  publisher={Verein zur F{\"o}rderung des Open Access Publizierens in den Quantenwissenschaften},
doi = {10.22331/q-2019-10-07-191},
url = {https://doi.org/10.22331/q-2019-10-07-191}
}

@article{Qi2019determininglocal,
  doi = {10.22331/q-2019-07-08-159},
  url = {https://doi.org/10.22331/q-2019-07-08-159},
  title = {Determining a local {H}amiltonian from a single eigenstate},
  author = {Qi, Xiao-Liang and Ranard, Daniel},
  journal = {{Quantum}},
  issn = {2521-327X},
  publisher = {{Verein zur F{\"{o}}rderung des Open Access Publizierens in den Quantenwissenschaften}},
  volume = {3},
  pages = {159},
  month = jul,
  year = {2019}
}

@article{PhysRevLett.132.160401,
  title = {Parent Hamiltonian Reconstruction via Inverse Quantum Annealing},
  author = {Rattacaso, Davide and Passarelli, Gianluca and Russomanno, Angelo and Lucignano, Procolo and Santoro, Giuseppe E. and Fazio, Rosario},
  journal = {Phys. Rev. Lett.},
  volume = {132},
  issue = {16},
  pages = {160401},
  numpages = {6},
  year = {2024},
  month = {Apr},
  publisher = {American Physical Society},
  doi = {10.1103/PhysRevLett.132.160401},
  url = {https://link.aps.org/doi/10.1103/PhysRevLett.132.160401}
}

@article{bethe1931theorie,
  title={Zur theorie der metalle: I. Eigenwerte und eigenfunktionen der linearen atomkette},
  author={Bethe, Hans},
  journal={Zeitschrift f{\"u}r Physik},
  volume={71},
  number={3},
  pages={205--226},
  year={1931},
  publisher={Springer},
doi={https://doi.org/10.1007/BF01341708},
url={https://doi.org/10.1007/BF01341708}
}

@article{PhysRev.128.2131,
  title = {Spin-Wave Spectrum of the Antiferromagnetic Linear Chain},
  author = {des Cloizeaux, Jacques and Pearson, J. J.},
  journal = {Phys. Rev.},
  volume = {128},
  issue = {5},
  pages = {2131--2135},
  numpages = {0},
  year = {1962},
  month = {Dec},
  publisher = {American Physical Society},
  doi = {10.1103/PhysRev.128.2131},
  url = {https://link.aps.org/doi/10.1103/PhysRev.128.2131}
}

@article{Faddeev1981SpinWave,
  title     = {What is the spin of a spin wave?},
  author    = {Faddeev, L. D. and Takhtajan, L. A.},
  journal   = {Phys. Lett. A},
  volume    = {85},
  number    = {6},
  pages     = {375--377},
  year      = {1981},
  doi       = {10.1016/0375-9601(81)90335-2}
}

@article{PhysRevB.55.12510,
  title = {Two-spinon dynamic structure factor of the one-dimensional s= Heisenberg antiferromagnet},
  author = {Karbach, Michael and M\"uller, Gerhard and Bougourzi, A. Hamid and Fledderjohann, Andreas and M\"utter, Karl-Heinz},
  journal = {Phys. Rev. B},
  volume = {55},
  issue = {18},
  pages = {12510--12517},
  numpages = {0},
  year = {1997},
  month = {May},
  publisher = {American Physical Society},
  doi = {10.1103/PhysRevB.55.12510},
  url = {https://link.aps.org/doi/10.1103/PhysRevB.55.12510}
}

@article{lake2010confinement,
  title={Confinement of fractional quantum number particles in a condensed-matter system},
  author={Lake, Bella and Tsvelik, Alexei M and Notbohm, Susanne and Alan Tennant, D and Perring, Toby G and Reehuis, Manfred and Sekar, Chinnathambi and Krabbes, Gernot and B{\"u}chner, Bernd},
  journal={Nat. Phys.},
  volume={6},
  number={1},
  pages={50--55},
  year={2010},
  publisher={Nature Publishing Group UK London},
doi={https://doi.org/10.1038/nphys1462},
url={https://doi.org/10.1038/nphys1462}
}

@article{PhysRevA.84.022304,
  title = {Statistical mechanics of the cluster Ising model},
  author = {Smacchia, Pietro and Amico, Luigi and Facchi, Paolo and Fazio, Rosario and Florio, Giuseppe and Pascazio, Saverio and Vedral, Vlatko},
  journal = {Phys. Rev. A},
  volume = {84},
  issue = {2},
  pages = {022304},
  numpages = {12},
  year = {2011},
  month = {Aug},
  publisher = {American Physical Society},
  doi = {10.1103/PhysRevA.84.022304},
  url = {https://link.aps.org/doi/10.1103/PhysRevA.84.022304}
}

@article{PhysRevResearch.4.L022020,
  title = {Crossing a topological phase transition with a quantum computer},
  author = {Smith, Adam and Jobst, Bernhard and Green, Andrew G. and Pollmann, Frank},
  journal = {Phys. Rev. Res.},
  volume = {4},
  issue = {2},
  pages = {L022020},
  numpages = {8},
  year = {2022},
  month = {Apr},
  publisher = {American Physical Society},
  doi = {10.1103/PhysRevResearch.4.L022020},
  url = {https://link.aps.org/doi/10.1103/PhysRevResearch.4.L022020}
}

@article{PhysRevB.96.165124,
  title = {One-dimensional symmetry protected topological phases and their transitions},
  author = {Verresen, Ruben and Moessner, Roderich and Pollmann, Frank},
  journal = {Phys. Rev. B},
  volume = {96},
  issue = {16},
  pages = {165124},
  numpages = {23},
  year = {2017},
  month = {Oct},
  publisher = {American Physical Society},
  doi = {10.1103/PhysRevB.96.165124},
  url = {https://link.aps.org/doi/10.1103/PhysRevB.96.165124}
}

@article{PhysRevB.92.195139,
  title = {Pair correlations in doped Hubbard ladders},
  author = {Dolfi, Michele and Bauer, Bela and Keller, Sebastian and Troyer, Matthias},
  journal = {Phys. Rev. B},
  volume = {92},
  issue = {19},
  pages = {195139},
  numpages = {9},
  year = {2015},
  month = {Nov},
  publisher = {American Physical Society},
  doi = {10.1103/PhysRevB.92.195139},
  url = {https://link.aps.org/doi/10.1103/PhysRevB.92.195139}
}

@article{aspuru2018,
author = {Benjamin Sanchez-Lengeling  and Alán Aspuru-Guzik },
title = {Inverse molecular design using machine learning: Generative models for matter engineering},
journal = {Science},
volume = {361},
number = {6400},
pages = {360-365},
year = {2018},
doi = {10.1126/science.aat2663},
URL = {https://www.science.org/doi/abs/10.1126/science.aat2663}}

@Article{ma15051811,
AUTHOR = {Wang, Jia and Wang, Yingxue and Chen, Yanan},
TITLE = {Inverse Design of Materials by Machine Learning},
JOURNAL = {Materials},
VOLUME = {15},
YEAR = {2022},
NUMBER = {5},
ARTICLE-NUMBER = {1811},
URL = {https://www.mdpi.com/1996-1944/15/5/1811},
PubMedID = {35269043},
ISSN = {1996-1944},
DOI = {10.3390/ma15051811}
}

@article{Inui2023,
  author  = {Inui, Koji and Motome, Yukitoshi},
  title   = {Inverse Hamiltonian design by automatic differentiation},
  journal = {Commun. Phys.},
  volume  = {6},
  number  = {1},
  pages   = {37},
  year    = {2023},
  month   = mar,
  doi     = {10.1038/s42005-023-01132-0},
  issn    = {2399-3650},
  url     = {https://doi.org/10.1038/s42005-023-01132-0}
}

@misc{bakshi2023learningquantumhamiltonianstemperature,
      title={Learning quantum Hamiltonians at any temperature in polynomial time}, 
      author={Ainesh Bakshi and Allen Liu and Ankur Moitra and Ewin Tang},
      year={2023},
      eprint={2310.02243},
      archivePrefix={arXiv},
      primaryClass={quant-ph},
      url={https://arxiv.org/abs/2310.02243}, 
}

@article{PhysRevLett.122.020504,
  title = {Learning a Local Hamiltonian from Local Measurements},
  author = {Bairey, Eyal and Arad, Itai and Lindner, Netanel H.},
  journal = {Phys. Rev. Lett.},
  volume = {122},
  issue = {2},
  pages = {020504},
  numpages = {5},
  year = {2019},
  month = {Jan},
  publisher = {American Physical Society},
  doi = {10.1103/PhysRevLett.122.020504},
  url = {https://link.aps.org/doi/10.1103/PhysRevLett.122.020504}
}

@article{SCHOLLWOCK201196,
title = {The density-matrix renormalization group in the age of matrix product states},
journal = {Annals of Physics},
volume = {326},
number = {1},
pages = {96-192},
year = {2011},
note = {January 2011 Special Issue},
issn = {0003-4916},
doi = {https://doi.org/10.1016/j.aop.2010.09.012},
url = {https://www.sciencedirect.com/science/article/pii/S0003491610001752},
author = {Ulrich Schollwöck}
}

@article{PhysRevX.8.031029,
  title = {Computational Inverse Method for Constructing Spaces of Quantum Models from Wave Functions},
  author = {Chertkov, Eli and Clark, Bryan K.},
  journal = {Phys. Rev. X},
  volume = {8},
  issue = {3},
  pages = {031029},
  numpages = {12},
  year = {2018},
  month = {Jul},
  publisher = {American Physical Society},
  doi = {10.1103/PhysRevX.8.031029},
  url = {https://link.aps.org/doi/10.1103/PhysRevX.8.031029}
}

@article{debreu1954representation,
  title={Representation of a preference ordering by a numerical function},
  author={Debreu, Gerard},
  journal={Decision processes},
  volume={3},
  pages={159--165},
  year={1954},
  publisher={Wiley New York},
  url={https://archive.org/details/decisionprocesse033215mbp/page/n172/mode/1up}
}

@article{bluvstein2025architectural,
  title={Architectural mechanisms of a universal fault-tolerant quantum computer},
  author={Bluvstein, Dolev and Geim, Alexandra A and Li, Sophie H and Evered, Simon J and Ataides, J and Baranes, Gefen and Gu, Andi and Manovitz, Tom and Xu, Muqing and Kalinowski, Marcin and others},
  journal={arXiv preprint arXiv:2506.20661},
  year={2025}
}

@article{PRXQuantum.6.020310,
  title = {Computational Complexity of Isometric Tensor-Network States},
  author = {Malz, Daniel and Trivedi, Rahul},
  journal = {PRX Quantum},
  volume = {6},
  issue = {2},
  pages = {020310},
  numpages = {16},
  year = {2025},
  month = {Apr},
  publisher = {American Physical Society},
  doi = {10.1103/PRXQuantum.6.020310},
  url = {https://link.aps.org/doi/10.1103/PRXQuantum.6.020310}
}

@article{google2025observation,
  title={Observation of constructive interference at the edge of quantum ergodicity},
  journal={Nature},
  volume={646},
  number={8086},
  pages={825--830},
  year={2025},
  publisher={Nature Publishing Group UK London},
doi = {10.1038/s41586-025-09526-6},
url = {https://doi.org/10.1038/s41586-025-09526-6}
}

@article{PhysRevResearch.3.033002,
  title = {Holographic quantum algorithms for simulating correlated spin systems},
  author = {Foss-Feig, Michael and Hayes, David and Dreiling, Joan M. and Figgatt, Caroline and Gaebler, John P. and Moses, Steven A. and Pino, Juan M. and Potter, Andrew C.},
  journal = {Phys. Rev. Res.},
  volume = {3},
  issue = {3},
  pages = {033002},
  numpages = {12},
  year = {2021},
  month = {Jul},
  publisher = {American Physical Society},
  doi = {10.1103/PhysRevResearch.3.033002},
  url = {https://link.aps.org/doi/10.1103/PhysRevResearch.3.033002}
}

@article{Bernien2017,
	author = {Bernien, Hannes and Schwartz, Sylvain and Keesling, Alexander and Levine, Harry and Omran, Ahmed and Pichler, Hannes and Choi, Soonwon and Zibrov, Alexander S. and Endres, Manuel and Greiner, Markus and Vuleti{\'c}, Vladan and Lukin, Mikhail D.},
	date = {2017/11/01},
	doi = {10.1038/nature24622},
	id = {Bernien2017},
	isbn = {1476-4687},
	journal = {Nature},
	number = {7682},
	pages = {579--584},
	title = {Probing many-body dynamics on a 51-atom quantum simulator},
	url = {https://doi.org/10.1038/nature24622},
	volume = {551},
	year = {2017}}

@article{Semeghini2021,
author = {G. Semeghini  and H. Levine  and A. Keesling  and S. Ebadi  and T. T. Wang  and D. Bluvstein  and R. Verresen  and H. Pichler  and M. Kalinowski  and R. Samajdar  and A. Omran  and S. Sachdev  and A. Vishwanath  and M. Greiner  and V. Vuletić  and M. D. Lukin },
title = {Probing topological spin liquids on a programmable quantum simulator},
journal = {Science},
volume = {374},
number = {6572},
pages = {1242-1247},
year = {2021},
doi = {10.1126/science.abi8794},
URL = {https://www.science.org/doi/abs/10.1126/science.abi8794}}

@article{Ebadi256Nature,
	author = {Ebadi, Sepehr and Wang, Tout T. and Levine, Harry and Keesling, Alexander and Semeghini, Giulia and Omran, Ahmed and Bluvstein, Dolev and Samajdar, Rhine and Pichler, Hannes and Ho, Wen Wei and Choi, Soonwon and Sachdev, Subir and Greiner, Markus and Vuleti{\'c}, Vladan and Lukin, Mikhail D.},
	date = {2021/07/01},
	doi = {10.1038/s41586-021-03582-4},
	id = {Ebadi2021},
	isbn = {1476-4687},
	journal = {Nature},
	number = {7866},
	pages = {227--232},
	title = {Quantum phases of matter on a 256-atom programmable quantum simulator},
	url = {https://doi.org/10.1038/s41586-021-03582-4},
	volume = {595},
	year = {2021}}

@misc{juliàfarré2025hybridquantumclassicalanalogsimulation,
      title={Hybrid quantum-classical analog simulation of two-dimensional Fermi-Hubbard models with neutral atoms}, 
      author={Sergi Julià-Farré and Antoine Michel and Christophe Domain and Joseph Mikael and Jacques-Charles Lafoucriere and Joseph Vovrosh and Ahmed Chahlaoui and Dorian Claveau and Guillaume Villaret and Julius de Hond and Loïc Henriet and Antoine Browaeys and Thomas Ayral and Alexandre Dauphin},
      year={2025},
      eprint={2510.05897},
      archivePrefix={arXiv},
      primaryClass={quant-ph},
      url={https://arxiv.org/abs/2510.05897}, 
}

@article{Léséleuc2019,
author = {Sylvain de Léséleuc  and Vincent Lienhard  and Pascal Scholl  and Daniel Barredo  and Sebastian Weber  and Nicolai Lang  and Hans Peter Büchler  and Thierry Lahaye  and Antoine Browaeys },
title = {Observation of a symmetry-protected topological phase of interacting bosons with Rydberg atoms},
journal = {Science},
volume = {365},
number = {6455},
pages = {775-780},
year = {2019},
doi = {10.1126/science.aav9105},
URL = {https://www.science.org/doi/abs/10.1126/science.aav9105},
eprint = {https://www.science.org/doi/pdf/10.1126/science.aav9105}}

@misc{kendrick2025pseudogapfermihubbardquantumsimulator,
      title={Pseudogap in a Fermi-Hubbard quantum simulator}, 
      author={Lev Haldar Kendrick and Anant Kale and Youqi Gang and Alexander Dennisovich Deters and Martin Lebrat and Aaron W. Young and Markus Greiner},
      year={2025},
      eprint={2509.18075},
      archivePrefix={arXiv},
      primaryClass={cond-mat.quant-gas},
      url={https://arxiv.org/abs/2509.18075}, 
}

@article{MuqingNature2025,
	author = {Xu, Muqing and Kendrick, Lev Haldar and Kale, Anant and Gang, Youqi and Feng, Chunhan and Zhang, Shiwei and Young, Aaron W. and Lebrat, Martin and Greiner, Markus},
	date = {2025/06/01},
	doi = {10.1038/s41586-025-09112-w},
	id = {Xu2025},
	isbn = {1476-4687},
	journal = {Nature},
	number = {8069},
	pages = {909--915},
	title = {A neutral-atom Hubbard quantum simulator in the cryogenic regime},
	url = {https://doi.org/10.1038/s41586-025-09112-w},
	volume = {642},
	year = {2025}}

@article{ImpertroNatPhys2025,
	author = {Impertro, Alexander and Huh, SeungJung and Karch, Simon and Wienand, Julian F. and Bloch, Immanuel and Aidelsburger, Monika},
	date = {2025/06/01},
	doi = {10.1038/s41567-025-02890-0},
	id = {Impertro2025},
	isbn = {1745-2481},
	journal = {Nature Physics},
	number = {6},
	pages = {895--901},
	title = {Strongly interacting Meissner phases in large bosonic flux ladders},
	url = {https://doi.org/10.1038/s41567-025-02890-0},
	volume = {21},
	year = {2025}}

@article{
HartkeScience23,
author = {Thomas Hartke  and Botond Oreg  and Carter Turnbaugh  and Ningyuan Jia  and Martin Zwierlein },
title = {Direct observation of nonlocal fermion pairing in an attractive Fermi-Hubbard gas},
journal = {Science},
volume = {381},
number = {6653},
pages = {82-86},
year = {2023},
doi = {10.1126/science.ade4245},
URL = {https://www.science.org/doi/abs/10.1126/science.ade4245}}

@article{IqbalNature2024,
	author = {Iqbal, Mohsin and Tantivasadakarn, Nathanan and Verresen, Ruben and Campbell, Sara L. and Dreiling, Joan M. and Figgatt, Caroline and Gaebler, John P. and Johansen, Jacob and Mills, Michael and Moses, Steven A. and Pino, Juan M. and Ransford, Anthony and Rowe, Mary and Siegfried, Peter and Stutz, Russell P. and Foss-Feig, Michael and Vishwanath, Ashvin and Dreyer, Henrik},
	date = {2024/02/01},
	doi = {10.1038/s41586-023-06934-4},
	id = {Iqbal2024},
	isbn = {1476-4687},
	journal = {Nature},
	number = {7999},
	pages = {505--511},
	title = {Non-Abelian topological order and anyons on a trapped-ion processor},
	url = {https://doi.org/10.1038/s41586-023-06934-4},
	volume = {626},
	year = {2024}}

@article{RevModPhys.93.025001,
  title = {Programmable quantum simulations of spin systems with trapped ions},
  author = {Monroe, C. and Campbell, W. C. and Duan, L.-M. and Gong, Z.-X. and Gorshkov, A. V. and Hess, P. W. and Islam, R. and Kim, K. and Linke, N. M. and Pagano, G. and Richerme, P. and Senko, C. and Yao, N. Y.},
  journal = {Rev. Mod. Phys.},
  volume = {93},
  issue = {2},
  pages = {025001},
  numpages = {57},
  year = {2021},
  month = {Apr},
  publisher = {American Physical Society},
  doi = {10.1103/RevModPhys.93.025001},
  url = {https://link.aps.org/doi/10.1103/RevModPhys.93.025001}
}

@article{JoshiKokailNature2023,
	author = {Joshi, Manoj K. and Kokail, Christian and van Bijnen, Rick and Kranzl, Florian and Zache, Torsten V. and Blatt, Rainer and Roos, Christian F. and Zoller, Peter},
	date = {2023/12/01},
	doi = {10.1038/s41586-023-06768-0},
	id = {Joshi2023},
	isbn = {1476-4687},
	journal = {Nature},
	number = {7992},
	pages = {539--544},
	title = {Exploring large-scale entanglement in quantum simulation},
	url = {https://doi.org/10.1038/s41586-023-06768-0},
	volume = {624},
	year = {2023}}

@article{Satzinger2021,
author = {K. J. Satzinger  and Y.-J Liu  and A. Smith  and C. Knapp  and M. Newman  and C. Jones  and Z. Chen  and C. Quintana  and X. Mi  and A. Dunsworth  and C. Gidney  and I. Aleiner  and F. Arute  and K. Arya  and J. Atalaya  and R. Babbush  and J. C. Bardin  and R. Barends  and J. Basso  and A. Bengtsson  and A. Bilmes  and M. Broughton  and B. B. Buckley  and D. A. Buell  and B. Burkett  and N. Bushnell  and B. Chiaro  and R. Collins  and W. Courtney  and S. Demura  and A. R. Derk  and D. Eppens  and C. Erickson  and L. Faoro  and E. Farhi  and A. G. Fowler  and B. Foxen  and M. Giustina  and A. Greene  and J. A. Gross  and M. P. Harrigan  and S. D. Harrington  and J. Hilton  and S. Hong  and T. Huang  and W. J. Huggins  and L. B. Ioffe  and S. V. Isakov  and E. Jeffrey  and Z. Jiang  and D. Kafri  and K. Kechedzhi  and T. Khattar  and S. Kim  and P. V. Klimov  and A. N. Korotkov  and F. Kostritsa  and D. Landhuis  and P. Laptev  and A. Locharla  and E. Lucero  and O. Martin  and J. R. McClean  and M. McEwen  and K. C. Miao  and M. Mohseni  and S. Montazeri  and W. Mruczkiewicz  and J. Mutus  and O. Naaman  and M. Neeley  and C. Neill  and M. Y. Niu  and T. E. O’Brien  and A. Opremcak  and B. Pató  and A. Petukhov  and N. C. Rubin  and D. Sank  and V. Shvarts  and D. Strain  and M. Szalay  and B. Villalonga  and T. C. White  and Z. Yao  and P. Yeh  and J. Yoo  and A. Zalcman  and H. Neven  and S. Boixo  and A. Megrant  and Y. Chen  and J. Kelly  and V. Smelyanskiy  and A. Kitaev  and M. Knap  and F. Pollmann  and P. Roushan },
title = {Realizing topologically ordered states on a quantum processor},
journal = {Science},
volume = {374},
number = {6572},
pages = {1237-1241},
year = {2021},
doi = {10.1126/science.abi8378},
URL = {https://www.science.org/doi/abs/10.1126/science.abi8378}}

@misc{alam2025_1,
      title={Fermionic dynamics on a trapped-ion quantum computer beyond exact classical simulation}, 
      author={Faisal Alam and Jan Lukas Bosse and Ieva Čepaitė and Adrian Chapman and Laura Clinton and Marcos Crichigno and Elizabeth Crosson and Toby Cubitt and Charles Derby and Oliver Dowinton and Paul K. Faehrmann and Steve Flammia and Brian Flynn and Filippo Maria Gambetta and Raúl García-Patrón and Max Hunter-Gordon and Glenn Jones and Abhishek Khedkar and Joel Klassen and Michael Kreshchuk and Edward Harry McMullan and Lana Mineh and Ashley Montanaro and Caterina Mora and John J. L. Morton and Dhrumil Patel and Pete Rolph and Raul A. Santos and James R. Seddon and Evan Sheridan and Wilfrid Somogyi and Marika Svensson and Niam Vaishnav and Sabrina Yue Wang and Gethin Wright},
      year={2025},
      eprint={2510.26300},
      archivePrefix={arXiv},
      primaryClass={quant-ph},
      url={https://arxiv.org/abs/2510.26300}, 
}

@misc{alam2025_2,
      title={Programmable digital quantum simulation of 2D Fermi-Hubbard dynamics using 72 superconducting qubits}, 
      author={Faisal Alam and Jan Lukas Bosse and Ieva Čepaitė and Adrian Chapman and Laura Clinton and Marcos Crichigno and Elizabeth Crosson and Toby Cubitt and Charles Derby and Oliver Dowinton and Paul K. Faehrmann and Steve Flammia and Brian Flynn and Filippo Maria Gambetta and Raúl García-Patrón and Max Hunter-Gordon and Glenn Jones and Abhishek Khedkar and Joel Klassen and Michael Kreshchuk and Edward Harry McMullan and Lana Mineh and Ashley Montanaro and Caterina Mora and John J. L. Morton and Dhrumil Patel and Pete Rolph and Raul A. Santos and James R. Seddon and Evan Sheridan and Wilfrid Somogyi and Marika Svensson and Niam Vaishnav and Sabrina Yue Wang and Gethin Wright},
      year={2025},
      eprint={2510.26845},
      archivePrefix={arXiv},
      primaryClass={quant-ph},
      url={https://arxiv.org/abs/2510.26845}, 
}

@article{GonzalezPNAS,
author = {D. González-Cuadra  and D. Bluvstein  and M. Kalinowski  and R. Kaubruegger  and N. Maskara  and P. Naldesi  and T. V. Zache  and A. M. Kaufman  and M. D. Lukin  and H. Pichler  and B. Vermersch  and Jun Ye  and P. Zoller },
title = {Fermionic quantum processing with programmable neutral atom arrays},
journal = {Proceedings of the National Academy of Sciences},
volume = {120},
number = {35},
pages = {e2304294120},
year = {2023},
doi = {10.1073/pnas.2304294120},
URL = {https://www.pnas.org/doi/abs/10.1073/pnas.2304294120},
eprint = {https://www.pnas.org/doi/pdf/10.1073/pnas.2304294120}}

@article{gebhartreview,
	author = {Gebhart, Valentin and Santagati, Raffaele and Gentile, Antonio Andrea and Gauger, Erik M. and Craig, David and Ares, Natalia and Banchi, Leonardo and Marquardt, Florian and Pezz{\`e}, Luca and Bonato, Cristian},
	date = {2023/03/01},
	doi = {10.1038/s42254-022-00552-1},
	id = {Gebhart2023},
	isbn = {2522-5820},
	journal = {Nature Reviews Physics},
	number = {3},
	pages = {141--156},
	title = {Learning quantum systems},
	url = {https://doi.org/10.1038/s42254-022-00552-1},
	volume = {5},
	year = {2023}}

@article{anuragNatPhys21,
	author = {Anshu, Anurag and Arunachalam, Srinivasan and Kuwahara, Tomotaka and Soleimanifar, Mehdi},
	date = {2021/08/01},
	doi = {10.1038/s41567-021-01232-0},
	id = {Anshu2021},
	isbn = {1745-2481},
	journal = {Nature Physics},
	number = {8},
	pages = {931--935},
	title = {Sample-efficient learning of interacting quantum systems},
	url = {https://doi.org/10.1038/s41567-021-01232-0},
	volume = {17},
	year = {2021}}

@article{PhysRevLett.94.170201,
  title = {Computational Complexity and Fundamental Limitations to Fermionic Quantum Monte Carlo Simulations},
  author = {Troyer, Matthias and Wiese, Uwe-Jens},
  journal = {Phys. Rev. Lett.},
  volume = {94},
  issue = {17},
  pages = {170201},
  numpages = {4},
  year = {2005},
  month = {May},
  publisher = {American Physical Society},
  doi = {10.1103/PhysRevLett.94.170201},
  url = {https://link.aps.org/doi/10.1103/PhysRevLett.94.170201}
}

@article{xu2024coexistence,
  title={Coexistence of superconductivity with partially filled stripes in the Hubbard model},
  author={Xu, Hao and Chung, Chia-Min and Qin, Mingpu and Schollw{\"o}ck, Ulrich and White, Steven R and Zhang, Shiwei},
  journal={Science},
  volume={384},
  number={6696},
  pages={eadh7691},
  year={2024},
  publisher={American Association for the Advancement of Science}
}

@article{PhysRevLett.98.140506,
  title = {Computational Complexity of Projected Entangled Pair States},
  author = {Schuch, Norbert and Wolf, Michael M. and Verstraete, Frank and Cirac, J. Ignacio},
  journal = {Phys. Rev. Lett.},
  volume = {98},
  issue = {14},
  pages = {140506},
  numpages = {4},
  year = {2007},
  month = {Apr},
  publisher = {American Physical Society},
  doi = {10.1103/PhysRevLett.98.140506},
  url = {https://link.aps.org/doi/10.1103/PhysRevLett.98.140506}
}

@INPROCEEDINGS{BergamaschiProc2024,
  author={Bergamaschi, Thiago and Chen, Chi-Fang and Liu, Yunchao},
  booktitle={2024 IEEE 65th Annual Symposium on Foundations of Computer Science (FOCS)}, 
  title={Quantum Computational Advantage with Constant-Temperature Gibbs Sampling}, 
  year={2024},
  volume={},
  number={},
  pages={1063-1085},
  doi={10.1109/FOCS61266.2024.00071}}

@article{BollScience2016,
author = {Martin Boll  and Timon A. Hilker  and Guillaume Salomon  and Ahmed Omran  and Jacopo Nespolo  and Lode Pollet  and Immanuel Bloch  and Christian Gross },
title = {Spin- and density-resolved microscopy of antiferromagnetic correlations in Fermi-Hubbard chains},
journal = {Science},
volume = {353},
number = {6305},
pages = {1257-1260},
year = {2016},
doi = {10.1126/science.aag1635},
URL = {https://www.science.org/doi/abs/10.1126/science.aag1635},
eprint = {https://www.science.org/doi/pdf/10.1126/science.aag1635}}

@article{PhysRevLett.110.216405,
  title = {Superconductivity and the Pseudogap in the Two-Dimensional Hubbard Model},
  author = {Gull, Emanuel and Parcollet, Olivier and Millis, Andrew J.},
  journal = {Phys. Rev. Lett.},
  volume = {110},
  issue = {21},
  pages = {216405},
  numpages = {5},
  year = {2013},
  month = {May},
  publisher = {American Physical Society},
  doi = {10.1103/PhysRevLett.110.216405},
  url = {https://link.aps.org/doi/10.1103/PhysRevLett.110.216405}
}

@article{3nx4-bnyy,
  title = {Programming Optical-Lattice Fermi-Hubbard Quantum Simulators},
  author = {Tabares, Cristian and Kokail, Christian and Zoller, Peter and Gonz\'alez-Cuadra, Daniel and Gonz\'alez-Tudela, Alejandro},
  journal = {PRX Quantum},
  volume = {6},
  issue = {3},
  pages = {030356},
  numpages = {26},
  year = {2025},
  month = {Sep},
  publisher = {American Physical Society},
  doi = {10.1103/3nx4-bnyy},
  url = {https://link.aps.org/doi/10.1103/3nx4-bnyy}
}

@article{PhysRevB.83.054508,
  title = {Enhanced pairing in the checkerboard Hubbard ladder},
  author = {Karakonstantakis, George and Berg, Erez and White, Steven R. and Kivelson, Steven A.},
  journal = {Phys. Rev. B},
  volume = {83},
  issue = {5},
  pages = {054508},
  numpages = {8},
  year = {2011},
  month = {Feb},
  publisher = {American Physical Society},
  doi = {10.1103/PhysRevB.83.054508},
  url = {https://link.aps.org/doi/10.1103/PhysRevB.83.054508}
}

@article{NOACK1996281,
title = {The ground state of the two-leg Hubbard ladder a density-matrix renormalization group study},
journal = {Physica C: Superconductivity},
volume = {270},
number = {3},
pages = {281-296},
year = {1996},
issn = {0921-4534},
doi = {https://doi.org/10.1016/S0921-4534(96)00515-1},
url = {https://www.sciencedirect.com/science/article/pii/S0921453496005151},
author = {R.M. Noack and S.R. White and D.J. Scalapino}
}

@article{PhysRevLett.87.047003,
  title = {Band-Structure Trend in Hole-Doped Cuprates and Correlation with ${\mathit{T}}_{\mathit{c}\mathrm{max}}$},
  author = {Pavarini, E. and Dasgupta, I. and Saha-Dasgupta, T. and Jepsen, O. and Andersen, O. K.},
  journal = {Phys. Rev. Lett.},
  volume = {87},
  issue = {4},
  pages = {047003},
  numpages = {4},
  year = {2001},
  month = {Jul},
  publisher = {American Physical Society},
  doi = {10.1103/PhysRevLett.87.047003},
  url = {https://link.aps.org/doi/10.1103/PhysRevLett.87.047003}
}

@article{PhysRevA.109.L050402,
  title = {Isometric tensor network optimization for extensive Hamiltonians is free of barren plateaus},
  author = {Miao, Qiang and Barthel, Thomas},
  journal = {Phys. Rev. A},
  volume = {109},
  issue = {5},
  pages = {L050402},
  numpages = {8},
  year = {2024},
  month = {May},
  publisher = {American Physical Society},
  doi = {10.1103/PhysRevA.109.L050402},
  url = {https://link.aps.org/doi/10.1103/PhysRevA.109.L050402}
}

@misc{slattery2021quantumcircuitstwodimensionalisometric,
      title={Quantum Circuits For Two-Dimensional Isometric Tensor Networks}, 
      author={Lucas Slattery and Bryan K. Clark},
      year={2021},
      eprint={2108.02792},
      archivePrefix={arXiv},
      primaryClass={quant-ph},
      url={https://arxiv.org/abs/2108.02792}, 
}

@article{HuangVariationalResponse,
	author = {Huang, Kaixuan and Cai, Xiaoxia and Li, Hao and Ge, Zi-Yong and Hou, Ruijuan and Li, Hekang and Liu, Tong and Shi, Yunhao and Chen, Chitong and Zheng, Dongning and Xu, Kai and Liu, Zhi-Bo and Li, Zhendong and Fan, Heng and Fang, Wei-Hai},
	journal = {The Journal of Physical Chemistry Letters},
	month = {10},
	number = {39},
	pages = {9114--9121},
	title = {Variational Quantum Computation of Molecular Linear Response Properties on a Superconducting Quantum Processor},
	volume = {13},
	year = {2022}}

@article{PhysRevResearch.2.033043,
  title = {Linear-response functions of molecules on a quantum computer: Charge and spin responses and optical absorption},
  author = {Kosugi, Taichi and Matsushita, Yu-ichiro},
  journal = {Phys. Rev. Res.},
  volume = {2},
  issue = {3},
  pages = {033043},
  numpages = {16},
  year = {2020},
  month = {Jul},
  publisher = {American Physical Society},
  doi = {10.1103/PhysRevResearch.2.033043},
  url = {https://link.aps.org/doi/10.1103/PhysRevResearch.2.033043}
}

@misc{huang2025generativequantumadvantageclassical,
      title={Generative quantum advantage for classical and quantum problems}, 
      author={Hsin-Yuan Huang and Michael Broughton and Norhan Eassa and Hartmut Neven and Ryan Babbush and Jarrod R. McClean},
      year={2025},
      eprint={2509.09033},
      archivePrefix={arXiv},
      primaryClass={quant-ph},
      url={https://arxiv.org/abs/2509.09033}, 
}

@article{jiang2019superconductivity,
  title={Superconductivity in the doped Hubbard model and its interplay with next-nearest hopping $t'$},
  author={Jiang, Hong-Chen and Devereaux, Thomas P},
  journal={Science},
  volume={365},
  number={6460},
  pages={1424--1428},
  year={2019},
  publisher={American Association for the Advancement of Science}
}

@article{chen2021anomalously,
  title={Anomalously strong near-neighbor attraction in doped 1D cuprate chains},
  author={Chen, Zhuoyu and Wang, Yao and Rebec, Slavko N and Jia, Tao and Hashimoto, Makoto and Lu, Donghui and Moritz, Brian and Moore, Robert G and Devereaux, Thomas P and Shen, Zhi-Xun},
  journal={Science},
  volume={373},
  number={6560},
  pages={1235--1239},
  year={2021},
  publisher={American Association for the Advancement of Science}
}

@article{PhysRevX.15.021049,
  title = {Beyond-Hubbard Pairing in a Cuprate Ladder},
  author = {Padma, Hari and Thomas, Jinu and TenHuisen, Sophia F. R. and He, Wei and Guan, Ziqiang and Li, Jiemin and Lee, Byungjune and Wang, Yu and Lee, Seng Huat and Mao, Zhiqiang and Jang, Hoyoung and Bisogni, Valentina and Pelliciari, Jonathan and Dean, Mark P. M. and Johnston, Steven and Mitrano, Matteo},
  journal = {Phys. Rev. X},
  volume = {15},
  issue = {2},
  pages = {021049},
  numpages = {10},
  year = {2025},
  month = {May},
  publisher = {American Physical Society},
  doi = {10.1103/PhysRevX.15.021049},
  url = {https://link.aps.org/doi/10.1103/PhysRevX.15.021049}
}

@article{scheie2025cooper,
  title={Cooper-pair localization in the magnetic dynamics of a cuprate ladder},
  author={Scheie, A and Laurell, P and Thomas, J and Sharma, V and Kolesnikov, AI and Granroth, GE and Zhang, Q and Lake, B and Mihalik Jr, M and Bewley, RI and others},
  journal={arXiv preprint arXiv:2501.10296},
  year={2025}
}

@article{KokailNatPhys,
	author = {Kokail, Christian and van Bijnen, Rick and Elben, Andreas and Vermersch, Beno{\^\i}t and Zoller, Peter},
	date = {2021/08/01},
	doi = {10.1038/s41567-021-01260-w},
	id = {Kokail2021},
	isbn = {1745-2481},
	journal = {Nature Physics},
	number = {8},
	pages = {936--942},
	title = {Entanglement Hamiltonian tomography in quantum simulation},
	url = {https://doi.org/10.1038/s41567-021-01260-w},
	volume = {17},
	year = {2021}}

@article{PRXQuantum.2.010102,
  title = {Theoretical and Experimental Perspectives of Quantum Verification},
  author = {Carrasco, Jose and Elben, Andreas and Kokail, Christian and Kraus, Barbara and Zoller, Peter},
  journal = {PRX Quantum},
  volume = {2},
  issue = {1},
  pages = {010102},
  numpages = {11},
  year = {2021},
  month = {Mar},
  publisher = {American Physical Society},
  doi = {10.1103/PRXQuantum.2.010102},
  url = {https://link.aps.org/doi/10.1103/PRXQuantum.2.010102}
}

@article{KokailNature2018,
	author = {Kokail, C. and Maier, C. and van Bijnen, R. and Brydges, T. and Joshi, M. K. and Jurcevic, P. and Muschik, C. A. and Silvi, P. and Blatt, R. and Roos, C. F. and Zoller, P.},
	date = {2019/05/01},
	doi = {10.1038/s41586-019-1177-4},
	id = {Kokail2019},
	isbn = {1476-4687},
	journal = {Nature},
	number = {7756},
	pages = {355--360},
	title = {Self-verifying variational quantum simulation of lattice models},
	url = {https://doi.org/10.1038/s41586-019-1177-4},
	volume = {569},
	year = {2019}}

@misc{rudolph2025paulipropagationcomputationalframework,
      title={Pauli Propagation: A Computational Framework for Simulating Quantum Systems}, 
      author={Manuel S. Rudolph and Tyson Jones and Yanting Teng and Armando Angrisani and Zoë Holmes},
      year={2025},
      eprint={2505.21606},
      archivePrefix={arXiv},
      primaryClass={quant-ph},
      url={https://arxiv.org/abs/2505.21606}, 
}

@misc{maskara2025fastsimulationfermionsreconfigurable,
      title={Fast simulation of fermions with reconfigurable qubits}, 
      author={Nishad Maskara and Marcin Kalinowski and Daniel Gonzalez-Cuadra and Mikhail D. Lukin},
      year={2025},
      eprint={2509.08898},
      archivePrefix={arXiv},
      primaryClass={quant-ph},
      url={https://arxiv.org/abs/2509.08898}, 
}

@book{altland2010condensed,
  title={Condensed matter field theory},
  author={Altland, Alexander and Simons, Ben D},
  year={2010},
  publisher={Cambridge university press}
}

\clearpage

\small

\section*{Methods}

\subsection*{IQS Protocol} \label{meth::iqsprotocol}
The first central step of IQS is to construct a cost functional $\mathcal{C}$ that translates a ``wish list'' of desired material properties into measurable targets~\cite{debreu1954representation}. These design objectives may include enhancing specific observables, stabilizing fragile quantum phases, shaping dynamical spectral responses, or enforcing constraint-like requirements such as low energy with respect to a reference Hamiltonian [Fig.~\ref{fig::Fig1}(b)]. Multiple goals can be incorporated simultaneously by assigning them weighted contributions within a single functional. Importantly, all components of $\mathcal{C}$ are evaluated directly from expectation values and correlators that are measured on the quantum state prepared on hardware, thus providing a quantitative figure of merit that guides the optimization toward states realizing the desired properties.

Once the cost functional is specified, IQS proceeds by variationally training parameterized quantum circuits acting on a pre-prepared—potentially highly correlated—initial state. In the Hubbard example, one natural realization employs native fermions, using fermionic atoms trapped in optical lattices as an analog or programmable quantum simulator. In this setting we start from a low-entropy state 
$\hat \rho_0 = e^{-\beta \hat H_0}/Z$, obtained, e.g., from cooling or adiabatic state preparation within the correlated regime \cite{BollScience2016, MuqingNature2025, kendrick2025pseudogapfermihubbardquantumsimulator}, and employ short-depth variational circuits \( \hat U(\boldsymbol{\theta}) \), constructed from the native operations of the programmable fermionic device \cite{3nx4-bnyy}, to explore the local neighborhood $\hat U(\boldsymbol{\theta})\hat \rho_0\hat U^\dagger(\boldsymbol{\theta})$ and enhance pairing correlations. To reach phases inaccessible to shallow deformations of the initial state, we additionally consider the optimization of deep sequential circuits incorporating mid-circuit measurement and qubit reset, which enable transitions across topological boundaries and entry into genuinely distinct quantum phases.

In the third and final step of IQS, we perform Hamiltonian learning to reconstruct an effective local model 
\(\hat{H}_{\mathrm{opt}} = \sum_i c_i \hat{h}_i\),
where the operator set \(\{\hat{h}_i\}\) is systematically extended 
to identify the interactions that best reproduce the optimized state.
The employed reconstruction techniques apply to both pure and finite-temperature Gibbs states \cite{Qi2019determininglocal, PhysRevX.8.031029, PhysRevLett.122.020504} and require only a polynomial number of samples independent of temperature \cite{anuragNatPhys21, bakshi2023learningquantumhamiltonianstemperature}.
To verify the learned models, we further discuss validation strategies such as adiabatic unpreparation, which provide quantitative benchmarks for the fidelity and consistency of the reconstruction [SI Sec.~\ref{sec::HL_additional}].

\subsection*{State Preparation Techniques for Hubbard Systems}
To enhance superconducting pair correlations, we use variational state-preparation protocols that are naturally suited to implementations with fermionic atoms in optical lattices, where local tunneling patterns can be engineered with optical mirror devices
, while remaining directly applicable to digital platforms through efficient fermion-to-qubit mappings. 

When enhancing $d$-wave correlations in the fermionic Hubbard ladder example, the variational state is generated by applying a parameterized unitary circuit
\begin{align}
    \hat U(\boldsymbol{\theta})
    = e^{-i\theta_d \hat{H}_R^{(d)}} \, e^{-i\theta_{d-1} \hat{H}_R^{(d-1)}} \cdots e^{-i\theta_1 \hat{H}_R^{(1)}},
    \label{eqn::main_unitary_VQE}
\end{align}
to an initial many-body state $\hat{\rho}_0$, yielding
\begin{equation} \label{eq::methrhotheta}
    \hat{\rho}(\boldsymbol{\theta}) = \hat U(\boldsymbol{\theta}) \, \hat{\rho}_0 \, \hat U^\dagger(\boldsymbol{\theta}).
\end{equation}
The resource operators $\hat{H}_R^{(k)}$ are Hubbard Hamiltonians [Eq.~(\ref{eqn::main_HU})] with fixed horizontal hopping $t_{j,\,j+\hat{x}}^{(k)}$, vertical hopping between ladder legs $t_{j,\,j+\hat{y}}^{(k)}$, and on-site interaction $U^{(k)}$.

The initial state $\hat{\rho}_0$ serves as the resource state and is taken, unless stated otherwise, as the ground state of a reference Hubbard Hamiltonian $\hat{H}_0$. Importantly, $\hat{\rho}_0$ may also be chosen as a finite-temperature Gibbs state, which is already accessible in fermionic optical-lattice experiments and is explicitly considered in SI Sec~\ref{sec::HL_additional}.

Variational parameters are updated using a local Hessian analysis that steers optimization along unstable directions of the cost landscape (see SI Sec.~\ref{sec::AdaptVQE}), with gradients and Hessians estimated from local observables at circuit depth $k$. Even with such elementary Hubbard quenches as resource operations, increasing circuit depth yields a systematic and pronounced enhancement of $d$-wave correlations, as shown, for example, in Fig.~\ref{fig::Fig2}(a).

We now briefly comment on the choice of the cost function.
While one might want to directly maximize a superconducting order parameter, it is not particularly meaningful in the effectively 1D Hubbard ladder, where true long-range order is forbidden by the Mermin–Wagner theorem~\cite{altland2010condensed}.  
For the same reason, the 1D nature of our system precludes any claim of enhancing $T_c$.
Experimentally, however, a phase transition is typically identified not by the onset of long-range order but by a  divergence (strong enhancement) of the susceptibility associated with the relevant instability near the critical temperature. 
Our cost functions are designed to amplify low-momentum $d$-wave fluctuations, thereby promoting the emergence of a superconducting instability.  
As such, and combined with the results in Fig.~\ref{fig::Fig2}(d)and Fig.~\ref{fig::Fig_temperature}(d), this suggests an enhanced $T_c$ when the learned Hamiltonian is applied to real materials.  
Moreover, the ultimate physical quantity of interest is the zero-frequency conductivity, whose enhancement does not require long-range coherence but only strong superconducting fluctuations -- a phenomenon known as paraconductivity -- underscoring the relevance of our simplified simulations.

\subsection*{Hamiltonian Learning}
For the Hamiltonian learning step of the IQS protocol,
we adopt a modification of the method proposed in Ref.~\cite{Qi2019determininglocal}, which we formulate as a \emph{linear regression} problem. Given a quantum state \( \ket{\psi} \) (in our case, corresponding to the optimized state with parameters $\theta_{\rm opt}$, cf. Eq.~\eqref{eq::methrhotheta}), our objective is to determine a Hamiltonian of the form
\begin{align} \label{eq::linansatz}
    \hat{H} = \hat{H}_0 + \sum_i c_i \hat{h}_i,
\end{align} 
for which \( \ket{\psi} \) is the ground state. The term \( \hat{H}_0 \) denotes a fixed reference Hamiltonian -- such as one containing known kinetic energy and on-site interaction terms -- while the coefficients \( c_i \) are variational parameters to be determined.

An important insight of Ref.~\cite{Qi2019determininglocal} is that if \(\ket{\psi}\) is an eigenstate of $\hat{H}$ then the variance
\begin{align}
    \langle \hat{H}^2 \rangle - \langle \hat{H} \rangle^2 = 0
\end{align}
has to be zero. Since the variance is positive, naturally one determines the coefficients $c_i$ by minimizing
\begin{align}
\begin{split}
    {\cal L}[\bm c] & \equiv \langle \hat{H}^2 \rangle - \langle \hat{H} \rangle^2 \\&= \sum_{ij} G_{ij} c_i c_j - 2 \sum_i c_i \Big( \langle \hat{H}_0 \rangle \langle\hat{h}_i\rangle - \frac{1}{2}\langle \{\hat{H}_0,\hat{h}_i \} \rangle \Big)  \\
    & + \langle \hat{H}_0^2 \rangle - \langle \hat{H}_0 \rangle^2,\label{eqn::L_vHL}
    \end{split}
\end{align}
where 
\begin{align}
    G_{ij} \equiv \frac{1}{2} \langle \{ \hat{h}_i, \hat{h}_j \} \rangle - \langle \hat{h}_i \rangle\langle \hat{h}_j \rangle
    \label{eqn::G_corr_mat}
\end{align}
defines a real symmetric correlation matrix.
Minimization of ${\cal L}[\bm c]$ then gives:
\begin{align}
    \sum_j G_{ij} c_j =  \langle \hat{H}_0 \rangle \langle\hat{h}_i\rangle - \frac{1}{2}\langle \{\hat{H}_0,\hat{h}_i \} \rangle  =: v_i. \label{eqn:QR_ext_eq}
\end{align}

Before proceeding, we note that since the correlation matrix $G$ is real and symmetric, it is diagonalizable; additionally, its eigenvalues are non-negative. 
For the linear equation 
\begin{align}
    G \bm c = \bm v,
\end{align}
to have a solution, it is required that $\bm v \perp \text{ker}(G)$. This condition is indeed satisfied in our case:
\begin{proof}
Suppose $\bm c \in \text{ker}(G)$, i.e., it is an eigenvector of $G$ with zero eigenvalue. 
This immediately implies that $\ket{\psi}$ is an eigenstate of $\sum_i c_i \hat{h}_i$, i.e., $\sum_i c_i \hat{h}_i\ket{\psi} = {\cal E}\ket{\psi}$ for some constant \( \mathcal{E} \). We this it is easy to show that
\begin{align}
    \bm v \cdot \bm c & = \sum_i c_i \Big(\langle \hat{H}_0 \rangle \langle\hat{h}_i\rangle - \frac{1}{2}\langle \{\hat{H}_0,\hat{h}_i \} \rangle \Big) = 0
\end{align}
This establishes that $\bm v \perp \text{ker}(G)$.
\end{proof}

We therefore conclude that Eq.~\eqref{eqn:QR_ext_eq} has a solution $\bm c_{\mathrm{inhom}}$. In fact, any solution can be written as 
\begin{align}
    \bm c = \bm c_{\mathrm{inhom}} + \sum_a \alpha_a \bm u_a,
\end{align}
where the vectors $\{ \bm u_a \}$ form an orthonormal basis of $\ker(G)$.

In practice, we often find that \( \ker(G) \) has nontrivial dimensionality, i.e., \( \dim[\ker(G)] \neq 0 \). Moreover, the learned Hamiltonian -- despite exhibiting zero variance -- may yield \( \ket{\psi} \) as an excited state rather than the ground state. 
Both of these issues can be partially addressed by modifying the variance cost function as follows:
\begin{align}
    \mathcal{L}[\bm{c}] \to \mathcal{L}[\bm{c}] 
    &= \langle \hat{H}^2 \rangle - \langle \hat{H} \rangle^2 + 2\alpha \langle \hat{H} \rangle.
\end{align}
This additional linear term often significantly improves the performance of the learning algorithm by biasing toward lower-energy solutions. 
The biasing is especially effective when \( \hat{H}_0 = 0 \) and a normalization condition such as \( \sum_i c_i^2 = 1 \) is imposed. In this case, we find that it helpful to further restrict the energy biasing to the linear subspace \( \ker(G) \).

We also introduce a second regularization term to control the magnitude of the coefficients:
\begin{align}
    {\cal L}[\bm c] & = \langle \hat{H}^2 \rangle - \langle \hat{H} \rangle^2 + 2 \alpha \langle H\rangle + \sum_i w_i c_i^2.
    \label{eqn::HL_full_cost_func}
\end{align}
The weights \( w_i \) ensure that the coefficients \( c_i \) remain bounded; they can also encode practical considerations such as the experimental cost or difficulty of implementing the corresponding terms \( \hat{h}_i \).
Minimization of the cost function ${\cal L}$ is nothing but the ridge regression and gives:
\begin{align}
    \big[ G_{ij} + w_i \delta_{ij} \big]c_j = v_i - \alpha \langle \hat h_i\rangle.
\end{align}

We note that in the CPHL algorithm (see SI Sec.~\ref{sup::cphl}), we set \( \alpha = 0 \) to ensure that if the initial state \( \ket{\psi} \) is an eigenstate (ground or excited) of \( \hat{H}_0 \), the resulting vector \( v_i = 0 \) and the learned coefficients \( c_i = 0 \), thereby halting further updates. This choice guarantees that the algorithm converges back to the original Hamiltonian \( \hat{H}_0 \), as expected.

As a remark, we emphasize that additional linear constraints can be imposed straightforwardly within the presented Hamiltonian learning framework. 
%
Since the correlation matrix $G$ in Eq.~\eqref{eqn::G_corr_mat} is positive-semidefinite, this guarantees that the unconstrained optimization problem to determine the coefficients $c_i$ is convex. Imposing linear constraints on $c_i$ restricts the feasible set to convex regions, and since the intersection of convex manifolds is itself convex, the constrained optimization remains a convex quadratic program. The resulting problem can therefore be efficiently solved using standard quadratic-programming solvers. In Fig.~\ref{fig::Fig2}, we exploit this flexibility to restrict the density-density interaction terms in the fermionic ladder Hamiltonian to be purely repulsive by enforcing non-negativity of the corresponding coupling coefficients.

\subsection*{Optimizing Response Functions}

We begin by writing down the equations of motion for the case where the variational manifold of quantum states being explored is parametrized by variables \( \theta_0, \theta_1, \dots, \theta_n \) [Fig.~\ref{fig::Figure_dyn}(a)]:
\begin{align}
    \ket{\psi} = e^{-i\theta_n \hat{A}_n}e^{-i\theta_{n - 1} \hat{A}_{n - 1}}\dots e^{-i\theta_1 \hat{A}_1}  e^{-i \theta_0 \hat{I}} \ket{\psi_0}. \label{eqn::ref_wf_chi}
\end{align}
Given a (time-dependent) Hamiltonian $\hat{H}$, the McLachan time-dependent variational principle yields~\cite{yuan2019theory, PRXQuantum.2.030307}:
\begin{align}
     {\cal M}_{ij}\dot{ \theta_{j}}  = 
    \text{Re}\big[ \!
    \bra{\psi} \hat{H} \ket{u_i} - \braket{\psi|\hat{H}|\psi}\!\braket{\psi|u_i}\!
    \big],  \label{eqn::var_dyn}
\end{align}
where  $\ket{u_i} \equiv i\partial_{\theta_i}\!\ket{\psi}$, and the metric matrix \( \mathcal{M} \) is given by
\begin{align}
    {\cal M}_{ij}  \equiv \text{Re}\big[\! 
\braket{u_i|u_j} - \braket{\psi| u_i}\!\braket{u_j |\psi} \!\big].
\end{align}
Note the inclusion of the global phase \( \theta_0 \) in Eq.~\eqref{eqn::ref_wf_chi}: although this phase does not influence physical observables, its dynamics does implicitly affect the variational equation of motion~\eqref{eqn::var_dyn}. In Eq.~\eqref{eqn::var_dyn} and below, we have ``integrated out'' this degree of freedom and track only the dynamics of the physical parameters \( \theta_1, \dots, \theta_n \).
We also remark that since \( \braket{\psi | u_i} \in \mathbb{R} \), the matrix \( \mathcal{M} \) is real and symmetric. 
The variational state in Eq.~\eqref{eqn::ref_wf_chi} is not intended for physical implementation on a quantum circuit, and thus the number of parameters $n$ can scale extensively with system size.

If the reference state $\ket{\psi_0}$ in Eq.~\eqref{eqn::var_dyn} does not correspond to the ground state, one first variationally minimizes a reference Hamiltonian $\hat{H}_0$ -- the corresponding optimal parameters are denoted by \( \bar{\theta}_j \). 
The system is then subjected to a weak, time-dependent perturbation \(\delta\hat{V}(t)\), and we aim to track the induced response in a target observable \(\hat{\mathcal{O}}\). Throughout, we adopt the convention
\begin{align}
    \hat{H}(t) = \hat{H}_0 +  \delta v \big( \hat{V} e^{-i\omega t} + \hat{V}^\dagger e^{i \omega t}\big),
\end{align}
where \(\delta v\) denotes the perturbation strength and is assumed to be weak.

Under this time-dependent perturbation, the variational parameters $\theta_j(t)$ evolve according to Eq.~\eqref{eqn::var_dyn}; to linear order in $\delta v$ we can expand:
\begin{align}
    \theta_j(t) \approx \bar{\theta}_j + \delta \theta_j(t).
\end{align}
Evaluation of the observable of interest is then understood as:
\begin{align}
    {\cal O} & \equiv \bra{\psi_0}e^{i\theta_1 (t) \hat{A}_1}\dots  e^{i\theta_n(t) \hat{A}_n} \hat{\mathcal{O}}e^{-i\theta_n(t) \hat{A}_n}\dots e^{-i\theta_1(t) \hat{A}_1} \ket{\psi_0}  \notag\\
    & \approx 
    {\cal O}_0 + \sum_{j = 1}^n  {\cal O}_j  \delta \theta_j(t),
    \label{eqn::lin_response_v0}
\end{align}
where ${\cal O}_j = i [ \braket{u_j|\hat{\cal O} | \psi} -\braket{\psi|\hat{\cal O} | u_j} ]$. Upon further linearizing Eq.~\eqref{eqn::var_dyn} with respect to the applied perturbation \(\delta \hat{V}\), we finally obtain $\delta{\cal O}(\omega) = \chi_{\rm var}(\omega) \delta v$ with:
\begin{align}
   \chi_{\rm var}(\omega) = \sum_{ij}  {\cal O}_i \Big( \frac{1}{ -i\omega {\cal M} +  {\cal K}}\Big)_{ij}  v_j,
    \label{eqn::var_chi_full}
\end{align}
where $v_j \equiv ( \braket{\psi| \hat{V} - V_\psi  | u_j} +  \braket{u_j| \hat{V} - V_\psi  | \psi}) / 2$ and 
\begin{align}
    {\cal K}_{ij} \equiv \text{Im}\big[\! \braket{u_j |\hat{H}_0 - E_\psi | u_i} - 
\braket{\psi |\hat{H}_0 - E_\psi  | u_{ij}} \!
\big].
\end{align}
Here $V_\psi \equiv \braket{\psi | \hat{V}| \psi}$, $E_\psi \equiv \braket{\psi| \hat{H}_0 |\psi}$, and $\ket{u_{ij}} \equiv i \partial_{\theta_j}\ket{u_i} = -\partial_{\theta_i\theta_j}^2\ket{\psi}$. 
All quantities are evaluated on top of the unperturbed state $\ket{\psi}$ with $\bar{\theta}_j$.

We compare the variational linear response in Eq.~\eqref{eqn::var_chi_full} to the exact one, which is evaluated using the Kubo formula:
\begin{equation}
    \chi(\omega) = \sum_n \Big[ \frac{ \braket{0|\hat{\cal O}|n}\!\braket{n|\hat{V}|0} }{\omega + i\delta - ({\cal E}_n - {\cal E}_0)} - \frac{ \braket{0|\hat{V}|n}\!\braket{n|\hat{\cal O}|0} }{\omega + i\delta + ({\cal E}_n - {\cal E}_0)}\Big].
    \label{eqn::Kubo}
\end{equation}

Written in the form of Eq.~\eqref{eqn::var_chi_full}, the evaluation of linear response functions on quantum hardware reduces to state preparation and subsequent measurement of overlaps such as \( \braket{\psi | u_i} \) and \( \braket{\psi | \hat{H}_0 | u_i} \). 
These overlaps are, in general, complex-valued, and extracting both their real and imaginary components could pose a significant practical challenge for quantum devices.
Although dedicated measurement circuits can be constructed to estimate these quantities~\cite{yuan2019theory,PRXQuantum.2.030307}, they might be impractical for the state-of-the-art quantum machines. 
It is therefore highly desirable to develop approaches that avoid such circuits altogether.

To this end, we note that a significant simplification arises when the reference state \(\ket{\psi_0}\) in Eq.~\eqref{eqn::ref_wf_chi} is already the ground state of \(\hat{H}_0\).
In this case every stationary parameter vanishes,
\(\bar{\theta}_j = 0\), and each TDVP tangent vector becomes
$\ket{u_j} = \hat{A}_j \ket{\psi_0}.$
As a result, all matrix elements required for the linear-response evaluations reduce to expectation values of Hermitian operators, which are accessible using standard measurement circuits.
For example,
\begin{align}
    \operatorname{Re}\!\bigl[\langle u_i | u_j \rangle\bigr]
    \;=\;
    \frac{1}{2}\,
    \langle\psi_0 | \hat{A}_i \hat{A}_j + \hat{A}_j \hat{A}_i | \psi_0\rangle.
\end{align}
(We also note an additional simplification:
\(
\langle \psi_0 | (\hat{H}_0 - E_0) | u_{ij} \rangle = 0.
\))
In this simplified setting, one can systematically add Hermitian operators \( \hat{A}_k \) to the ansatz in Eq.~\eqref{eqn::ref_wf_chi} until the resulting linear response converges.

Once the TDVP elements have been measured, the rest of the calculation of variational susceptibility $\chi_{\mathrm{var}}(\omega)$ is done on a classical computer and, thus, is computationally inexpensive.
Optimization of the dynamical properties reduces to learning a Hamiltonian of the form
\begin{align}
    \hat{H}_0 = \sum_{j} c_{j}\hat{h}_{j}
    \label{eq:H_expansion}
\end{align}
from a wave function that minimizes the response‐based cost functional $\mathcal{L}[{\bm \alpha}]$, which we further discuss in SI Sec.~\ref{subsec::dyn_Cost}.   

\clearpage
\onecolumngrid
\beginsupplement

\title{ Supplemental Information for ``Inverse Quantum Simulation for Quantum Material Design'' }

\author{Christian~Kokail}
\thanks{These authors contributed equally to this work\\
\href{mailto:christian.kokail@cfa.harvard.edu}{christian.kokail@cfa.harvard.edu}\\
\href{mailto:p_dolgirev@g.harvard.edu}{p\_dolgirev@g.harvard.edu}}
\affiliation{\ITAMP}
\affiliation{\Harvard}
\affiliation{\QuEra}

\author{Pavel~E.~Dolgirev}
\thanks{These authors contributed equally to this work\\
\href{mailto:christian.kokail@cfa.harvard.edu}{christian.kokail@cfa.harvard.edu}\\
\href{mailto:p_dolgirev@g.harvard.edu}{p\_dolgirev@g.harvard.edu}}
\affiliation{\Harvard}
\affiliation{\QuEra}

\author{Rick~van~Bijnen}
\affiliation{\InnsbruckTh}
\affiliation{\InnsbruckQO}
\affiliation{\PlanQC}

\author{Daniel Gonz\'alez-Cuadra}
\affiliation{\Harvard}
\affiliation{\CSIC}

\author{Mikhail~D.~Lukin}
\affiliation{\Harvard}

\author{Peter~Zoller}
\affiliation{\InnsbruckTh}
\affiliation{\InnsbruckQO}

\date{\today}

\maketitle

\beginsupplement

\section{Hessian augmented ADAPT-VQE}
\label{sec::AdaptVQE}

Motivated by the prospect of implementing quantum circuits that directly optimize the observable of interest (here, the energy), we adopt a VQE-type variational ansatz of the form  
\begin{align}
    \ket{\psi} = e^{-i\theta_n \hat{A}_n} \, e^{-i\theta_{n-1} \hat{A}_{n-1}} \dots e^{-i\theta_1 \hat{A}_1} \ket{\psi_0},
\end{align}  
where the Hermitian operators $\hat{A}_j$ are drawn from a fixed pool of $N$ operators $\{ \hat{A}_1, \hat{A}_2, \dots, \hat{A}_N\}$, and $\ket{\psi_0}$ is a reference state (often chosen here as the ground state of a target fermionic Hubbard Hamiltonian).

Following Ref.~\cite{grimsley2019adaptive}, we employ the ADAPT-VQE method. Starting from a current state $\ket{\psi}$ with $n$ operators already appended and optimized, the goal is to select the next most relevant operator to extend this ansatz. To this end, we consider the auxiliary state 
\begin{align}
    \ket{\psi(\bm \alpha)} = \exp\big( - i\sum_{i = 1}^N \alpha_i \hat{A}_i \big) \ket{\psi}, \label{eqn::var_state}
\end{align} 
which involves a superposition over all pool operators (while such a superposition is not necessarily directly implementable on quantum hardware, it is useful for our analysis). Expanding the energy yields  
\begin{align}
    E(\bm \alpha)  = \bra{\psi} \exp\big( i\sum_{i} \alpha_i \hat{A}_i \big)  \hat{H} \exp\big( - i\sum_{i} \alpha_i \hat{A}_i \big)\ket{\psi} \approx \langle \hat{H} \rangle + i \sum_i \alpha_i \langle [\hat{A}_i,\hat{H}] \rangle .
\end{align}
The central idea of ADAPT-VQE is to select the operator associated with the largest energy gradient,  
\begin{align}
    \frac{\partial E(\bm \alpha)}{\partial \alpha_i}\Big|_{\bm \alpha=0} = i \langle [\hat{A}_i, \hat{H}] \rangle,  
    \label{eqn::grad_ADAPTVQE}
\end{align}  
which can be directly estimated on quantum hardware via measurements of commutators. This operator is then appended to the circuit as an additional variational unitary, and the full ansatz is re-optimized.

\begin{figure}[t!]
    \centering
    \includegraphics[width=0.85\linewidth]{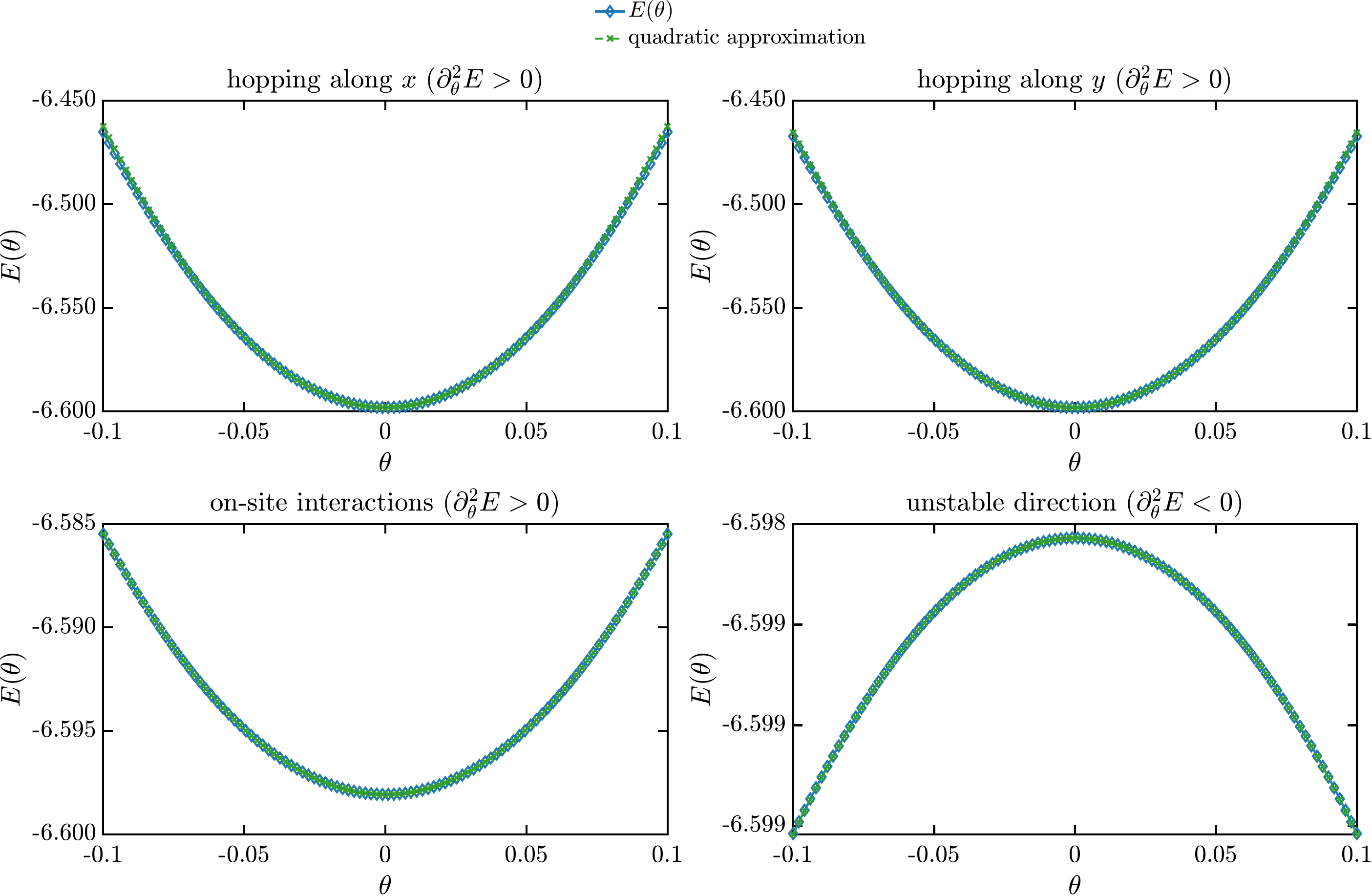}
    \caption{ Energy of the variational state $\exp(-i\theta \hat{A})\ket{\psi_0}$  as a function of $\theta$ for the Hubbard model with $t_x = t_y = -1$ (these set the unit of energy) and $U=8$.
    The reference state $\ket{\psi_0}$ is the ground state of the $U=4$ Hubbard Hamiltonian with the same hoppings.
    For the three natural Hubbard operators in the pool—hopping along $x$, hopping along $y$, and on-site interaction—we find that the gradient~\eqref{eqn::grad_ADAPTVQE} vanishes, while the second derivative is positive ($\partial^2_\theta E>0$),
    which makes the state $\ket{\psi_0}$ appear optimized.
    However, Hessian analysis [Eq.~\eqref{eqn::Hessian}] reveals a saddle point with an unstable direction ($\partial^2_\theta E<0$). Quadratic expansions (green crosses) are overlaid with exact numerics (blue diamonds), thereby providing a benchmark for the Hessian calculations.
    }
    \label{fig::Hessian_Hubbard}
\end{figure}

During our work on the Hubbard model, we found that the ADAPT-VQE gradient can vanish even when the variational state is not at a local minimum  -- see Fig.~\ref{fig::Hessian_Hubbard} for an example. 
In general, this situation arises when the reference state $\ket{\psi_0}$ is real (e.g., the ground state of a real Hamiltonian), and when both the Hamiltonian and all operators in the pool $\hat{A}_j$ are real.  
In this case, the expectation value $\langle [\hat{A}_i, \hat{H}] \rangle$  must be real, as it involves products of real matrices and vectors, while the commutator $[\hat{A}_i,\hat{H}]$ is anti-Hermitian and therefore has a purely imaginary expectation value. 
The only consistent resolution is that $\langle [\hat{A}_i, \hat{H}] \rangle = 0$, implying that the ADAPT-VQE gradient~\eqref{eqn::grad_ADAPTVQE} vanishes and the method may falsely appear to have converged, even though the state $\ket{\psi_0}$ can still be far from optimal (Fig.~\ref{fig::Hessian_Hubbard}).

To address this issue, we expand energy to second order in $\bm \alpha$
\begin{align}
    E(\bm \alpha)  \approx \langle \hat{H} \rangle + i \sum_i \alpha_i \langle [\hat{A}_i,\hat{H}] \rangle - \frac{1}{2} \sum_{i,j} \alpha_i\alpha_j \langle [\hat{A}_i,[\hat{A}_j,\hat{H}]]\rangle
\end{align}
and evaluate the Hessian:
\begin{align}
    \frac{\partial^2 E(\bm \alpha)}{\partial \alpha_i \partial \alpha_j}\Big|_{\bm \alpha = 0} = - \frac{1}{2} \big( \langle [\hat{A}_i,[\hat{A}_j,\hat{H}]]\rangle + \langle [\hat{A}_j,[\hat{A}_i,\hat{H}]]\rangle \big).
    \label{eqn::Hessian}
\end{align}
By diagonalizing this matrix and examining its eigenvalues, we can determine whether a true local minimum has been reached, indicated by all eigenvalues being positive. If the minimum has not been achieved, the eigenvectors can reveal unstable operator directions -- see Fig.~\ref{fig::Hessian_Hubbard}. 
Additionally, if physical constraints prevent the addition of superpositions of operators, all pertinent operators can be added simultaneously in the current iteration step.
Importantly, evaluating the Hessian on quantum hardware reduces to measuring Hermitian operators, a task already within reach of current quantum devices.

\emph{Gradient evaluation routine}---For completeness, we mention that in order to facilitate the optimization of the energy
\begin{align}
    E(\bm \theta) = \bra{\psi_0}  e^{i\theta_1 \hat{A}_1}\dots  e^{i\theta_n \hat{A}_n} \hat{H} e^{-i\theta_n \hat{A}_n}\dots e^{-i\theta_1 \hat{A}_1} \ket{\psi_0}
\end{align}
with respect to the variational parameters $\theta_1,\dots,\theta_n$, 
we supplement our numerical routines with the gradient $\partial E/\partial \theta_j$. 
It can be evaluated as (see Ref.~\cite{shi2018variational} for additional details and for a hardware-oriented discussion) 
\begin{align}
    \frac{\partial E(\bm \theta)}{\partial \theta_i} = 2 \,\text{Im}\big[ \braket{\varphi_i| \hat{A}_i | \psi_i } \big], \label{eqn::E_grad_an}
\end{align}
where 
\begin{align}
    \ket{\psi_i}  \equiv  e^{-i\theta_i \hat{A}_i}\dots e^{-i\theta_1 \hat{A}_1} \ket{\psi_0}, \qquad \ket{\varphi_i} \equiv e^{i\theta_{i + 1} \hat{A}_{i + 1}}\dots e^{i\theta_n \hat{A}_n} \hat{H} \ket{\psi}.
\end{align}
In practice, the gradient in Eq.~\eqref{eqn::E_grad_an} is conveniently computed in reverse order, starting from $i=n$ down to $i=1$.  
At the final layer, we initialize $\ket{\psi_n} = \ket{\psi}$ and $\ket{\varphi_n} = \hat{H}\ket{\psi}$, and then update iteratively via  
\begin{align}
    \ket{\psi_i} = e^{i\theta_{i+1}\hat{A}_{i+1}} \ket{\psi_{i+1}}, \qquad
    \ket{\varphi_i} = e^{i\theta_{i+1}\hat{A}_{i+1}} \ket{\varphi_{i+1}}.
\end{align}

\section{Finite temperature routines}
\label{sec::finite_T}

Consider initializing not just a wave-function but a thermal density matrix:
\begin{align}
    \hat{\rho}_0 = \frac{1}{{\cal Z}} e^{ - \beta \hat{H}_0} = \sum_\alpha p_\alpha \ket{\psi^\alpha_0}\!\bra{\psi^\alpha_0},
\end{align}
where ${\cal Z} = \tr \big\{ e^{ - \beta \hat{H}_0} \big\}$ denotes the partition function, $\ket{\psi_0^\alpha}$ represents the energy eigenstates of $\hat{H}_0$, and $p_\alpha = e^{-\beta E_0^\alpha}/{\cal Z}$ are the corresponding Boltzmann weights.

On a quantum machine, we can prepare a density matrix of the form
\begin{align}
    \hat{\rho}(\bm \theta) = e^{-i\theta_n \hat{A}_n}e^{-i\theta_{n - 1} \hat{A}_{n - 1}}\dots e^{-i\theta_1 \hat{A}_1} \hat{\rho}_0  e^{i\theta_1 \hat{A}_1}\dots  e^{i\theta_n \hat{A}_n},
\end{align}
and our goal is to minimize the expectation value of some Hermitian observable $\hat{\cal D}$ of interest:
\begin{align}
    {\cal L}(\bm \theta) \equiv \tr\big\{ \hat{\rho}(\bm \theta) \hat{\cal D} \big\} = \sum_\alpha p_\alpha \bra{\psi_0^\alpha} e^{i\theta_1 \hat{A}_1}\dots  e^{i\theta_n \hat{A}_n}  \hat{\cal D}
    e^{-i\theta_n \hat{A}_n}e^{-i\theta_{n - 1} \hat{A}_{n - 1}}\dots e^{-i\theta_1 \hat{A}_1} \ket{\psi_0^\alpha}.
\end{align}

\emph{ADAPT-VQE at finite $T$}---Equation~\eqref{eqn::grad_ADAPTVQE} is now understood as:
\begin{align}
    \frac{\partial {\cal L}}{\partial \alpha_i} = i \sum_\alpha p_\alpha \braket{\psi^\alpha|[\hat{A}_i,\hat{\cal D}] |\psi^\alpha},
\end{align}
and a similar generalization holds for the Hessian.

\emph{Gradient evaluation routine}---Evaluation of the gradient is analogous to Eq.~\eqref{eqn::E_grad_an}:
\begin{align}
    \frac{\partial {\cal L}(\bm \theta)}{\partial \theta_i} =
     2 \sum_\alpha p_\alpha\text{Im}\big[ \braket{\varphi_i^\alpha| \hat{A}_i | \psi_i^\alpha } \big],
\end{align}
where 
\begin{align}
    \ket{\psi_i^\alpha}  & \equiv  e^{-i\theta_i \hat{A}_i}\dots e^{-i\theta_1 \hat{A}_1} \ket{\psi_0^\alpha}, \qquad \ket{\varphi_i^\alpha} \equiv e^{i\theta_{i + 1} \hat{A}_{i + 1}}\dots e^{i\theta_n \hat{A}_n} \hat{\cal D} \ket{\psi^\alpha}.
\end{align}

\emph{Hamiltonian learning from finite-temperature Gibbs states}---Once the variational states $\hat{\rho}(\bm \theta_\text{opt})$ have been prepared, the next step is to learn the parent Hamiltonian $\hat{H}_\text{opt}$ that underlies the state 
\[
\hat{\rho}(\bm \theta_\text{opt}) = \frac{1}{Z}\, e^{-\beta \hat{H}_\text{opt}}.
\]
There are several ways to infer $\hat{H}_\text{opt}$. Most directly, if the variational circuit $U(\bm\theta)$ prepares the state from a reference Gibbs state of a known Hamiltonian $\hat{H}_0$, one has
\[
\hat{H}_\text{opt} = U(\bm\theta_\text{opt})\, \hat{H}_0\, U^{\dagger}(\bm\theta_\text{opt}).
\]
If $\hat{H}_0$ is local, and $U(\bm\theta)$ is shallow and composed of local gates, this conjugation can even be computed classically by successively transforming the individual terms of $\hat{H}_0$. In the case of long-range gates or deep circuits, however, local terms may spread under the circuit, leading to interactions of beyond a desired range. To mitigate this, one may truncate the operator growth -- for instance using schemes based on variational projection back to the space of $\ell$-local Hamiltonians. Such an approach is for instance discussed in Ref.~\cite{PhysRevLett.132.160401}, where the projection is performed via a time-dependent variational principle that leads to local Hamiltonians that minimize the trace distance to the current operator.

In scenarios involving deep circuits, such a consecutive conjugation approach becomes infeasible. In these cases, one can instead infer $\hat{H}_\text{opt}$ directly from the prepared state using Hamiltonian learning techniques. It has been shown that local Hamiltonians can be efficiently learned from Gibbs states using a polynomial number of samples for arbitrary temperatures~\cite{bakshi2023learningquantumhamiltonianstemperature}. Practical  approaches for learning such Hamiltonians have been proposed -- for example in Ref.~\cite{PhysRevLett.122.020504}, where the ansatz coefficients are constrained using expectation values derived from the Ehrenfest theorem. In previous work \cite{PRXQuantum.2.010102}, it was shown that Hubbard-type Hamiltonians can be learned from low-entropy states using this method in a scalable manner. 

In our fermionic Hubbard example, the optimized circuits remain shallow, so the corresponding Hamiltonians could in principle be handled efficiently through direct conjugation. However, for the purposes of comparing the resulting Gibbs states with the exact results, we instead use a straightforward reconstruction approach based on Gibbs-state Hamiltonian tomography~\cite{KokailNatPhys} to infer local parent Hamiltonians associated with $\hat{\rho}(\boldsymbol{\theta}_\text{opt})$. We introduce a local Hamiltonian ansatz $\hat{H}(\boldsymbol{c}) = \sum_i c_i \hat{h}_i$ which we use to generate a parametrization of the corresponding Gibbs state $\hat{\rho}(\boldsymbol{c}) = 1/Z(\boldsymbol{c}) e^{-\hat{H}(\boldsymbol{c})}$. The optimal coefficients $\boldsymbol{c}_\text{opt}$ are obtained by minimizing the Hilbert--Schmidt distance
\[
\mathfrak{D}\!\left[\hat{\rho}(\boldsymbol{c}), \hat{\rho}(\boldsymbol{\theta}_\text{opt})\right]
= \big\| \hat{\rho}(\boldsymbol{c}) - \hat{\rho}(\boldsymbol{\theta}_\text{opt}) \big\|_{\mathrm{HS}}^2 .
\]
The results of this reconstruction procedure are shown in Fig.~\ref{fig::Fig_temperature}(c), where we plot the expectation value of the cost function in the learned state $\hat{\rho}(\boldsymbol{c}_\text{opt})$ and compare it with the exact results for the variationally optimized state in Fig.~\ref{fig::Fig_temperature}(b).

\section{ Extended discussion and additional results on fermionic Hubbard ladders}
\label{sec::HL_additional}

\subsection{Learning maps and locality of the Hamiltonian anzatz}

In this section, we analyze how the learning maps introduced in Fig.~2 of the main text depend on the locality of the Hamiltonian ansatz. Specifically, we demonstrate that systematically extending the ansatz to include additional interaction terms enlarges the region of low Hamiltonian variance -- i.e., the regime in which faithful learning is possible -- toward larger values of both $\lambda$ and the circuit depth $d$.

\begin{figure}
    \centering
    \includegraphics[width=0.9\linewidth]{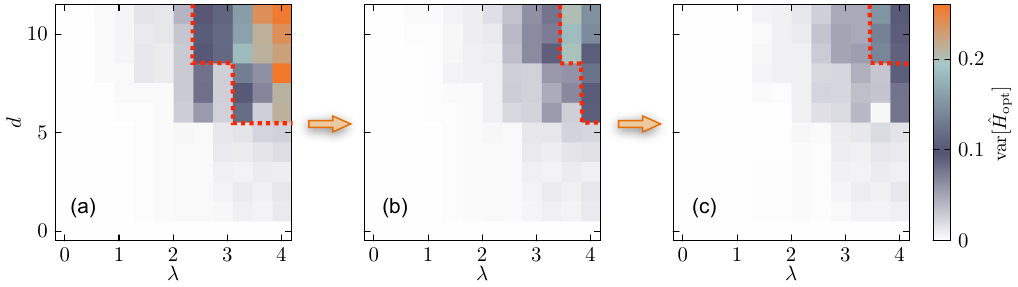}
    \caption{ \textbf{Extension of the learning maps for the Hamiltonian variance by adding additional ansatz terms.} (a) The learning map generated from the extended Hubbard Hamiltonian containing the nearest-neighbor 
    and next-nearest-neighbor hopping terms as well as the onsite density-density interaction.
     (b) Improved learning map via adding offsite density-density interaction $\sum_\sigma \hat{n}_{i, \sigma} \hat{n}_{i + \hat{y}, \sigma}$ along the vertical direction.
    (c) In addition to the terms in (b), further improvement can be achieved by including the diagonal density-density interaction $\sum_\sigma \hat{n}_{i, \sigma} \hat{n}_{i + \hat{x} + \hat{y}, \sigma}$, the next-next nearest-neighbor hopping $\hat{c}_{j, \sigma}^{\dagger} \hat{c}_{j + 2\hat{x} + \hat{y}} + \text{H.c.}$, and the next-next nearest-neighbor density-density interaction $\sum_\sigma \hat{n}_{i, \sigma} \hat{n}_{i + 2\hat{x} + \hat{y}, \sigma}$. 
    The red dashed lines indicate the region upon which the Hamiltonian variance $\text{var} [\hat{H}_\text{opt}]$ exceeds 0.1.
    }
    \label{fig::Fig_suppmaps}
\end{figure}

The results are summarized in Fig.~\ref{fig::Fig_suppmaps}, where we perform Hamiltonian learning on variationally optimized states without imposing constraints on the Hamiltonian coefficients. In Fig.~\ref{fig::Fig_suppmaps}(a), the ansatz consists of an extended Hubbard Hamiltonian that includes next-nearest-neighbor hopping terms. In this case, the learned Hamiltonians exhibit a significantly reduced hopping amplitude between ladder rungs -- an effect that enhances the $d$-wave pair correlator (see also Fig.~2 of the main text) -- as well as a plaquette-diagonal hopping with opposite sign relative to the nearest-neighbor terms. Incorporating off-site density--density interactions further reduces the high-variance region and shifts it toward larger values of $\lambda$, as shown in Fig.~\ref{fig::Fig_suppmaps}(b). Finally, the inclusion of longer-range interaction terms, specified in the figure caption~\ref{fig::Fig_suppmaps}, leads to an additional, though more modest, improvement, as illustrated in Fig.~\ref{fig::Fig_suppmaps}(c).










\subsection{Verification via adiabatic unpreparation}

\begin{figure}
    \centering
    \includegraphics[width=0.4\linewidth]{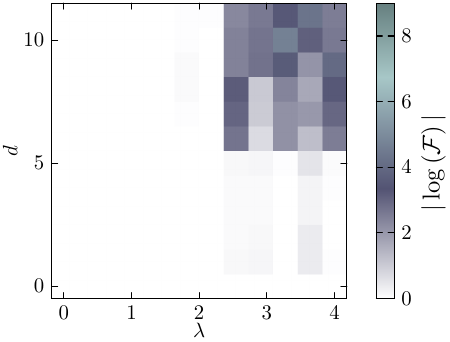}
    \caption{ \textbf{Learning map determined via adiabatic unpreparation.} Logarithm of the return probability when the variational states $\ket{\psi(\boldsymbol{\theta}_\text{opt})}$ of Fig.~2 (b) (main text) are mapped back to $\ket{\psi_0}$ via an adiabatic sweep.
    }
    \label{fig::Unprep}
\end{figure}

A central question in IQS on pure states is how to verify that the learned Hamiltonian has the optimized state as its ground state. Hamiltonian learning based on variance minimization [Methods] reconstructs a parent Hamiltonian for which the learned state is an approximate energy eigenstate, but does not, in general, guarantee that this eigenstate is the ground state. While alternative learning strategies exist that can enforce the ground-state condition by construction (see \cite{PhysRevLett.132.160401} and Sec.~\ref{sec::finite_T}), here we instead describe an independent verification procedure based on directly measuring the fidelity presented in the learning maps of Fig.~2 of the main text.

The protocol is based on a reverse adiabatic sweep from the learned Hamiltonian $\hat H(\boldsymbol{\theta}_{\mathrm{opt}})$ back to the parent Hamiltonian $\hat H_0$ of the initial state,
\begin{align}
\ket{\psi_0}
=
\mathcal{T}
\exp\!\Big[ -i
\int_0^T \! d\tau \,
\bigl( (1-\tau)\,\hat H(\boldsymbol{\theta}_{\mathrm{opt}})
+ \tau\, \hat H_0 \bigr)
\Big]
\ket{\psi(\boldsymbol{\theta}_{\mathrm{opt}})} .
\end{align}
Provided that no phase transition is crossed during the sweep, adiabatic evolution for sufficiently large $T$ returns the state to the original ground state $\ket{\psi_0}$. By subsequently reversing the initial state preparation to a known product state, the many-body fidelity can be accessed directly by measuring the return probability,
\begin{align}
\mathcal{F} = \Big|
\bra{\psi_0}
\mathcal{T}
\exp\!\Big[ -i
\int_0^T \! d\tau \,
\bigl( (1-\tau)\,\hat H(\boldsymbol{\theta}_{\mathrm{opt}})
+ \tau\, \hat H_0 \bigr)
\Big]
U(\boldsymbol{\theta}_{\mathrm{opt}})
\ket{\psi_0}
\Big|^2 .
\end{align}

To support the feasibility of this verification protocol, in Fig.~\ref{fig::Unprep} we numerically simulate such an adiabatic unpreparation procedure for the learning protocol discussed around Fig.~2 in the main text. These simulations demonstrate that, for the considered parameter regimes, adiabatic unpreparation is possible and faithfully reproduces the corresponding learning map.

\subsection{Enhancing $d$-wave pair correlations at finite temperature}

\begin{figure}
    \centering
    \includegraphics[width=1\linewidth]{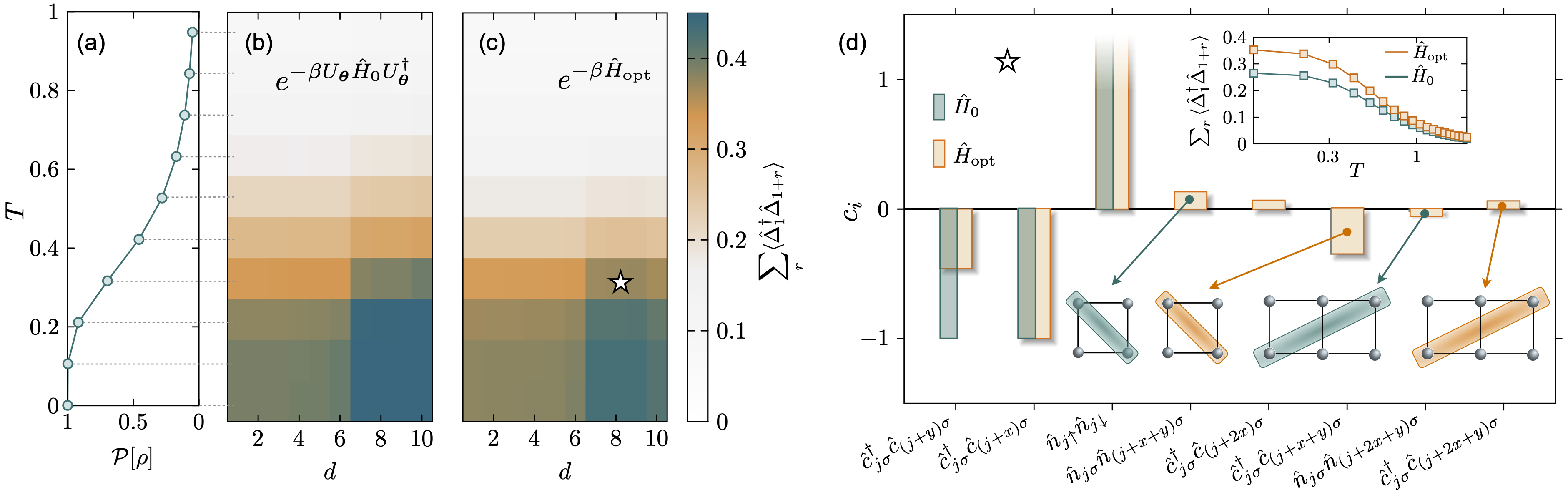}
    \caption{ \textbf{Finite-temperature enhancement of target properties.} (a) Purity, ${\cal P}[\rho] = {\rm tr} \rho^2$, of the four-rung Hubbard ladder decreases with increasing $T$. 
    (b) Optimization of the cost function (at $\lambda = 2$ in Eq.~(2) in the main text) for different $T$ and circuit depth $d$; during optimization, the purity remains fixed. 
    (c) Computed cost function of Hamiltonians learned at finite temperature, in good agreement with (b). 
    Notably, the optimized circuit at relatively high temperature $T = 0.32$ (star symbol) achieves a cost function comparable to the zero-temperature Hubbard model. 
    (d) Relevant terms of the star Hamiltonian (rescaled so that the $x$-hopping is set to one). The inset shows $d$-wave correlations as a function of $T$, with a clear enhancement in the learned model relative to the reference $\hat{H}_0$.  
    }
    \label{fig::Fig_temperature}
\end{figure}

While Fig.~2(d) in the main text shows that our results extend across a broad range of hole dopings, a central challenge for high-$T_c$ superconductivity is to stabilize and enhance fermion pairing at elevated temperatures. 
This is practically crucial because cold-atom experiments intrinsically prepare finite-$T$ Gibbs states, with current temperatures likely exceeding the regime of robust pairing, while still providing natural initial states for our optimization. 
We therefore extend the framework to finite $T$ and optimize directly on thermal density matrices,
$\hat\rho[ \bm \theta] = \hat{U}_{\bm \theta} \exp\big(- \hat{H}_0 / T\big)\hat{U}_{\bm \theta}^\dagger$, where $\hat{U}_{\bm \theta}$ represents the circuit operations. 
Apart from this substitution, both the optimization and learning steps proceed similarly as before [Sec.~\ref{sec::finite_T}].
During optimization, the purity ${\cal P}[\rho] = {\rm tr}\,\rho^2$ is preserved, while decreasing with increasing $T$ [Fig.~\ref{fig::Fig_temperature}(a)].  
Remarkably, a circuit optimized at relatively high temperature $T\simeq 0.32$ (star symbol) yields $d$-wave correlations compatible with the zero-temperature Hubbard model $\hat{H}_0$ [Fig.~\ref{fig::Fig_temperature}(b)]. Local Hamiltonians can be reconstructed efficiently from such states with polynomial sampling complexity as detailed in Sec.~\ref{sec::finite_T}.  
Figure~\ref{fig::Fig_temperature}(c) confirms that the reconstructed Hamiltonians across $(d,\, T)$ reproduce the optimized cost function of Fig.~\ref{fig::Fig_temperature}(b).  
Strikingly, the `star' circuit, which was optimized at a single finite temperature, yields a Hamiltonian $\hat{H}_{\text{opt}}$ [rescaled such that the $x$-hopping term is set to one, 
Fig.~\ref{fig::Fig_temperature}(d)], that outperforms $\hat{H}_0$ across the entire temperature range [inset of Fig.~\ref{fig::Fig_temperature}(d)].

\subsection{Optimizing pair correlations with a finite measurement budget }






\begin{figure}
    \centering
    \includegraphics[width=0.9\linewidth]{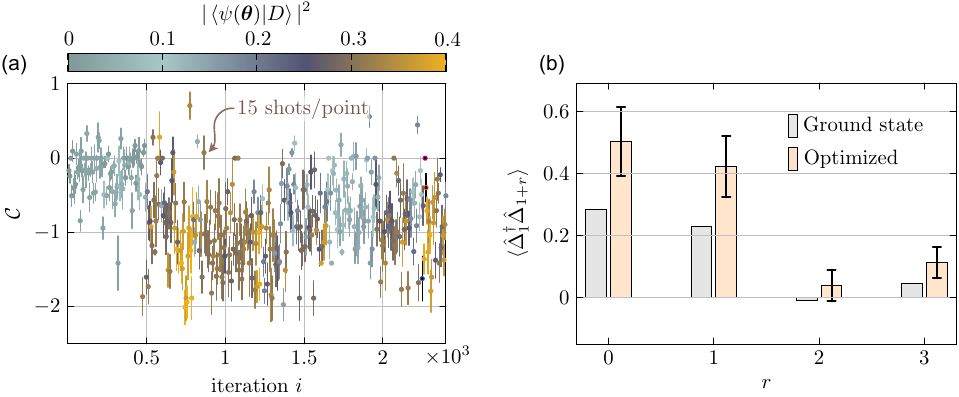}
    \caption{ \textbf{Optimizing the cost function under a finite measurement budget.} (a) Representative optimization trajectory of the sum of $d$-wave pair correlations $\mathcal{C} = \sum_r \braket{\hat{\Delta}_1^{\dagger} \hat{\Delta}_{1+r}}$ for a circuit with depth $d = 5$ (see Eq.~(\ref{eqn::main_wf_VQE})). The color scale indicates the overlap with the maximally pair-correlated target state $\ket{D}$, e.g., the ground state of $\mathcal{C}$. The total number of measurement shots for the entire optimization was limited to $3\times10^3$.
(b) Bar plot showing the improvement of pair correlations in the final optimized state, averaged over 50 independent trajectories. The error bars indicate the standard deviation across all trajectories (not the standard error).} 
    \label{fig::noise}
\end{figure}

Here, we provide an additional noise analysis to assess the robustness of optimizing the cost function in IQS  under a finite measurement budget. Specifically, we study the effect of shot-noise on the optimization procedure in the context of the Fermi-Hubbard example discussed around Fig.~2 of the main text and discuss strategies to minimize the total number of shots. The purpose of this analysis is to illustrate that the optimization remains stable, even in regimes operating with a remarkably small number of experimental runs.

To model this scenario, we evaluate the cost function using a finite number of samples drawn from the variational wave function $\ket{\psi(\boldsymbol{\theta})}$ and employ a pattern-search optimization strategy that accounts for uncertainty in the measured expectation values (see Ref.~\cite{KokailNature2018} for details). Figure~\ref{fig::noise}(a) shows a representative optimization trajectory where the total number of shots has been restricted to be smaller than $3\times 10^4$. The color scale encodes the fidelity with a maximally pair-correlated target state $\ket{D}$, defined as the ground state of the long-range pairing correlator $\sum_r \hat{\Delta}_{1}^{\dagger} \hat{\Delta}_{1 + r}$, and serves as a visual indicator of optimization progress.

Figure~\ref{fig::noise}(b) shows the resulting pair-correlation order parameter as a function of distance $r$ both in the exact Hubbard ground state and after averaging over 50 independent noisy optimization runs. The error bars in Fig.~\ref{fig::noise}(b) represent the sample variance across these independent runs, rather than the standard error of the mean, and therefore quantify the variability induced by finite sampling between the runs rather than statistical uncertainty in the averaged result.

In order to optimize the measurement requirements, we treat the minimum number of measurement shots per cost-function evaluation as a hyperparameter. Optimizing this parameter yields an optimum of approximately 15 shots per evaluation, allowing a complete optimization trajectory to be executed with a total measurement budget of the order $3 \times 10^4$
experimental runs. This analysis shows that the required measurement budget is compatible with modern ultracold quantum gas experiments with fermions in optical lattices, a platform that is particularly well suited for the inverse design of correlated fermionic superconductors.

Finally, we note that the required number of shots per cost-function evaluation is not expected to increase significantly with system size. This follows from a self-averaging effect in the pair-correlation observables. Larger lattices provide more correlators at a given distance $r$, which can be used to retrieve more accurate estimates of the cost function components. As a result, the overall measurement cost is expected to scale favorably with the system size.

\section{Continuous phase Hamiltonian learning}
\label{sup::cphl}

In the Hubbard ladder example of Fig.~2 of the main text and Fig.~\ref{fig::Fig_temperature}, we amplified $d$-wave fluctuations at a single point in the phase diagram, i.e., for fixed doping and temperature.  
While this was sufficient to enhance the desired property across the entire phase diagram,
here, we generalize this idea to what we term \emph{continuous-phase Hamiltonian learning} (CPHL), which targets not a single parent Hamiltonian for an isolated point but a whole (continuous) class of parent Hamiltonians for an extended region of parameter space.
We discretize the chosen region into a grid of points and amplify the desired property across the entire grid, while enforcing \emph{continuity} -- the learned Hamiltonian coefficients vary smoothly across the domain.  
Formally, we write
\begin{align}
    \hat{H}[\bm g] = \hat{H}_0[\bm{g}] + \sum_i c_i(\bm{g}) \, \hat{h}_i,
    \label{eqn::cphl_gen}
\end{align}
where $\bm{g}$ denotes the control parameters (e.g., interaction strength, magnetic and/or electric fields, pressure, etc.) and $c_i(\bm{g})$ are continuous functions over the targeted region.  

By optimizing across a phase rather than a single point, CPHL provides a route to \emph{quantum phase design} -- engineering Hamiltonians that retain desired properties over wide parameter ranges -- making the approach inherently more robust than pointwise Hamiltonian learning. 
The resulting continuous family of Hamiltonians is also interpretable and generalizable.

Such quantum phase design is particularly valuable for stabilizing \emph{fragile topological order} in programmable quantum simulators.  
Here, tailored corrections can enhance the persistence of the target phase, facilitating its preparation, detection, and exploration.  
As a concrete demonstration, we apply CPHL to the cluster Ising model (CIM)~\cite{PhysRevA.84.022304, PhysRevB.96.165124, PhysRevResearch.4.L022020}, a minimal one-dimensional setting where topological order competes with ferromagnetism.  
While the CIM serves as a clean testbed, the method is general and can be applied to a broad class of systems, from frustrated magnets to unconventional superconductors.

We also discuss the variational staircase circuit in Fig.~3(c) of the main text, which can naturally capture topological phases~\cite{PhysRevResearch.4.L022020} -- including the entire phase diagram of CIM and its CPHL extension -- by generating long-range entanglement across the system.
The corresponding states admit straightforward holographic implementation~\cite{PhysRevResearch.3.033002} involving circuits with periodic measurement and qubit reset, thereby enabling constant-entropy preparation of such states with logical qubits~\cite{bluvstein2025architectural} and natural extensions to higher-dimensional architectures. We show that in the 1D case, expectation values can be evaluated efficiently on a classical computer for arbitrary system sizes by concatenating a sequence of quantum channels, or equivalently, by contracting the tensor network representing the expectation value.

\subsection{Application example: CIM and its extensions}
\label{subsec::CIM}

Here, the CIM describes a spin-$\frac12$ chain of length \(N\) with open boundary conditions, and its microscopic Hamiltonian depends on a single tunable parameter \(g\):
\begin{align}
    \hat{H}(g) &= - \frac{1 - g}{2} \Big[ \sum_{i = 2}^{N - 1} \hat{Z}_{i-1} \hat{X}_i \hat{Z}_{i + 1} 
    + \hat{X}_1 \hat{Z}_2 + \hat{Z}_{N - 1} \hat{X}_{N} \Big]
    - \frac{1 + g}{2} \sum_{i = 1}^{N - 1} \hat{Z}_i \hat{Z}_{i + 1}. \label{eqn::CIM_H}
\end{align}
At $g = -1$, the ground state corresponds to the cluster state. A distinctive feature of this state is that it maximizes the expectation value of the nonlocal string operator
\begin{align}
    \hat{\cal O} = (-1)^N \hat{Z}_1 \hat{Y}_2 \Big[ \prod_{i = 3}^{N - 2}  \hat{X}_i   \Big]\hat{Y}_{N - 1} \hat{Z}_N .
\end{align}
In contrast, at \(g = 1\) the ground state is ferromagnetic; the ground-state selection (such as the product state $\ket{\uparrow\uparrow\dots\uparrow}$) can explicitly break the \(\mathbb{Z}_2\) symmetry and thus mimic spontaneous symmetry breaking, while \(\langle \hat{\mathcal{O}} \rangle = 0\).
Tuning \(g\) from \(-1\) to \(1\) therefore drives a quantum phase transition from a topologically ordered phase with \(\langle \hat{\mathcal{O}} \rangle \neq 0\) to a ferromagnetic phase with \(\langle \hat{\mathcal{O}} \rangle = 0\) -- see also Fig.~3 of the main text.

Within the quantum phase design framework, we use Hamiltonian learning to reshape the Hamiltonian family in Eq.~\eqref{eqn::CIM_H} so that the cluster and \(ZZ\)-ferromagnetic terms remain fixed, while the phase boundary is shifted to enlarge the topological phase.
To this end, we specialize Eq.~\eqref{eqn::cphl_gen} to the CIM:
\begin{align}
    \hat{H}_g[\mathbf{c}] = - \frac{1 - g}{2} \Big[ \sum_{i = 2}^{N - 1} \hat{Z}_{i-1} \hat{X}_i \hat{Z}_{i + 1} 
    + \hat{X}_1 \hat{Z}_2 + \hat{Z}_{N - 1} \hat{X}_{N} \Big]
    - \frac{1 + g}{2} \sum_{i = 1}^{N - 1} \hat{Z}_i \hat{Z}_{i + 1} + \sum_a c_a(g) \,\hat{h}_a ,\label{eqn::CIM_expansion_H}
\end{align}
where the coefficient functions \(c_a(g)\) are required to be continuous and vanish at both endpoints: 
\begin{align}
    c_a(-1) = c_a(1) = 0.
\end{align}
To fulfill these conditions, we expand these coefficients $c_a(g)$ in a \emph{Fourier harmonic basis}:
\begin{align}
    c_a(g) = \sum_{m = 1}^{M_{\mathrm{max}}} \alpha_{a,m} \,\sin\Big[\frac{m \pi (g + 1)}{2}\Big],
    \label{eqn::FH}
\end{align}
so that the design task reduces to determining the harmonic amplitudes \(\alpha_{a,m}\).
The parameter \( M_{\rm max} \) sets the harmonic cutoff and thus controls the smoothness of the resulting Hamiltonians.

\subsection{CPHL algorithm}
\label{subsec::cphl_alg}

We turn to discuss how to algorithmically determine the harmonic coefficients \( \alpha_{a,m} \) -- see also Fig.~3(a) of the main text for a schematic pipeline.

At iteration \(k\) of the algorithm, we have a set of Hamiltonians \(\hat{H}^{[k]}_{g_j}\), defined on a uniform grid of points \( g_j \in [-1, 1]\) with $ j = 1, 2, \dots , N_g$, and  characterized by the current values of \( \alpha^{[k]}_{a,m} \).
Initially, all harmonic coefficients are set to zero.  
Following our material design framework, we perform a point-by-point optimization over \(g_j\), treating each value independently, by minimizing the cost function  
\begin{align}
    \mathcal{C}(g_j) = \braket{\psi[\bm{\theta}_{g_j}] | \hat{H}^{[k]}_{g_j} | \psi[\bm{\theta}_{g_j}]} 
    - \lambda_k \braket{\psi[\bm{\theta}_{g_j}] | \hat{\mathcal{O}} | \psi[\bm{\theta}_{g_j}]},\label{eqn::cphl_cf}
\end{align}
where \(\ket{\psi[\bm{\theta}_{g_j}]}\) denotes a variational quantum state for each \({g_j}\), prepared by a parameterized quantum circuit (in our main text numerical simulations, we employ MPSs and optimize via DMRG but we could have also used variational staircase circuits as discussed below).  
The second term, proportional to \(\lambda_k\), biases the optimization towards states with larger string-order expectation value, thereby promoting the emergence of topological order beyond the original phase boundary.

From each optimized variational wave function \(\ket{\psi[\bm{\theta}_{g_j}]}\), and using the Hamiltonian learning method based on variance minimization detailed in Methods, we extract corrections \( \delta {c}^{[k]}_a(g_j) \) to the current coefficients $c^{[k]}_a(g_j)$.
These corrections \(\delta {c}^{[k]}_a(g_j)\) are not, in general, continuous in \(g\).  
From them, we estimate the corresponding Fourier harmonics \(\delta {\alpha}^{[k]}_{a,m}\) as in Eq.~\eqref{eqn::FH}, as described below, cf. Eq.~\eqref{eqn::FH_approx}; this acts as a smoothing step for the otherwise discrete set of points. 
The harmonic coefficients are then updated using a damped iterative rule with weight \(\delta_k\), which may depend on the iteration index \(k\):
\begin{align}
    \alpha_{a,m}^{[k + 1]} =  \alpha_{a,m}^{[k]} + \delta_k \, \delta {\alpha}^{[k]}_{a,m}.
\end{align}
This procedure is repeated until convergence.

\begin{figure}[b!]
    \centering
    \includegraphics[width=0.6\linewidth]{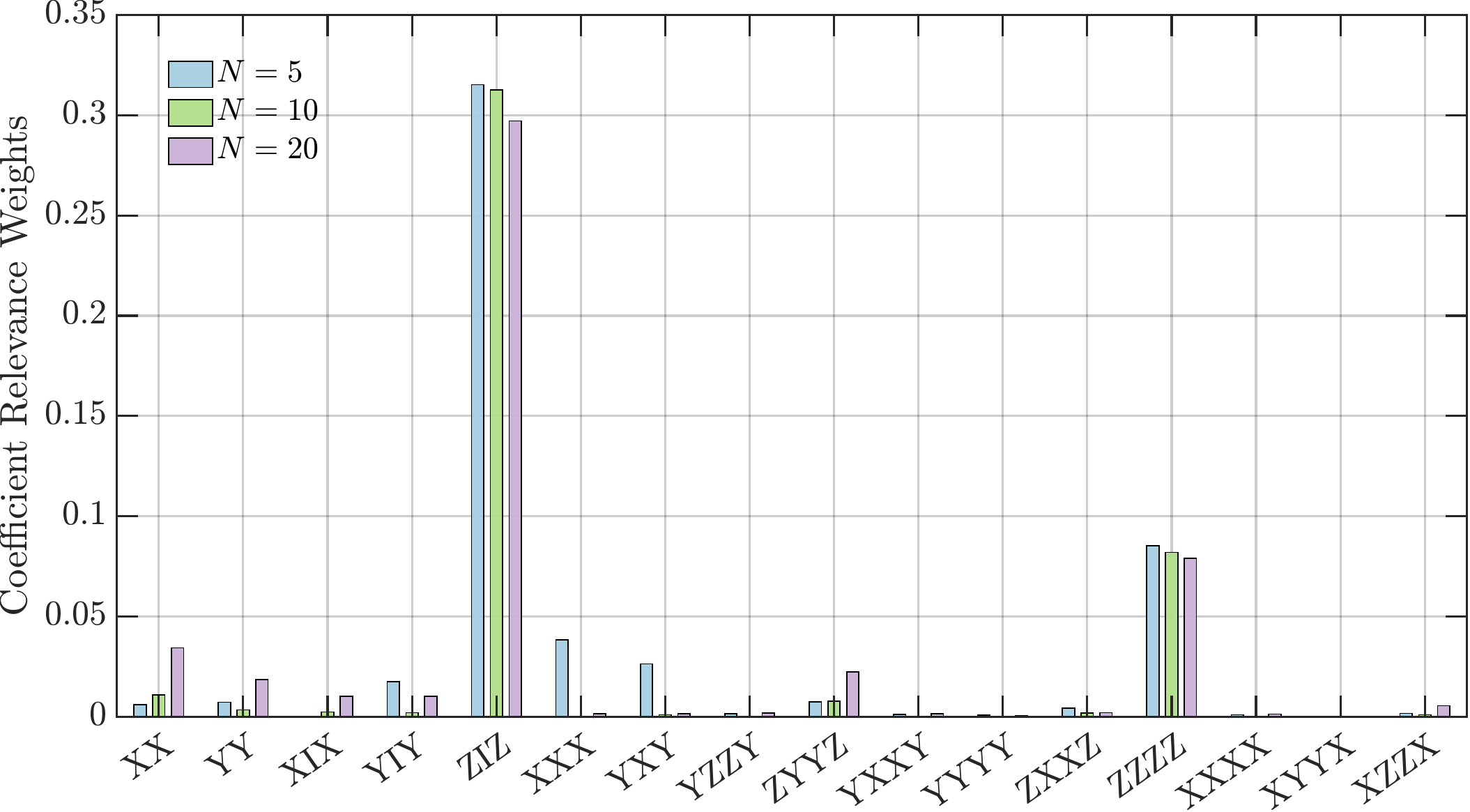}
    \caption{ Coefficient relevance weights for the CIM obtained via CPHL, showing only weak dependence on the system size \(N\).  
    Each weight is defined as \(\max_g |c_a(g)|\) for the corresponding coefficient in the expansion~\eqref{eqn::CIM_expansion_H}.
    }
    \label{fig::cphl_weights}
\end{figure}

A few additional comments on the algorithm:
\begin{itemize}
    \item The Fourier smoothing step to the coefficients \(\delta {c}^{[k]}_a(g_j)\), which may contain irregularities -- particularly in the presence of degeneracies in the learned Hamiltonian -- incorporates information from all grid points \(g_j\), thereby effectively averaging out any local discontinuities.  
    This approach enhances the robustness of the Hamiltonian-learning procedure across the entire parameter range.
    \item   Given the set of values \( c(g_j) \) for a fixed index \( a \), the corresponding harmonic coefficients \( \alpha_m \) can be estimated by minimizing the regularized least-squares cost:
    \begin{align}
        \left\| A \bm{\alpha} - \bm{c} \right\|^2 + \varkappa \left\| \bm{\alpha} \right\|^2,
    \end{align}
    where \( A_{j,m} = \sin(\pi m (g_j +1)/2) \) defines the design matrix and \( \varkappa \) is a regularization parameter. This is a standard ridge regression problem, and its closed-form solution is given by:
    \begin{align}
        \bm{\alpha} = \left( A^\top A + \varkappa I \right)^{-1} A^\top \bm{c}. \label{eqn::FH_approx}
    \end{align}
    At later stages of the algorithm, as the HL becomes more and more robust, we might want to reduce $\varkappa$ to zero.
    \item To ensure convergence of the algorithm, we impose a decay on the topological weight,  \( \lambda_k = \lambda_0 / (1 +  k \beta) \), where \(\beta\) is the damping rate and \(\lambda_0\) is the initial strength.  
    When \(\lambda = 0\), the learning algorithm (see Methods) halts automatically: \(\delta {c}^{[k]}_a(g_j) = \delta {\alpha}^{[k]}_{a,m} = 0\), and no further updates occur.
    \item In practice, we define the current error as \(\left\| \delta \alpha^{[k]} \right\|\).  
    If this error increases, or if any of the values \(|c_a(g_j)|\) exceed a set threshold, we reduce the iterative weight by a constant factor to improve convergence: $\delta_{k + 1} = \delta_k / (1 + r)$.
\end{itemize}

\subsection{CPHL results on the CIM}
\label{subsec::cphl_res}

The main CPHL results for the CIM are shown in Fig.~3 of the main text; here we provide additional details and insights.  

In the Hamiltonian ansatz of Eq.~\eqref{eqn::CIM_expansion_H}, we include only translationally invariant terms (as if we had the \(N \to \infty\) limit, where boundary corrections are unimportant) and impose locality, allowing for two-, three-, and four-spin interactions as indicated in Fig.~\ref{fig::cphl_weights}.  
We further require that the bulk of each of these terms commutes with the \(X\)-string \(\prod_i \hat{X}_i\) and that each term is real, i.e., \(\hat Y\)-operators appear only in pairs.  
These constraints preserve the \(\mathbb{Z}_2 \times \mathbb{Z}_2\) symmetry that protects the topological order~\cite{PhysRevA.84.022304}.  
For the quartic interactions, the list of terms considered is not exhaustive: we include only those of the form \(\hat{A}\hat{B}\hat{B}\hat{A}\).  
This choice is motivated by the fact that multiplying two consecutive cluster stabilizers produces the \(\hat{Z}\hat{Y}\hat{Y}\hat{Z}\) term; the same structure also contains \(\hat{Z}\hat{Z}\hat{Z}\hat{Z}\), which can influence the \(z\)-axis ferromagnetic order.  
Restricting to this subset is not essential, but it simplifies the ansatz by limiting the total number of terms -- here, 16 in total, as listed in Fig.~\ref{fig::cphl_weights}. 
Finally, to mimic plausible experimental challenges in engineering interactions, we assign weights in Eq.~18 in Methods such that three-body (four-body) terms are penalized by a factor of two (four) relative to two-body ones.

While the main CPHL results on the CIM -- showing an appreciable extension of the topological phase via newly introduced continuous terms -- are presented in the main text, here we highlight that the resulting model depends only weakly on the system size \(N\), as shown in Fig.~\ref{fig::cphl_weights}.  
This suggests that the phase extension is a genuine physical effect: the relevant Hamiltonian terms can be identified, and their relative importance assessed, even from rather small systems. 
We find that for the CIM phase diagram, the most important term is the next-nearest-neighbour AFM coupling \(\hat{Z}\hat{I}\hat{Z}\), which significantly suppresses the emergence of ferromagnetic correlations (see Fig.~3 of the main text) while leaving the topological order largely intact, thereby extending the topological phase.  
The next most relevant term is \(\hat{Z}\hat{Z}\hat{Z}\hat{Z}\), which has a similar effect.  
All other terms appear unimportant in comparison.  
These observations strengthen our confidence that CPHL produces not only effective but also interpretable Hamiltonians.

\subsection{Constant-entropy computation with staircase circuits}

Figure~3(c) of the main text illustrates the variational staircase circuit, where all spins are initialized in $\ket{0}$, a Hadamard gate is applied to the first qubit, and a sequence of identical two-qubit gates is then applied along the chain.
The two-qubit gate is parametrized by 15 real parameters,
\begin{align}
    \hat{U}(\bm \theta) = \exp \Big[ - i \Big( \sum_{a, b} \theta_{ab} \hat{\sigma}^a_1\otimes \hat{\sigma}^b_2  + \sum_a \theta^{(1)}_a  \hat{\sigma}^a_1\otimes \hat{I}_2 + \sum_a \theta^{(2)}_a  \hat{I}_2\otimes \hat{\sigma}^a \Big)\Big].\label{eqn::U_theta_v0}
\end{align}
By suitable choice of parameters, the circuit can generate both the cluster and GHZ states, indicating that the entire CIM and its CPHL extension can be captured with these simplified circuits. 
To verify this capability, Fig.~3(b1) of the main text shows results for staircase circuits optimized to maximize fidelities with the DMRG ground states, yielding high overlaps ($\gtrsim 0.3$ for $N=20$ and $\gtrsim 0.85$ for $N=5$) and confirming the topological phase extension.
We note that applying the staircase circuits within the CPHL algorithm [Eq.~\eqref{eqn::cphl_cf}] would, in addition, yield learned Hamiltonians whose ground states are accurately represented by MPSs of bond dimension $\chi = 2$.

The staircase circuit shown in Fig.~3(c, left panel) of the main text can be reformulated in a holographic form [Fig.~3(c, right panel)].  
Specifically, the measurement outcome of the first qubit depends only on the two-qubit unitary $U_{1,2}$; thus, the first qubit can be measured immediately after applying $U_{1,2}$, reset to $\ket{0}$, and reused for qubit~3.  
Repeating this procedure sequentially across the chain transforms the original staircase circuit into the holographic representation depicted in Fig.~3(c, right panel) of the main text.

The mechanism by which the holographic circuit in Fig.~3(c, right panel) operates can be understood as a sequence of completely positive trace-preserving (CPTP) maps \(\mathcal{M}(\cdot)\) that periodically transform the initial state \(\rho_0\).
After the first gate, observing a measurement outcome \(\ket{m}\) followed by qubit reset, the state collapses to
\begin{align}
\frac{1}{p_m}\!\left(   \ket{0}\!\bra{m}\otimes \mathbb{1} \right) U_{1,2}\,\rho_0\,U_{1,2}^{\dagger} \left( \ket{m}\!\bra{0} \otimes \mathbb{1}\right)
= \frac{1}{p_m} K_m \rho_0 K_m^{\dagger}.
\end{align}
Averaging over these outcomes yields the Kraus map
\begin{align}
\mathcal{M}(\rho_0)
= \sum_m K_m \rho_0 K_m^{\dagger}
= \operatorname{Tr}_{1}\!\left[ U\,\rho_0\,U^{\dagger} \right] \left(\ket{0}\!\bra{0} \otimes \mathbb{1}\right),
\end{align}
where \(\operatorname{Tr}_1\) denotes the partial trace over the first qubit.

In the context of early fault-tolerant hardware, computations such as those depicted in Fig.~3(c) are critical for keeping entropy during state preparation approximately constant \cite{bluvstein2025architectural}.
In our setting, measuring the first qubit teleports the relevant logical information onto the second qubit in the form of the Kraus map \(\bra{m}U\,\rho_0\,U^{\dagger}\ket{m}\).
At the same time, the measured qubit can be replaced by a fresh code block that is reset, recooled, and reinitialized.
In such an approach, with appropriate syndrome processing and periodic reinitialization, it can be ensured that errors are suppressed after a constant number of maps, while the logical information continues to propagate through the circuit~\cite{bluvstein2025architectural}.

\begin{figure}[t!]
    \centering
    \includegraphics[width=1\linewidth]{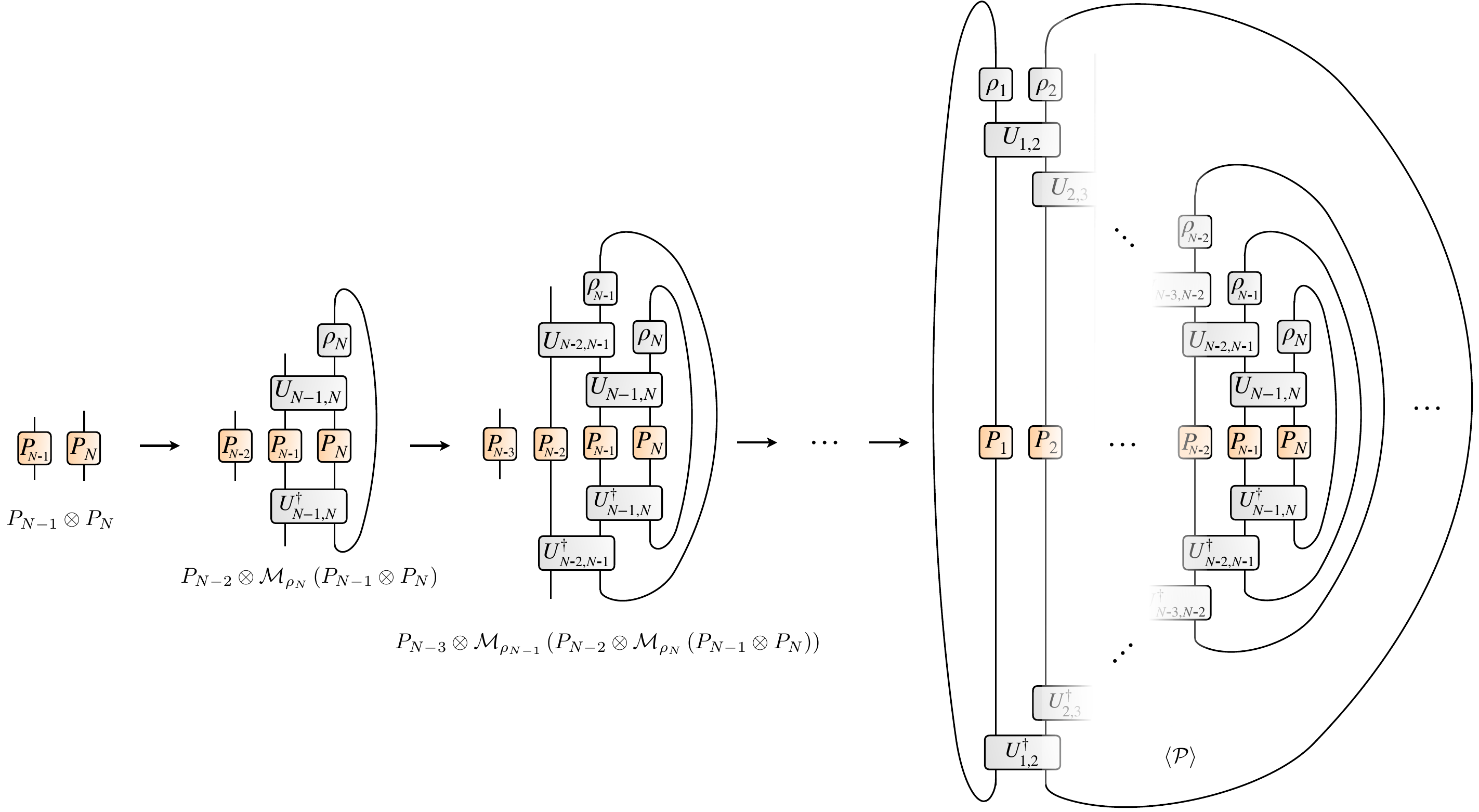}
    \caption{
    \textbf{Efficient evaluation of string-like expectation values in a staircase circuit.}
The tensor network representing the expectation value of a Pauli string is built sequentially from right to left.
    }
    \label{fig::staircase_eval_1}
\end{figure}

\emph{Efficient evaluation of Pauli expectation values}---Any staircase circuit composed of two-qubit unitaries as in Fig.~3(c) of the main text generates an MPS with bond dimension \(\chi=2\).
Consequently, expectation values at arbitrary system size can be computed efficiently by contracting the corresponding tensor network.
Below we briefly explain how this contraction relates to the holographic circuit picture and how it can be carried out by concatenating CPTP maps. This procedures is closely related to the recently introduced framework of Pauli propagation~\cite{rudolph2025paulipropagationcomputationalframework}.
For general discussions of holographic state preparation and relation to MPSs, the reader is referred to Ref.~\cite{PhysRevResearch.3.033002}.

Our goal is to evaluate the expectation value
\begin{align}
 \braket{\mathcal{P} } =  \text{Tr} \Big[U_{1, 2}^{\dagger} U_{2, 3}^{\dagger} \cdots U_{N-1, N}^{\dagger} \ \mathcal{P} \ U_{N-1, N} U_{N-2, N-1} \cdots U_{1, 2} \ \rho_0\Big], 
\end{align}
where $\mathcal{P} = \bigotimes_{j = 1}^N P_j$ is the Pauli string of interest and $\rho_0 = \bigotimes_{j = 1}^N \rho_j$ an initial product state. Due to the staircase structure of the circuit, the trace can be rewritten as a chain of nested traces:

\begin{align} \label{eq:nested}
\begin{split}
\braket{\mathcal{P}} = & \, \text{Tr}_1 \Big[ \text{Tr}_2 \Big( U_{1, 2}^{\dagger} \text{Tr}_3 \Big( U_{2, 3}^{\dagger} \dots \\
&\dots \text{Tr}_{N-1} \Big( U^{\dagger}_{N-2, N-1} \text{Tr}_N \Big( U^{\dagger}_{N-1, N} P_{N-1} \otimes P_N U_{N-1, N} \rho_N \Big) \otimes P_{N-2} U_{N-2, N-1} \rho_{N-1} \Big)\\
& \dots \otimes  P_1 U_{1, 2} \rho_1 \Big) \Big].
\end{split}
\end{align}
Defining the auxiliary CPTP map 
\begin{align}
\mathcal{M}_{\rho_j} [ \cdot ] = \text{Tr}_j \Big(U^{\dagger}_{j-1, j} \  \Big[\cdot \Big] \ U_{j-1, j} \ \rho_j \Big) 
\end{align}
we can write Eq.~(\ref{eq:nested}) in a more compact form:
\begin{align} \label{eq:nestedmaps}
\braket{\mathcal{P}} = \text{Tr} \left[ \mathcal{M}_{\rho_2} \left( P_1 \otimes \dots  \mathcal{M}_{\rho_{N-2}} \left( P_{N - 3} \otimes \mathcal{M}_{\rho_{N - 1}} \left( P_{N - 2} \otimes \mathcal{M}_{\rho_N} \left( P_{N-1} \otimes P_N \right) \right) \right) \dots \right) \right].
\end{align}
The interpretation of Eqs.~(\ref{eq:nested}) and (\ref{eq:nestedmaps}) is illustrated in Fig.~\ref{fig::staircase_eval_1}. We apply the inverse staircase circuit to the Pauli string $\mathcal{P}$, proceeding from right to left. 
We start with the two rightmost factors, $P_{N-1} \otimes P_N$, and propagate this two-qubit Pauli through $U_{N-1,N}$. 
Since no subsequent gate acts on site $N$, the $N$-th qubit can then be traced out, implementing a CPTP map on the residual operator supported on site $N-1$ (with dependence on $\rho_N$). 
We then start the next step by tensoring in $P_{N-2}$ and repeating the procedure: propagate with $U_{N-2,N-1}$, trace out site $N-1$, and thereby apply the map $\mathcal{M}_{\rho_{N-1}}$. 
Iterating this right-to-left reduction yields the desired nested composition of local maps (Fig.~\ref{fig::staircase_eval_1}).

\begin{figure}
    \centering
    \includegraphics[width=1\linewidth]{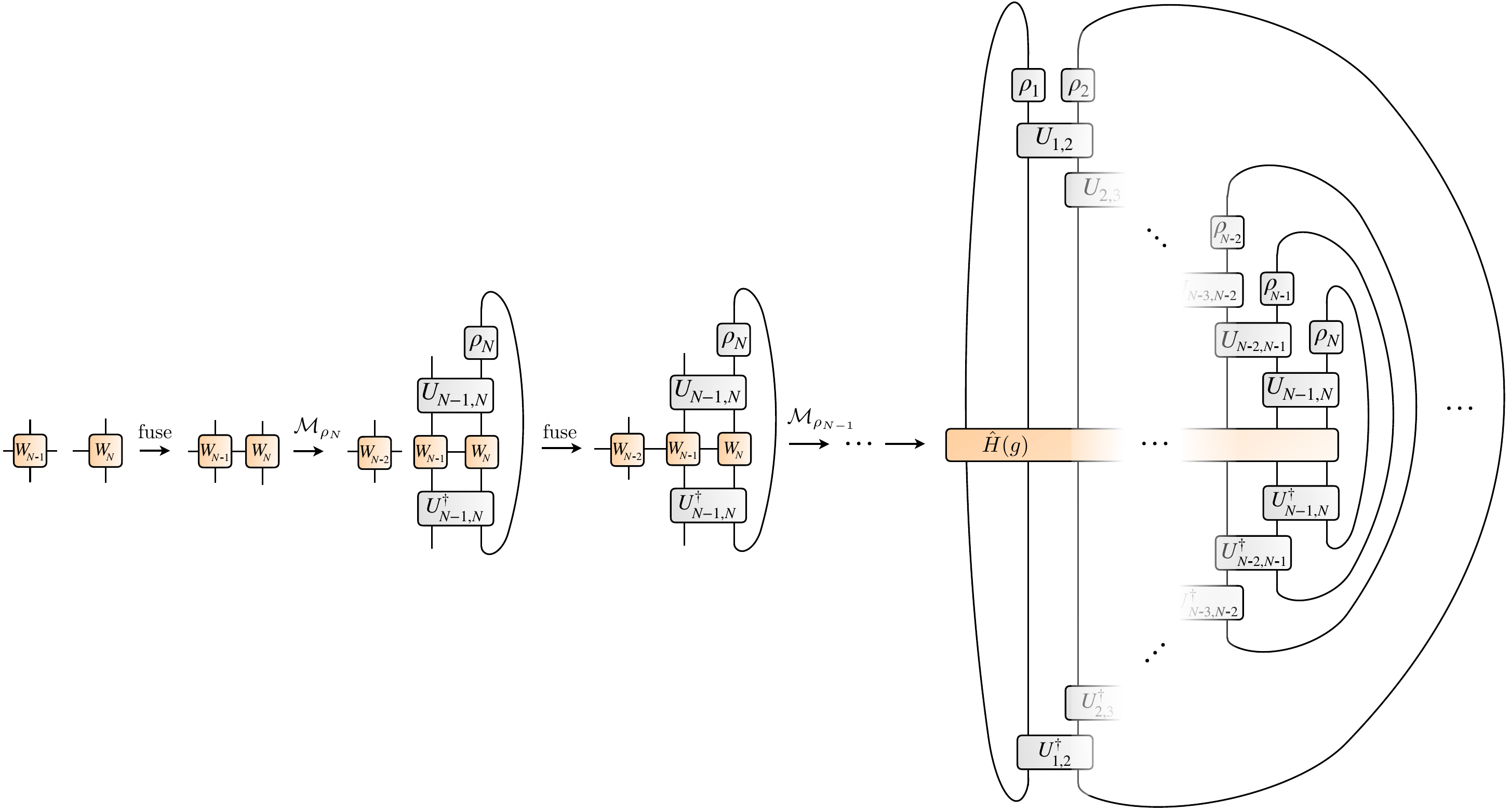}
    \caption{ \textbf{Efficient contraction of an MPO in a staircase circuit.} The tensor network representing the expectation value of $\hat{H}_I$ is built sequentially from right to left.
    }
    \label{fig::staircase_eval_2}
\end{figure}

\emph{Efficient evaluation of expectation values of local Hamiltonians}---Evaluating the cost function may require expectation values of Hamiltonians expressed as sums of local Pauli operators.   
A straightforward approach would apply the above procedure term by term, requiring $\mathcal{O}(N)$ repetitions of the contraction in Fig.~\ref{fig::staircase_eval_1}.  
A more efficient alternative is to represent the full operator as a \textit{matrix product operator} (MPO), $\hat{W}^{[1]} \hat{W}^{[2]} \cdots \hat{W}^{[N]}$, and evaluate its expectation value in a single sweep, as illustrated in Fig.~\ref{fig::staircase_eval_2}.  
In our case, this procedure can be carried out without explicitly storing an MPS.

We illustrate the construction for the CIM Hamiltonian in Eq.~\eqref{eqn::CIM_H}.  
Following the approach of Ref.~\cite[p.~142--143]{SCHOLLWOCK201196}, we define four intermediate states of the Pauli string at a given link connecting two sites:
\begin{enumerate}[label=State~\arabic*:, leftmargin=4.5em]
\item Idle (nothing happened yet).
\item after the first $Z$.
\item after the second $X$.
\item done (only $\mathbb{1}$ operators after this stage).
\end{enumerate}
The transitions between these states define the MPO matrices:
\begin{align*}
&1 \overset{\mathbb{1}}{\longrightarrow} 1 .\\
&1 \overset{Z}{\longrightarrow} 2 \overset{ \frac{1 - g}{2}X}{\longrightarrow} 3 \overset{Z}{\longrightarrow} 4 .\\
&1 \overset{Z}{\longrightarrow} 2  \overset{ -\frac{1 + g}{2}Z}{\longrightarrow} 4 .\\
&4 \overset{\mathbb{1}}{\longrightarrow} 4 .\\
\end{align*}
Accordingly, $\hat{H}(g)$ can be written as an MPO with bond dimension $\chi = 4$:
\begin{equation}
W^{[i]} =
\begin{pmatrix}
\mathbb{1} & Z & 0 & 0 \\
0 & 0 & \frac{1 - g}{2} X & -\frac{1 + g}{2} Z \\
0 & 0 & 0 & Z \\
0 & 0 & 0 & \mathbb{1}
\end{pmatrix}_i,
\qquad
W^{[1]} = \big( \mathbb{1},\, Z,\, \frac{1 - g}{2}X,\, 0 \big)_1,
\qquad
W^{[N]} =
\begin{pmatrix}
0 \\[2pt]
-\frac{1 + g}{2} Z + \frac{1 - g}{2} X \\[2pt]
Z \\[2pt]
\mathbb{1}
\end{pmatrix}_N.
\label{eq:MPO_ZXZ_ZZ}
\end{equation}
Given such an MPO representation, expectation values can now be readily computed by applying the channels to the individual MPO matrix elements as depicted in Fig.~\ref{fig::staircase_eval_2} and by consecutively contracting the MPO tensors as we sweep from right to left.

\section{Spectral Hamiltonian Learning}
\label{sup::dyn}

Here, we introduce \emph{Spectral Hamiltonian Learning} (SHL), a framework for engineering and inferring Hamiltonians for optimizing and designing frequency-resolved dynamical properties.  
We highlight three key applications: (i) enhancing a system’s response at a target frequency (relevant to photochemical and molecular design), (ii) reconstructing a Hamiltonian from a given molecular spectrum, and (iii) learning a microscopic, translationally invariant Hamiltonian from frequency- and momentum-resolved condensed matter data.  
Crucially, SHL requires only state preparation and measurements of a fixed set of Hermitian operators on near-term quantum hardware, combined with efficient classical post-processing (see Methods).

In Sec.~\ref{subsec::dyn_Cost}, we discuss the practical meaning of optimizing frequency-dependent properties and propose corresponding cost functions whose minimization achieves this goal.  
Secs.~\ref{subsec::ex_1}–\ref{subsec::ex_3} demonstrate SHL through three representative examples. 
Finally, in Sec.~\ref{subsec::fin_dyn}, we summarize the method and outline possible extensions.

\subsection{Cost functions for dynamical properties optimization}
\label{subsec::dyn_Cost}

Depending on the application, one may choose between three distinct types of cost functions to optimize dynamical response properties.

When the goal is to enhance a specific frequency component -- such as maximizing the photoabsorption of a molecule at a target optical frequency \(\omega_*\), relevant for the design of organic lasers -- a natural choice is:
\begin{align}
    {\cal L} \equiv \text{Im}[\chi_{\rm var}(\omega_*)].
    \label{eqn::L_omega_opt}
\end{align}
In the photoabsorption context, \(\chi_{\rm var}(\omega)\) encodes the frequency-dependent transition dipole response of the molecule to an external oscillating electric field. Since \(\text{Im}[\chi_{\rm var}(\omega)] < 0\), minimizing \({\cal L}\) in Eq.~\eqref{eqn::L_omega_opt} effectively enhances the photoabsorption rate at $\omega_*$, given by \(-\text{Im}[\chi_{\rm var}(\omega_*)]\).

Another application is Hamiltonian inference by matching a target spectrum as closely as possible -- whether obtained from experiment or manually specified to reflect a desired frequency profile. In this case, a cost function could be introduced as:
\begin{align}
    {\cal L} \equiv - \int \frac{{\rm d}\omega}{2\pi} \, \text{Im}[\chi_{\rm var}(\omega)] \, \text{Im}[\chi_{\rm tar}(\omega)].
    \label{eqn::main_dyn_cost_func}
\end{align}
For more complex spectral features, such as Fano-like lineshapes, a more appropriate choice is:
\begin{align}
    {\cal L} \equiv - \int \frac{{\rm d}\omega}{2\pi} \, |\chi_{\rm var}(\omega) - \chi_{\rm tar}(\omega)|^2.
\end{align}

We envision applying these SHL ideas to analyze condensed-matter experimental data -- effectively learning a translationally invariant many-body Hamiltonian from frequency- and momentum-resolved measurements, such as those obtained in neutron scattering experiments (see Sec.~\ref{subsec::ex_3} for further discussion). In this case, the cost function is defined as:
\begin{align}
    {\cal L} \equiv - \sum_k \int \frac{{\rm d}\omega}{2\pi} \, \text{Im}[\chi_{\rm var} (\omega, k)] \, \text{Im}[\chi_{\rm tar} (\omega, k)],\label{eqn::L_dyn_k}
\end{align}
which aggregates contributions from each momentum mode \(k\).

In the examples that follow, we compare our variational response with the exact Kubo expression (Methods), which takes the form of a sum over sharply peaked Lorentzians. Accordingly, we adopt the cost function in Eq.~\eqref{eqn::main_dyn_cost_func} and evaluate the frequency integral analytically:
\begin{align}
    {\cal L} = - \int \frac{{\rm d}\omega}{2\pi} \text{Im}[\chi_{\rm var}(\omega)] \text{Im}[\chi(\omega)] \approx \sum_n |\!\braket{n| \hat{V}|0}\!|^2 \, \text{Im}[\chi_{\rm var}({\cal E}_n - {\cal E}_0)], 
\end{align}
where, for concreteness, we have assumed \(\hat{\mathcal{O}} = \hat{V}\).
One might argue that this cost function becomes ill-defined in the limit \(\delta \to 0^+\), since the variational response contains infinitely sharp poles. However, in realistic material systems, energy levels are never truly discrete. A small but finite broadening \(\delta\) is physically justified, as dissipative processes -- whether intrinsic or environmental -- naturally introduce linewidths that regularize the response.

\subsection{Example: two spins-$\frac 1 2$ in a ferromagnetic ground state}
\label{subsec::ex_1}

To illustrate the method and its capabilities, we consider a simple analytical model consisting of two spins-\(\tfrac{1}{2}\) coupled ferromagnetically:
\begin{align}
    \hat{H} = B_z (\hat{\sigma}^z_1 + \hat{\sigma}^z_2) - J_I \hat{\sigma}^z_1 \hat{\sigma}^z_2 - J_H \, \hat{\bm \sigma}_1 \cdot \hat{\bm \sigma}_2, \label{eqn::2spins_H}
\end{align}
where \(B_z, J_I, J_H > 0\).
A particular appeal of this model is that, regardless of the values of the Hamiltonian parameters, the ground state remains \(\ket{\downarrow\downarrow}\). 
This highlights a fundamental limitation of wavefunction-based Hamiltonian learning methods: the ground state alone does not always uniquely determine the underlying local Hamiltonian.
Here, we argue that incorporating dynamical response properties introduces additional constraints, potentially enabling full Hamiltonian reconstruction.

\emph{Exact susceptibility}---Diagonalization of the Hamiltonian in Eq.~\eqref{eqn::2spins_H} is straightforward. Its eigenstates are given by $\ket{0} = \ket{\downarrow\downarrow}$, $\ket{1} = (\ket{\uparrow\downarrow} + \ket{\downarrow\uparrow})/\sqrt{2}$, $\ket{2} = (\ket{\uparrow\downarrow} - \ket{\downarrow\uparrow})/\sqrt{2}$, and $\ket{3} = \ket{\uparrow\uparrow}$ and ${\cal E}_0 = -2B_z - J_I - J_H$, ${\cal E}_1 = J_I - J_H$, ${\cal E}_2 = J_I + 3 J_H$, ${\cal E}_3 = 2B_z - J_I - J_H$ are the corresponding eigenenergies.
The Kubo formula then yields, with the drive taken as a global magnetic field and the probe as the global magnetization:
\begin{align}
    \chi_{xx} = \frac{ 1 }{\omega + i\delta - 2(B_z + J_I)} - \frac{ 1 }{\omega + i\delta + 2(B_z + J_I)}, \quad \chi_{xy} = -\frac{ i }{\omega + i\delta - 2(B_z + J_I)} - \frac{ i }{\omega + i\delta + 2(B_z + J_I)}.
    \label{eqn::suscED_ferro_2spins}
\end{align}

\emph{Variational analysis}---To capture the system's response to time-dependent perturbations, it is sufficient to consider the ansatz of the form (see Methods):
\begin{align}
    \ket{\psi(\bm \theta)} = \exp(-i \theta_{1x} \hat{\sigma}_1^x )\exp(-i \theta_{1y} \hat{\sigma}_1^y )
    \exp(-i \theta_{2x} \hat{\sigma}_2^x )\exp(-i \theta_{2y} \hat{\sigma}_2^y ) \ket{\downarrow\downarrow}.\label{eqn::var_susc_ansatz_ferro_v0}
\end{align}
For this ansatz, we get $\ket{u_{1x}} = \hat{\sigma}_1^x\ket{\downarrow\downarrow}$, $\ket{u_{1y}} = \hat{\sigma}_1^y\ket{\downarrow\downarrow}$, $\ket{u_{2x}} = \hat{\sigma}_2^x\ket{\downarrow\downarrow}$, 
$\ket{u_{2y}} = \hat{\sigma}_2^y\ket{\downarrow\downarrow}$, and
\begin{align}
    {\cal M} = I_{4\times 4},\qquad {\cal K} = -2 (B_z + J_I + J_H) \begin{bmatrix}
        i\sigma_y & 0_{2\times 2} \\
         0_{2\times 2} & i\sigma_y
    \end{bmatrix}
    + 2 J_H \begin{bmatrix}
        0 & i\sigma_y\\
        i\sigma_y & 0
    \end{bmatrix}.\label{eqn::K_mat_ferro_sp}
\end{align}
We can further write:
\begin{align}
    -i {\cal K} = U \text{diag}(\lambda_+, - \lambda_+,\lambda_-, - \lambda_-) U^\dagger, \qquad U =\frac{1}{2}\begin{bmatrix}
        1 & 1 & 1 & 1\\
        -i & i & -i & i\\
        -1 & -1 & 1 & 1\\
        i & -i & -i & i
    \end{bmatrix},
\end{align}
where $\lambda_+ = 2  (B_z + J_I + 2 J_H)$ and $\lambda_- = 2 (B_z + J_I)$. This form implies that the variational susceptibility reads:
\begin{align}
    \chi_{\rm var}(\omega) = i  {\bm o}^\top U \, \text{diag}\Big[\frac{ 1 }{\omega + i\delta - \lambda_+}, \frac{ 1 }{\omega + i\delta + \lambda_+},\frac{ 1 }{\omega + i\delta - \lambda_-}, \frac{ 1 }{\omega + i\delta + \lambda_-} \Big]  U^{\dagger}  \bm v.
\end{align}
When the magnetic perturbation is applied along the \(x\)-axis, we have \(\bm{v}_x = \tfrac{1}{2}(0,\, 1,\, 0,\, 1)^\top\); for a perturbation along the \(y\)-axis, \(\bm{v}_y = \tfrac{1}{2}(1,\, 0,\, 1,\, 0)^\top\). We assume the observable is aligned with the \(x\)-axis, so that \(\bm{o}_x = -2(1,\, 0,\, 1,\, 0)^\top\). With this setup, the variational response functions take the form:
\begin{align}
    \chi^{xx}_{\rm var}(\omega) = \frac{ 1 }{\omega + i\delta - \lambda_-} - \frac{ 1 }{\omega + i\delta + \lambda_-}, \qquad \chi^{xy}_{\rm var}(\omega) = -\frac{ i }{\omega + i\delta - \lambda_-} - \frac{ i }{\omega + i\delta + \lambda_-}.
\end{align}
In other words, the variational susceptibilities reproduce the exact ones in Eq.~\eqref{eqn::suscED_ferro_2spins}.

This result reveals that the full information one can extract from these response functions is encoded in the single parameter \(\lambda_-\). While this does not completely lift the degeneracy inherent in the Hamiltonian learning problem, it significantly constrains the space of consistent models. For instance, if the value of \(B_z\) is known (often it is a tunable parameter in experiments) then optimization of the dynamical response functions uniquely determines the value of \(J_I\).

\emph{Extensions}---We note that access to additional types of response functions, beyond global magnetization drive and probe considered above, might be sufficient to fix the remaining unknown Hamiltonian parameters. 
For instance, the Heisenberg coupling $J_H$ could be probed via a response function involving a local perturbation (such as \(\hat{V} = \hat{\sigma}^x_2\)) and a local probe (for instance, $\hat{\cal O} = \hat{\sigma}^x_1$). 
Indeed, the exact susceptibility in this case is given by:
\begin{align}
    \chi_{X_1, X_2}(\omega) & = \frac{1}{2}\Big[ \frac{1}{\omega + i\delta - 2 (B_z + J_I)} - \frac{1}{\omega + i\delta + 2 (B_z + J_I)} \Big] \notag\\
    &- \frac{1}{2}\Big[ \frac{1}{\omega + i\delta - 2 (B_z + J_I + 2J_H)} - \frac{1}{\omega + i\delta + 2 (B_z + J_I+ 2J_H)} \Big].
\end{align}
We observe that the second term enables access to a pole at \(2(B_z + J_I + 2J_H)\). This, combined with the previous analysis, allows us to unambiguously extract the value of $J_H$. 
We remark that, as before, the variational susceptibility reproduces the exact linear response. 

At this stage, we can proceed assuming that \(B_z + J_I\) and \(J_H\) are both known. One additional dynamical response is required to disentangle \(B_z\) and \(J_I\) from each other.

The remaining susceptibility can be obtained by choosing \(\hat{V} = \hat{\mathcal{O}} = \hat{\sigma}_1^x \hat{\sigma}_2^x\), yielding
\begin{align}
    \chi_{X_1X_2,X_1X_2} = \frac{1}{\omega + i\delta - 4 B_z} - \frac{1}{\omega + i\delta + 4 B_z},\label{eqn::chi_ferro_ED_long}
\end{align}
which allows us to determine \(B_z\) directly, and thereby also fix \(J_I\). 
We note that the variational ansatz used previously in Eq.~\eqref{eqn::var_susc_ansatz_ferro_v0} is no longer sufficient to capture this response and requires an extension to include two-spin excitations:
\begin{align}
    \ket{\psi(\bm \theta)} = \exp(-i \theta_{xx} \hat{\sigma}_1^x\hat{\sigma}_2^x) \exp(-i \theta_{xy} \hat{\sigma}_1^x\hat{\sigma}_2^y) \exp(-i \theta_{1x} \hat{\sigma}_1^x )\exp(-i \theta_{1y} \hat{\sigma}_1^y )
    \exp(-i \theta_{2x} \hat{\sigma}_2^x )\exp(-i \theta_{2y} \hat{\sigma}_2^y ) \ket{\downarrow\downarrow}. \label{eqn::var_susc_ansatz_ferro_v1}
\end{align} 
For this extended ansatz, we have \({\cal M} =I_{6 \times 6} \) while the \({\cal K}\)-matrix takes a block-diagonal form:
\begin{align}
    {\cal K} = \begin{bmatrix}
        {\cal K}_{1p} & 0_{4\times 2} \\
        0_{2\times 4} & {\cal K}_{2p}
    \end{bmatrix},
\end{align}
where \({\cal K}_{1p}\) is the \(4 \times 4\) matrix defined in Eq.~\eqref{eqn::K_mat_ferro_sp}, capturing single-particle excitations, and \({\cal K}_{2p} = 4i B_z \sigma_y\) describes the two-spin excitation subspace. Within this extended ansatz, the variational susceptibility now reproduces the exact response function given in Eq.~\eqref{eqn::chi_ferro_ED_long}.

To conclude, access to dynamical response functions -- such as the global spin susceptibility \(\chi_{xx}(\omega)\), local cross-correlations like \(\chi_{X_1,X_2}(\omega)\), and two-spin probes such as \(\chi_{X_1X_2,X_1X_2}(\omega)\) -- can unambiguously determine the underlying Hamiltonian, even in scenarios where conventional Hamiltonian learning methods based solely on static wavefunction properties fail to provide useful information.

\subsection{Example: three spins-$\frac 1 2$ and learning the ring exchange interaction}
\label{subsec::ex_2}

A more complex benchmark for our method -- presented as the first example in Fig.~4 of the main text -- is a system of three spins-$1/2$ coupled via 
\begin{align}
    \hat{H} = B_z (\hat \sigma^z_1 + \hat \sigma^z_2 + \hat \sigma^z_3)  - J_H ( \hat{\bm \sigma}_1 \cdot  \hat{\bm\sigma}_2 + \hat{\bm \sigma}_2 \cdot  \hat{\bm\sigma}_3 + \hat{\bm \sigma}_3 \cdot  \hat{\bm\sigma}_1) + J_{\rm RE} \sum_{\alpha\beta\gamma} \varepsilon_{\alpha\beta\gamma} \hat{\sigma}^\alpha_1 \hat{\sigma}^\beta_2\hat{\sigma}^\gamma_3,\label{eqn::3_spins_H}
\end{align}
where the third term represents a ring-exchange interaction in the form of the scalar spin chirality $\sim \varepsilon_{\alpha\beta\gamma}$, with $\varepsilon_{\alpha\beta\gamma}$ the antisymmetric Levi--Civita tensor.

In our demonstration, we fix $B_z$ and $J_H$ -- these could either be tunable or approximately known in practice (for Fig.~4 of the main text, $B_z = 1$ sets the energy scale, $J_H = 0.3$, and the target value is $J_{\rm RE} = 0.5$) -- and aim to learn $J_{\rm RE}$ from a dynamical response function.  For concreteness, we take $\hat{\cal O} = \hat{V} = \hat{\sigma}^x_1 \hat{\sigma}^y_2 \hat{\sigma}^z_3$, chosen because the associated susceptibility is sensitive to $J_{\rm RE}$. 
In the variational ansatz to capture weak time-dependent perturbations (see Methods), we include all 9 single-spin operators $\hat{\sigma}^\alpha_j$ and all 27 two-spin operators $\hat{\sigma}^\alpha_i \hat{\sigma}^\beta_j$. We numerically verified that that this ansatz is sufficient as the variational susceptibility reproduces the exact one.

Learning the single unknown parameter $J_{\rm RE}$ converges rapidly, as shown in Fig.~4 of the main text. Remarkably, the initial susceptibility contains only a single peak, in stark contrast to the three-peak structure of the target response, underscoring the resolving power of the method.

\subsection{Example: learning Hamiltonian from the dynamical structure factor }
\label{subsec::ex_3}

One of the most promising applications of the SHL method is the reconstruction of a Hamiltonian from the measured spectrum of collective excitations.  Such spectra can be obtained experimentally with a variety of techniques, including angle-resolved photoemission spectroscopy (ARPES), resonant inelastic X-ray scattering (RIXS), Brillouin light scattering, and inelastic neutron scattering. 
Given such data, our objective is to infer the underlying many-body translationally invariant Hamiltonian. 
We illustrate this capability with two related examples, demonstrating how a quantum computer provides an ideal platform for solving this inverse problem -- see also the bottom panels of Fig.~4 of the main text.  
In addition, we show how to optimally construct the operators $\hat{A}_i$ in the dynamical variational ansatz (Methods) so as to explicitly preserve translational invariance.

To this end, we consider a one-dimensional spin-$\frac{1}{2}$ chain with periodic boundary conditions:
\begin{align}
    \hat{H} = B_z \sum_j \hat\sigma^z_j - J \sum_j \bm \sigma_j \cdot \bm \sigma_{j + 1}. \label{eqn::ferro_chain_H}
\end{align}

For $J > 0$ (or sufficiently large $B_z$), the ground state is ferromagnetic, $\ket{\downarrow\downarrow\dots\downarrow}$.  
The low-energy excitations in this case are magnons, corresponding to a single spin flip:
\begin{align}
    \ket{k} \equiv \frac{1}{\sqrt{N}} \sum_j e^{i k j}  \hat{\sigma}^+_j\ket{\downarrow\downarrow\dots\downarrow} = \sigma_k^+ \ket{\downarrow\downarrow\dots\downarrow},\quad k = 2\pi n/N. 
\end{align}
These states are exact eigenmodes of the Hamiltonian~\eqref{eqn::ferro_chain_H}, with analytically computable dispersion:
\begin{align}
    \omega_k = 2B_z  + 4 J (1 - \cos k).
\end{align}
The central idea is that, given only this spectrum, we use the SHL method to reconstruct the original Hamiltonian~\eqref{eqn::ferro_chain_H}.  
To this end, we adopt the cost function of Eq.~\eqref{eqn::L_dyn_k}, where for a given momentum $k$ we take $\hat{V} = \hat{\sigma}_k^+$ and $\hat{\cal O} = \hat{\sigma}_k^-$.  
The corresponding observable is nothing but the dynamical structure factor, which is directly measurable in neutron scattering experiments~\cite{lake2010confinement}.
For the ferromagnetic case, it exhibits a sharp peak at the magnon dispersion $\omega = \omega_k$. For the dynamical variational ansatz (Methods), it is sufficient to consider two operators for each $k$:
\begin{align}
    \ket{\psi_k} = \exp(-i \alpha_k (\hat\sigma_k^+ + \hat\sigma_k^-)) \exp(\beta_k (\hat\sigma_k^+ - \hat\sigma_k^-)) \ket{\downarrow\downarrow\dots\downarrow}.
\end{align}
While the ferromagnetic case can be solved analytically (see also Sec.~\ref{subsec::ex_1}), in Fig.~4 of the main text, we proceed numerically, considering a chain of $N=10$ spins to benchmark our numerical routines. Both $B_z$ and $J$ are treated as free parameters, and the learning procedure converges to the correct values ($B_z = 1$ and $J = 1.5$) as shown in the inset of Fig.~4(d) of the main text.

While the ferromagnetic case serves as a clean, analytically tractable benchmark, the antiferromagnetic regime ($J < 0$ and moderate $B_z$) poses a far more difficult challenge.
In this case, the dynamical structure factor displays striking complexity: a spin flip (which is a spin-1 excitation) fractionalizes into two spin-\(\frac12\) spinons, producing a continuum of excitations, as predicted by the Bethe ansatz~\cite{bethe1931theorie, PhysRev.128.2131, Faddeev1981SpinWave, PhysRevB.55.12510}.  
For \(B_z = 0\) in the thermodynamic limit, the two-spinon continuum is bounded by  
\begin{align}
\omega_{\mathrm{LB}}(q) = - 2\pi J \, |\sin q|, \quad
\omega_{\mathrm{UB}}(q) = - 4\pi J \, \big|\sin(q/2)\big|,
\end{align}
as indicated by the dashed lines in Fig.~4(e) of the main text.
Learning the Hamiltonian from such a highly nontrivial spectral profile provides a stringent test of the SHL method.

 \begin{figure}
    \centering
    \includegraphics[width=0.6\linewidth]{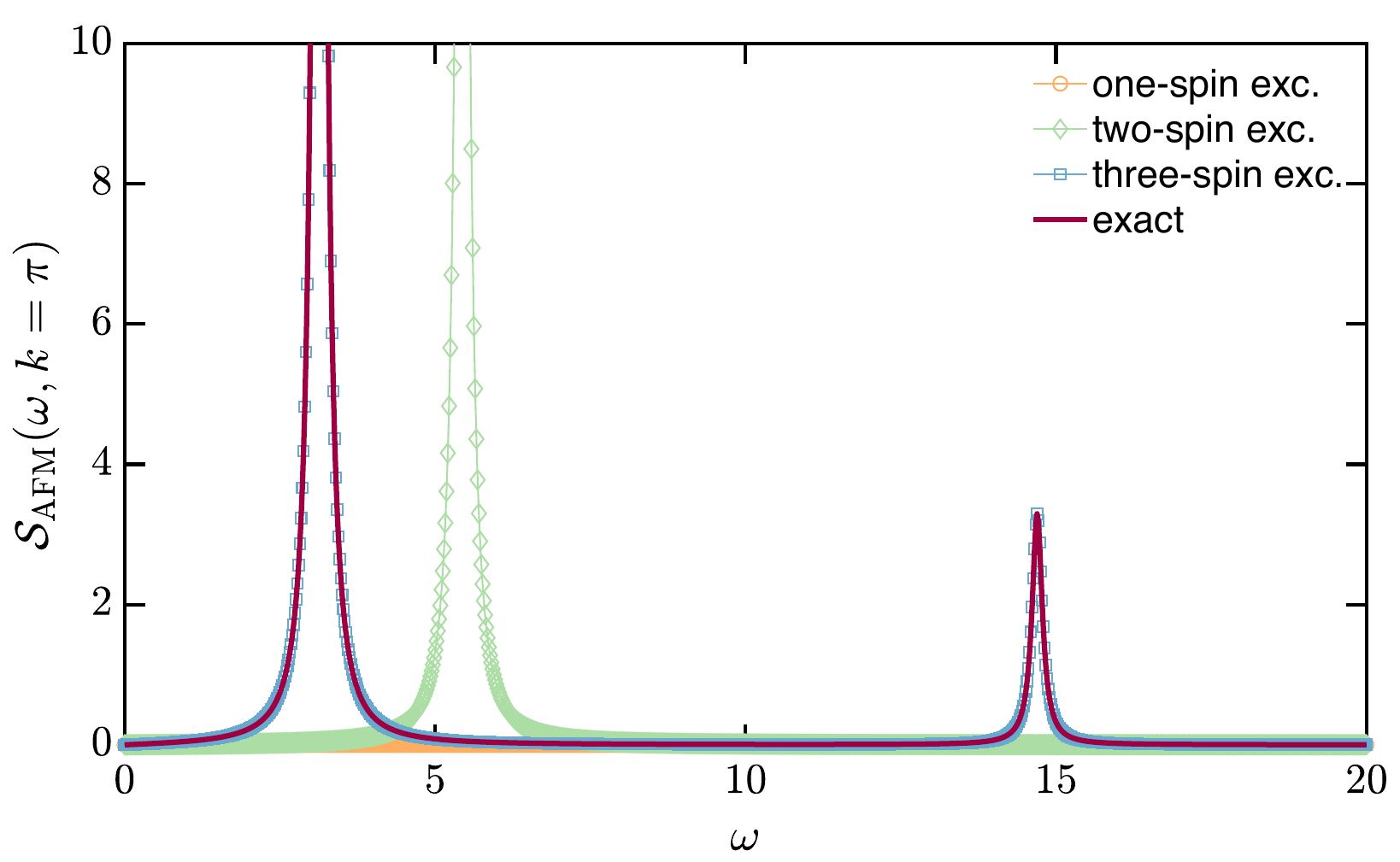}
    \caption{
    To accurately reproduce the exact dynamical structure factor \({\cal S}_{\rm AFM}(\omega, k)\) (red solid line) for the AFM case (\(B_z = 0\), \(J = -1.5\), \(N = 8\)), it is necessary to include, in addition to one-spin excitations, both two- and three-spin excitations, cf. Eq.~\eqref{eqn::AFM_ansatz}.
    Shown is the cut at $k = \pi$.
    }
    \label{fig::AFM_cuts}
\end{figure}

In our numerics, we compute the exact dynamical structure factor (specifically, its symmetrized form with \(\hat{V} = \hat{\mathcal{O}} = \hat{\sigma}_k^+ + \hat{\sigma}_k^-\)) for $N = 8$ spins using exact diagonalization and the Kubo formula (Methods).
The resulting spectrum is discrete and fills the region between the lower and upper branches (we fixed $B_z = 0$), as shown in Fig.~4(e) of the main text. 
Our goal is then to infer the original Hamiltonian from this spectrum alone.

When constructing the variational ansatz to describe small perturbations on top of the exactly computed ground state $\ket{\rm AFM}$, we find it necessary to include not only single-particle excitations, but also two- and three-particle contributions in order to fully capture the momentum-resolved spectra for $N = 8$, as shown in Fig.~\ref{fig::AFM_cuts}:
\begin{align}
    \ket{\psi_k}  = & \, \exp\Big(- i\sum_{k', k''} \theta_{k' k''} (\hat{\sigma}^z_{k'}\hat{\sigma}^z_{k''}\hat{\sigma}^+_{k - k' - k''}  + \text{h.c.}) \Big) \exp\Big(\sum_{k' ,k''} \tilde\theta_{k'k''} (\hat{\sigma}^z_{k'}\hat{\sigma}^z_{k''}\hat{\sigma}^+_{k - k' - k''}  - \text{h.c.}) \Big) \notag\\
    & \times \exp\Big(- i\sum_{k'} \theta_{k'} (\hat{\sigma}^z_{k'}\hat{\sigma}^+_{k - k'}  + \text{h.c.}) \Big) \exp\Big(\sum_{k'} \tilde\theta_{k'} (\hat{\sigma}^z_{k'}\hat{\sigma}^+_{k - k'}  - \text{h.c.}) \Big)
    \notag\\
    & \times \exp\big(-i \alpha_k (\hat\sigma_k^+ + \hat\sigma_k^-)\big) \exp\big(\beta_k (\hat\sigma_k^+ - \hat\sigma_k^-)\big) \ket{\rm AFM}.
    \label{eqn::AFM_ansatz}
\end{align}
For a given momentum \(k\), the number of two-particle (three-particle) operators entering this ansatz scales linearly with \(N\) (quadratically with \(N\)). We expect that as the chain length increases, an even larger set of excitations may be required. In practice, it will often be necessary to truncate the excitation set so that the variational dynamical structure factor retains the key spectral features while remaining computationally tractable.

Our results, shown in the inset of Fig.~4(e) of the main text, demonstrate that the parameters $B_z$ and $J$ converge to their correct target values ($B_z = 0$, $J = -1.5$).  
This highlights the power of SHL to reconstruct a microscopic, translationally invariant Hamiltonian even in cases where the spectrum is complex and non-intuitive.

\subsection{Concluding remarks}
\label{subsec::fin_dyn}

The SHL framework combines quantum state preparation with measurements of Hermitian operators to compute, optimize, and design frequency-resolved dynamical properties, via employing the time-dependent variational principle.
It can enhance a system’s response at a chosen frequency relevant to molecular design, reconstruct a Hamiltonian to match a target spectral profile, and infer translationally invariant models directly from frequency- and momentum-resolved experimental data.

Several natural extensions could further expand the scope of SHL. Incorporating nonlinear response would enable studies of advanced nonlinear optical materials, while integrating quantum master-equation dynamics in place of purely Schr{\"o}dinger evolution would make the method applicable to open quantum systems and driven quantum materials.



\end{document}